\documentclass[twoside,11pt]{article}

\usepackage{jmlr2e}

\usepackage{lastpage}

\ShortHeadings{Model Class Reliance}{Fisher, Rudin, and Dominici}
\firstpageno{1}

\usepackage{array}
\usepackage{bm}
\usepackage{amsmath}
\usepackage{setspace}
\usepackage{xcolor}

\PassOptionsToPackage{normalem}{ulem}
\usepackage{ulem}

\providecommand{\tabularnewline}{\\}

\usepackage{cleveref} 
\usepackage{url}
\usepackage{bbm}
\DeclareMathOperator*{\argmin}{arg\,min}
\DeclareMathOperator*{\argmax}{arg\,max}

\newtheorem{assumption}[theorem]{Assumption}
\newtheorem{condition}[theorem]{Condition}

\newcommand{\mred}[1]{\textcolor{black}{\ensuremath{#1}}}
\numberwithin{equation}{section}

\makeatother

\begin{document}

\title{All Models are Wrong, but \emph{Many} are Useful: Learning a Variable's Importance by Studying an Entire Class of Prediction Models Simultaneously}

\author{\name Aaron Fisher \email afishe27@alumni.jh.edu \\
       \addr Takeda Pharmaceuticals\\
       Cambridge, MA 02139, USA
       \AND
       \name Cynthia Rudin  \email cynthia@cs.duke.edu \\
       \addr Departments of Computer Science and Electrical and Computer Engineering\\
       Duke University\\
       Durham, NC 27708, USA
       \AND
       Francesca Dominici \email fdominic@hsph.harvard.edu \\
       \addr Department of Biostatistics\\
       Harvard T.H. Chan School of Public Health\\
       Boston, MA 02115, USA\\ \\
       (Authors are listed in order of contribution, with highest contribution listed first.)} 

\editor{ }

\maketitle

\begin{abstract}
Variable importance (VI) tools describe how much covariates contribute to a prediction model's accuracy. However, important variables for one well-performing model (for example, a linear model $f(\mathbf{x})=\mathbf{x}^{T}\beta$ with a fixed coefficient vector $\beta$) may be unimportant for another model. In this paper, we propose model class reliance (MCR) as the range of VI values across \emph{all} well-performing model in a prespecified class. Thus, MCR gives a more comprehensive description of importance by accounting for the fact that many prediction models, possibly of different parametric forms, may fit the data well. In the process of deriving MCR, we show several informative results for permutation-based VI estimates, based on the VI measures used in Random Forests. Specifically, we derive connections between permutation importance estimates for a \emph{single} prediction model, U-statistics, conditional variable importance, conditional causal effects, and linear model coefficients. We then give probabilistic bounds for MCR, using a novel, generalizable technique. We apply MCR to a public data set of Broward County criminal records to study the reliance of recidivism prediction models on sex and race. In this application, MCR can be used to help inform VI for unknown, proprietary models.
\end{abstract}


\section{Introduction}

Variable importance (VI) tools describe how much a prediction model's accuracy depends on the information in each covariate. For example, in Random Forests, VI is measured by the decrease in prediction accuracy when a covariate is permuted (\citealp{breiman2001random_forests,breiman2001statistical}; see also \citealp{strobl2008conditional_VI,altmann2010permutation,zhu2015reinforcement_trees,gregorutti2015grouped_VI,datta2016algorithmic_qii,gregorutti2017correlation_LM_VI_RF}). A similar ``Perturb'' VI measure has been used for neural networks, where noise is added to covariates \citep{recknagel1997artificial_perturb_sensitivity,yao1998forecasting,scardi1999developing_perturb_sensitivity,gevrey2003review}. Such tools can be useful for identifying covariates that must be measured with high precision, for improving the transparency of a ``black box'' prediction model (see also \citealp{Rudin19}), or for determining what scenarios may cause the model to fail.

However, existing VI measures do not generally account for the fact that many prediction models may fit the data almost equally well. In such cases, the model used by one analyst may rely on entirely different covariate information than the model used by another analyst. This common scenario has been called the \textquotedblleft Rashomon\textquotedblright{} effect of statistics (\citealp{breiman2001statistical}; see also \citealp{lecue2011interplay,statnikov2013multiple_markov_boundaries,tulabandhula_rudin2014robust_opt,nevo2015minimal_class,letham2016dynamical_systems}). The term is inspired by the 1950 Kurosawa film of the same name, in which four witnesses offer different descriptions and explanations for the same encounter. Under the Rashomon effect, how should analysts give comprehensive descriptions of the importance of each covariate? How well can one analyst recover the conclusions of another? Will the model that gives the best predictions necessarily give the most accurate interpretation? 

To address these concerns, we analyze the \emph{set }of prediction models that provide near-optimal accuracy, which we refer to as a \emph{Rashomon set}. This approach stands in contrast to training to select a \emph{single} prediction model, among a prespecified class of candidate models. Our motivation is that Rashomon sets (defined formally below) summarize the range of effective prediction strategies that an analyst might choose. Additionally, even if the candidate models do not contain the true data generating process, we may hope that some of these models function in similar ways to the data generating process. In particular, we may hope there exist well performing candidate models that place the same importance on a variable of interest as the underlying data generating process does. If so, then studying sets of well-performing models will allow us to deduce information about the data generating process.

Applying this approach to study variable importance, we define \emph{model class reliance} (MCR) as the highest and lowest degree to which any well-performing model within a given class may rely on a variable of interest for prediction accuracy. Roughly speaking, MCR captures the range of explanations, or mechanisms, associated with well-performing models. Because the resulting range summarizes many prediction models simultaneously, rather a single model, we expect this range to be less affected by the choices that an individual analyst makes during the model-fitting process. Instead of reflecting these choices, MCR aims to reflect the nature of the prediction problem itself.

We make several, specific technical contributions in deriving MCR. First, we review a core measure of how much an individual prediction model relies on covariates of interest for its accuracy, which we call \emph{model reliance }(MR). This measure is based on permutation importance measures for Random Forests \citep{breiman2001statistical,breiman2001random_forests}, and can be expanded to describe conditional importance (see Section \ref{sec:Connections-between-MR-causality}, as well as \citealt{strobl2008conditional_VI}). We draw a connection between permutation-based importance estimates (MR) and U-statistics, which facilitates later theoretical results. Additionally, we derive connections between MR, conditional causal effects, and coefficients for additive models.  Expanding on MR, we propose MCR, which generalizes the definition of MR for a \emph{class of models}. We derive finite-sample bounds for MCR, which motivate an intuitive estimator of MCR. Finally, we propose computational procedures for this estimator.

The tools we develop to study Rashomon sets are quite general, and can be used to make finite-sample inferences for arbitrary characteristics of well-performing models. For example, beyond describing variable importance, these tools can describe the range of risk predictions that well-fitting models assign to a particular covariate profile, or the variance of predictions made by well-fitting models. In some cases, these novel techniques may provide finite-sample confidence intervals (CIs) where none have previously existed (see Section \ref{sec:Connection-confidence-CIs}).

MCR and the Rashomon effect become especially relevant in the context of criminal recidivism prediction. Proprietary recidivism risk models trained from criminal records data are increasingly being used in U.S. courtrooms. One concern is that these models may be relying on information that would otherwise be considered unacceptable (for example, race, sex, or proxies for these variables), in order to estimate recidivism risk. The relevant models are often proprietary, and cannot be studied directly. Still, in cases where the predictions made by these models are publicly available, it may be possible to identify alternative prediction models that are sufficiently similar to the proprietary model of interest.

In this paper, we specifically consider the proprietary model COMPAS (Correctional Offender Management Profiling for Alternative Sanctions), developed by the company Northpointe Inc. (subsequently, in 2017, Northpointe Inc., Courtview Justice Solutions Inc., and Constellation Justice Systems Inc. joined together under the name Equivant). Our goal is to estimate how much COMPAS relies on either race, sex, or proxies for these variables not measured in our data set. To this end, we apply a broad class of flexible, kernel-based prediction models to predict COMPAS score. In this setting, the MCR interval reflects the highest and lowest degree to which any prediction model in our class can rely on race and sex while still predicting COMPAS score relatively accurately. Equipped with MCR, we can relax the common assumption of being able to correctly specify the unknown model of interest (here, COMPAS) up to a parametric form. Instead, rather than assuming that the COMPAS model itself is contained in our class, we assume that our class contains at least one well-performing alternative model that relies on sensitive covariates to the same degree that COMPAS does. Under this assumption, the MCR interval will contain the VI value for COMPAS. Applying our approach, we find that race, sex, and their potential proxy variables, are likely not the dominant predictive factors in the COMPAS score (see analysis and discussion in Section \ref{sec:Data-Analysis}).

The remainder of this paper is organized as follows. In Section \ref{sec:notation} we introduce notation, and give a high level summary of our approach, illustrated with visualizations. In Sections \ref{subsec:model-reliance} and \ref{subsec:Model-class-reliance} we formally present MR and MCR respectively, and derive theoretical properties of each. We also review related variable importance practices in the literature, such as retraining a model after removing one of the covariates. In Section \ref{sec:Connection-confidence-CIs}, we discuss general applicability of our approach for determining finite-sample CIs for other problems. In Section \ref{sec:Calculating-MCR}, we present a general procedure for computing MCR. In Section \ref{subsec:linear-additive-interpretation-computation}, we give specific implementations of this procedure for (regularized) linear models, and linear models in a reproducing kernel Hilbert space. We also show that, for additive models, MR can be expressed in terms of the model's coefficients. In Section \ref{sec:Connections-between-MR-causality} we outline connections between MR, causal inference, and conditional variable importance. In Section \ref{sec:Simulations-all}, we illustrate MR and MCR with a simulated toy example, to aid intuition. We also present simulation studies for the task of estimating MR for an unknown, underlying conditional expectation function, under misspecification. We analyze a well-known public data set on recidivism in Section \ref{sec:Data-Analysis}, described above. All proofs are presented in the appendices.

\section{Notation \& Technical Summary\label{sec:notation}}

The label of ``variable importance'' measure has been broadly used to describe approaches for either inference \citep{van2006statistical_inf_VI,diaz2015vi,williamson2017nonparametric_VI} or prediction. While these two goals are highly related, we primarily focus on how much prediction models rely on covariates to achieve accuracy. We use terms such as \textquotedblleft model reliance\textquotedblright{} rather than \textquotedblleft importance\textquotedblright{} to clarify this context.

In order to evaluate how much prediction models rely on variables, we now introduce notation for random variables, data, classes of prediction models, and loss functions for evaluating predictions. Let $Z=(Y,X_{1},X_{2})\in\mathcal{Z}$ be a random variable with outcome $Y\in\mathcal{Y}$ and covariates $X=(X_{1},X_{2})\in\mathcal{X}$, where the covariate subsets $X_{1}\in\mathcal{X}_{1}$ and $X_{2}\in\mathcal{X}_{2}$ may each be multivariate. We assume that observations of $Z$ are $iid$, that $n\geq2$, and that solutions to $\argmin$ and $\argmax$ operations exist whenever optimizing over sets mentioned in this paper (for example, in Theorem \ref{thm:mcr-conserve-bounds}, below). Our goal is to study how much different prediction models rely on $X_{1}$ to predict $Y$. 

We refer to our data set as $\mathbf{Z}=\left[\begin{array}{cc}
\mathbf{y} & \mathbf{X}\end{array}\right]$, a matrix composed of a $n$-length outcome vector $\mathbf{y}$ in the first column, and a $n\times p$ covariate matrix $\mathbf{X}=\left[\begin{array}{cc}
\mathbf{X}_{1} & \mathbf{X}_{2}\end{array}\right]$ in the remaining columns. In general, for a given vector $\mathbf{v}$, let $\mathbf{v}_{[j]}$ denote its $j^{th}$ element(s). For a given matrix $\mathbf{A}$, let $\mathbf{A}'$, $\mathbf{A}_{[i,\cdot]}$, $\mathbf{A}_{[\cdot,j]}$, and $\mathbf{A}_{[i,j]}$ respectively denote the transpose of $\mathbf{A}$, the $i^{th}$ row(s) of $\mathbf{A}$, the $j^{th}$ column(s) of $\mathbf{A}$, and the element(s) in the $i^{th}$ row(s) and $j^{th}$ column(s) of $\mathbf{A}$.

We use the term \emph{model class} to refer to a prespecified subset $\mathcal{F}\subset\{f\mid f:\mbox{\ensuremath{\mathcal{X}\rightarrow\mathcal{Y}}\}}$ of the measurable functions from $\mathcal{X}$ to $\mathcal{Y}$. We refer to member functions $f\in\mathcal{F}$ as \emph{prediction models, }or simply as \emph{models.} Given a model $f$, we evaluate its performance using a nonnegative \emph{loss function} $L:(\mathcal{F}\times\mathcal{Z})\rightarrow\mathbb{R}_{\geq0}$. For example, $L$ may be the squared error loss $L_{\text{se}}(f,(y,x_{1},x_{2}))=(y-f(x_{1},x_{2}))^{2}$ for regression, or the hinge loss $L_{\text{h}}(f,(y,x_{1},x_{2}))=(1-yf(x_{1},x_{2}))_{+}$ for classification. We use the term \emph{algorithm} to refer to any procedure $\mathcal{A}:\mathcal{Z}^{n}\rightarrow\mathcal{F}$ that takes a data set as input and returns a model $f\in\mathcal{F}$ as output.

\subsection{Summary of Rashomon Sets \& Model Class Reliance\label{subsec:Summary-of-Rashomon}}

Many traditional statistical estimates come from descriptions of a \emph{single, }fitted prediction model. In contrast, in this section, we summarize our approach for studying a \emph{set} of near-optimal models. To define this set, we require a prespecified ``reference'' model, denoted by $f_{\text{ref}}$, to serve as a benchmark for predictive performance. For example, $f_{\text{ref}}$ may come from a flowchart used to predict injury severity in a hospital's emergency room, or from another quantitative decision rule that is currently implemented in practice. Given a reference model $f_{\text{ref}}$, we define a\emph{ population} $\epsilon$-\emph{Rashomon set} as  the subset of models with expected loss no more than $\epsilon$ above that of $f_{\text{ref}}$. We denote this set as $\mathcal{R}(\epsilon):=\left\{ f\in\mathcal{F}\,:\,\mathbb{E}L(f,Z)\leq\mathbb{E}L(f_{\text{ref}},Z)+\epsilon\right\} $, where $\mathbb{E}$ denotes expectations with respect to the population distribution. This set can be thought of as representing models that might be arrived at due to differences in data measurement, processing, filtering, model parameterization, covariate selection, or other analysis choices (see Section \ref{subsec:Model-class-reliance}).
\begin{figure}[h]
\begin{centering}
\includegraphics[width=1\columnwidth]{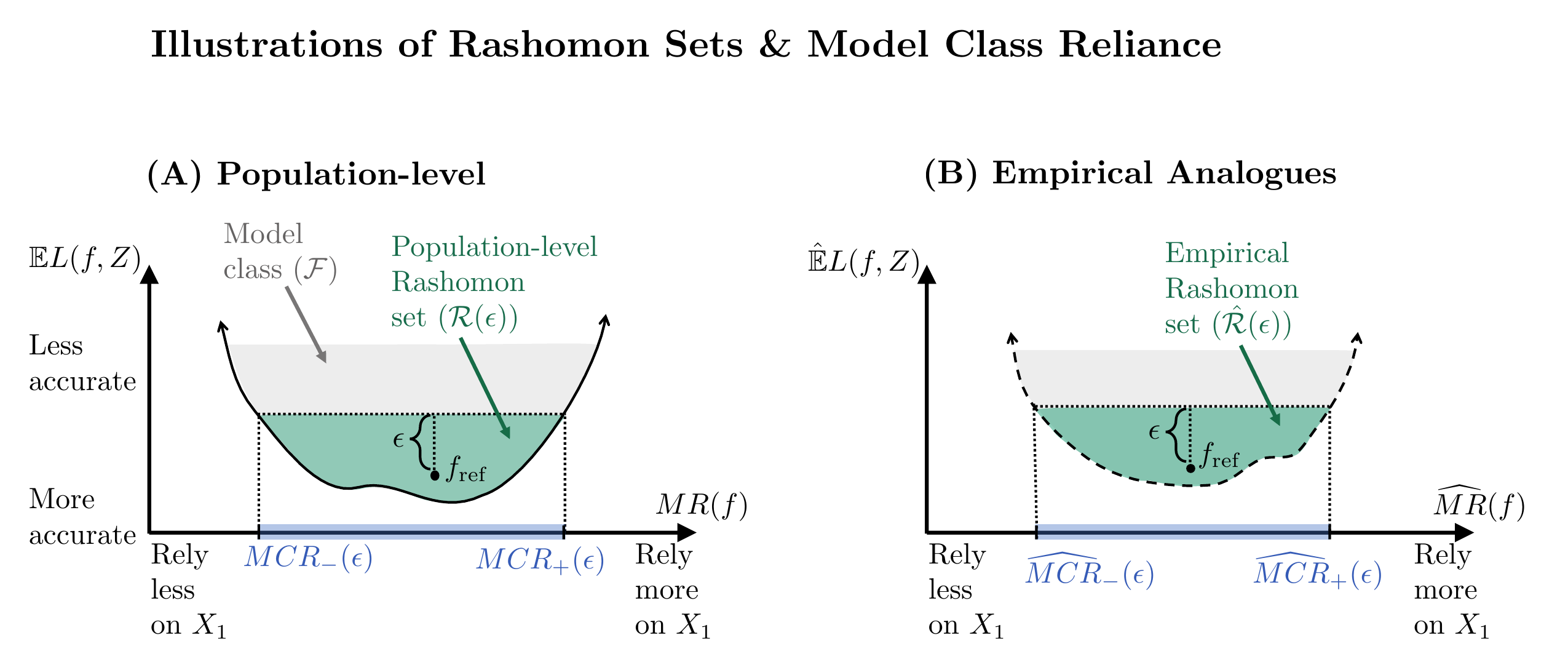}
\par\end{centering}
\raggedright{}\textcolor{black}{\caption{\textcolor{black}{\label{fig:A-Rashomon-set-emp-pop} Rashomon sets and model class reliance \textendash{} Panel (A) illustrates a hypothetical Rashomon set $\mathcal{R}(\epsilon)$, within a model class $\mathcal{F}$. The y-axis shows the expected loss of each model $f\in\mathcal{F}$, and the x-axis shows how much each model $f$ relies on $X_{1}$ (defined formally in Section \ref{subsec:model-reliance}). Along the x-axis, the population-level MCR range is highlighted in blue, showing the values of MR corresponding to well-performing models (see Section \ref{subsec:Model-class-reliance}). Panel (B) shows the in-sample analogue of Panel (A). Here, the y-axis denotes the in-sample loss, $\hat{\mathbb{E}}L(f,Z):=\frac{1}{n}\sum_{i=1}^{n}L(f,\mathbf{Z}_{[i,\cdot]})$; the x-axis shows the empirical model reliance of each model $f\in\mathcal{F}$ on $X_{1}$ (see Section \ref{subsec:model-reliance}); and the highlighted portion of the x-axis shows empirical MCR (see Section \ref{subsec:Model-class-reliance}).}}
}
\end{figure}

Figure \ref{fig:A-Rashomon-set-emp-pop}-A illustrates a hypothetical example of a population $\epsilon$-Rashomon set. Here, the y-axis shows the expected loss of each model $f\in\mathcal{F}$, and the x-axis shows how much each model relies on $X_{1}$ for its predictive accuracy. More specifically, given a prediction model $f$, the x-axis shows the percent increase in $f$'s expected loss when noise is added to $X_{1}$. We refer to this measure as the \emph{model reliance }(MR) of $f$ on $X_{1}$, written informally as
\begin{equation}
MR(f):=\frac{\text{Expected loss of }f\text{ under noise}}{\text{Expected loss of }f\text{ without noise}}.\label{eq:MR-informal-def}
\end{equation}
The added noise must satisfy certain properties, namely, it must render $X_{1}$ completely uninformative of the outcome $Y$, without altering the marginal distribution of $X_{1}$ (for details, see Section \ref{subsec:model-reliance}, as well as \citealp{breiman2001random_forests,breiman2001statistical}).

Our central goal is to understand how much, or how little, models may rely on covariates of interest ($X_{1}$) while still predicting well. In Figure \ref{fig:A-Rashomon-set-emp-pop}-A, this range of possible MR values is shown by the highlighted interval along the x-axis. We refer to an interval of this type as a\emph{ population-level model class reliance }(MCR) range (see Section \ref{subsec:Model-class-reliance}), formally defined as 
\begin{equation}
[MCR_{-}(\epsilon),\,\,MCR_{+}(\epsilon)]:=\left[\min_{f\in\mathcal{R}(\epsilon)}MR(f),\,\,\max_{f\in\mathcal{R}(\epsilon)}MR(f)\right].\label{eq:pop-min-max}
\end{equation}

To estimate this range, we use empirical analogues of the population $\epsilon$-Rashomon set, and of MR, based on observed data (Figure \ref{fig:A-Rashomon-set-emp-pop}-B). We define an \emph{empirical $\epsilon$-Rashomon set} as the set of models with \emph{in-sample} loss no more than $\epsilon$ above that of $f_{\text{ref}}$, and denote this set by $\hat{\mathcal{R}}(\epsilon)$. Informally, we define the \emph{empirical }MR of a model $f$ on $X_{1}$ as 
\begin{equation}
\widehat{MR}(f):=\frac{\text{In-sample loss of }f\text{ under noise}}{\text{In-sample loss of }f\text{ without noise}},\label{eq:informal-MR-hat}
\end{equation}
that is, the extent to which $f$ appears to rely on $X_{1}$ in a given sample (see Section \ref{subsec:model-reliance} for details). Finally, we define the \emph{empirical model class reliance} as the range of empirical MR values corresponding to models with strong in-sample performance (see Section \ref{subsec:Model-class-reliance}), formally written as
\begin{equation}
[\widehat{MCR}_{-}(\epsilon),\,\,\widehat{MCR}_{+}(\epsilon)]:=\left[\min_{f\in\hat{\mathcal{R}}(\epsilon)}\widehat{MR}(f),\,\,\max_{f\in\hat{\mathcal{R}}(\epsilon)}\widehat{MR}(f)\right].\label{eq:sample-min-max}
\end{equation}
In Figure \ref{fig:A-Rashomon-set-emp-pop}-B, the above range is shown by the highlighted portion of the x-axis.

\begin{figure}[h]
\begin{centering}
\includegraphics[width=1\columnwidth]{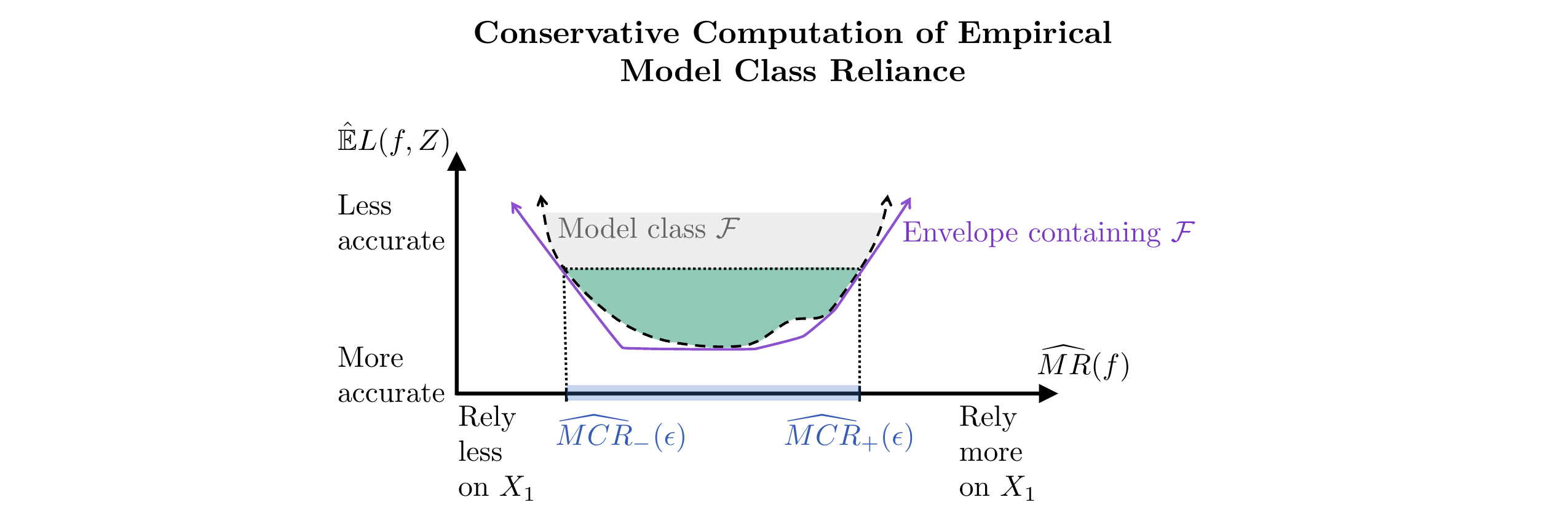}
\par\end{centering}
\raggedright{}\textcolor{black}{\caption{\textcolor{black}{\label{fig:A-Rashomon-set-calc}Illustration of output from our empirical MCR computational procedure \textendash{} Our computation procedure produces a closed-form, convex envelope that contains $\mathcal{F}$ (shown above as the solid, purple line), which bounds empirical MCR for any value of $\epsilon$ (see Eq \ref{eq:sample-min-max}). The procedure works sequentially, tightening these bounds as much as possible near the $\epsilon$ value of interest (Section \ref{sec:Calculating-MCR}). The results from our data analysis (Figure \ref{fig:Broward}) are presented in the same format as the above purple envelope.}}
}
\end{figure}

We make several technical contributions in the process of developing MCR.
\begin{itemize}
\item \textbf{Estimation of MR, and population-level MCR:} Given $f$, we show desirable properties of $\widehat{MR}(f)$ as an estimator of $MR(f)$, using results for U-statistics (Section \ref{subsec:Estimating-MR-u-stats} and Theorem \ref{thm:MR-unif}). We also derive finite sample bounds for population-level MCR, some of which require a limit on the complexity of $\mathcal{F}$ in the form of a covering number. These bounds demonstrate that, under fairly weak conditions, empirical MCR provides a sensible estimate of population-level MCR (see Section \ref{subsec:Model-class-reliance} for details).
\item \textbf{Computation of empirical MCR: }Although empirical MCR is fully determined given a sample, the minimization and maximization in Eq \ref{eq:sample-min-max} require nontrivial computations. To address this, we outline a general optimization procedure for MCR (Section \ref{sec:Calculating-MCR}). We give detailed implementations of this procedure for cases when the model class $\mathcal{F}$ is a set of (regularized) linear regression models, or a set of regression models in a reproducing kernel Hilbert space (Section \ref{subsec:linear-additive-interpretation-computation}). The output of our proposed procedure is a closed-form, convex envelope containing $\mathcal{F}$, which can be used to approximate empirical MCR for any performance level $\epsilon$ (see Figure \ref{fig:A-Rashomon-set-calc} for an illustration). Still, for complex model classes where standard empirical loss minimization is an open problem (for example, neural networks), computing empirical MCR remains an open problem as well.
\item \textbf{Interpretation of MR in terms of model coefficients, and causal effects:} We show that MR for an additive model can be written as a function of the model's coefficients (Proposition \ref{thm:Linear-models}), and that MR for a binary covariate $X_{1}$ can be written as a function of the conditional causal effects of $X_{1}$ on $Y$ (Proposition \ref{thm:VI_TRT}).
\item \textbf{Extensions to conditional importance:} We provide an extension of MR that is analogous to the notion of conditional importance \citep{strobl2008conditional_VI}. This extension describes how much a model relies on the specific information in $X_{1}$ that cannot otherwise be gleaned from $X_{2}$ (Section \ref{subsec:mcr-impute}). 
\item \textbf{Generalizations for Rashomon sets:} Beyond notions of variable importance, we also generalize our finite sample results for MCR to describe arbitrary characterizations of models in a population $\epsilon$-Rashomon set. As we discuss in concurrent work \citep{coker2018hacking_interval}, this generalization is analogous to the profile likelihood interval, and can, for example, be used to bound the range of risk predictions that well-performing prediction models may assign to a particular set of covariates (Section \ref{sec:Connection-confidence-CIs}).
\end{itemize}
We begin in the next section by formally reviewing model reliance.

\section{Model Reliance\label{subsec:model-reliance}}

To formally describe how much the expected accuracy of a fixed prediction model $f$ relies on the random variable $X_{1}$, we use the notion of a ``switched'' loss where $X_{1}$ is rendered uninformative. Throughout this section, we will treat $f$ as a pre-specified prediction model of interest (as in \citealp{hooker2007generalized}). Let $Z^{(a)}=(Y^{(a)},X_{1}^{(a)},X_{2}^{(a)})$ and $Z^{(b)}=(Y^{(b)},X_{1}^{(b)},X_{2}^{(b)})$ be independent random variables, each following the same distribution as $Z=(Y,X_{1},X_{2})$. We define 
\[
e_{\text{switch}}(f):=\mathbb{E}L\{f,(Y^{(b)},X_{1}^{(a)},X_{2}^{(b)})\}
\]
as representing the expected loss of model $f$ across pairs of observations $(Z^{(a)},Z^{(b)})$ in which the values of $X_{1}^{(a)}$ and $X_{1}^{(b)}$ have been switched. To see this interpretation of the above equation, note that we have used the variables $(Y^{(b)},X_{2}^{(b)})$ from $Z^{(b)}$, but we have used the variable $X_{1}^{(b)}$ from an independent copy $Z^{(b)}$. This is why we say that $X_{1}^{(a)}$ and $X_{1}^{(b)}$ have been switched; the values of $(Y^{(b)},X_{1}^{(a)},X_{2}^{(b)})$ do not relate to each other as they would if they had been chosen together. An alternative interpretation of $e_{\text{switch}}(f)$ is as the expected loss of $f$ when noise is added to $X_{1}$ in such a way that $X_{1}$ becomes completely uninformative of $Y$, but that the marginal distribution of $X_{1}$ is unchanged.

As a reference point, we compare $e_{\text{switch}}(f)$ against the standard expected loss when none of the variables are switched, ${e_{\text{orig}}(f):=\mathbb{E}L(f,(Y,X_{1},X_{2}))}$. From these two quantities, we formally define \emph{model reliance} (MR) as the ratio,
\begin{equation}
MR(f):=\frac{e_{\text{switch}}(f)}{e_{\text{orig}}(f)},\label{eq:MR-formal-def}
\end{equation}
as we alluded to in Eq \ref{eq:MR-informal-def}. Higher values of $MR(f)$ signify greater reliance of $f$ on $X_{1}$. For example, an $MR(f)$ value of 2 means that the model relies heavily on $X_{1}$, in the sense that its loss doubles when $X_{1}$ is scrambled. An $MR(f)$ value of 1 signifies no reliance on $X_{1}$, in the sense that the model's loss does not change when $X_{1}$ is scrambled. Models with reliance values strictly less than 1 are more difficult to interpret, as they rely less on the variable of interest than a random guess. Interestingly, it is possible to have models with reliance less than one. For instance, a model $f'$ may satisfy $MR(f')<1$ if it treats $X_{1}$ and $Y$ as positively correlated when they are in fact negatively correlated. However, in many cases, the existence of a model $f'\in\mathcal{F}$ satisfying $MR(f')<1$ implies the existence of another, better performing model $f''\in\mathcal{F}$ satisfying $MR(f'')=1$ and $e_{\text{orig}}(f'')\leq e_{\text{orig}}(f')$. That is, although models may exist with MR values less than 1, they will typically be suboptimal (see Appendix \ref{subsec:Model-reliance-leq-1}).

Model reliance could alternatively be defined as a difference rather than a ratio, that is, as $MR_{\text{difference}}(f):=e_{\text{switch}}(f)-e_{\text{orig}}(f)$. In Appendix \ref{sec:Discussion---Ratios}, we discuss how many of our results remain similar under either definition. 

\subsection{Estimating Model Reliance with U-statistics, and Connections to Permutation-based Variable Importance\label{subsec:Estimating-MR-u-stats}}

Given a model $f$ and data set $\mathbf{Z}=\left[\begin{array}{cc}
\mathbf{y} & \mathbf{X}\end{array}\right]$, we estimate $MR(f)$ by separately estimating the numerator and denominator of Eq \ref{eq:MR-formal-def}. We estimate $e_{\text{orig}}(f)$ with the standard empirical loss,
\begin{equation}
\hat{e}_{\text{orig}}(f):=\frac{1}{n}\sum_{i=1}^{n}L\{f,(\mathbf{y}_{[i]},\mathbf{X}_{1[i,\cdot]},\mathbf{X}_{2[i,\cdot]})\}.\label{eq:e-orig-hat}
\end{equation}
We estimate $e_{\text{switch}}(f)$ by performing a ``switch'' operation across all observed pairs, as in
\begin{equation}
\hat{e}_{\text{switch}}(f):=\frac{1}{n(n-1)}\sum_{i=1}^{n}\sum_{j\neq i}L\{f,(\mathbf{y}_{[j]},\mathbf{X}_{1[i,\cdot]},\mathbf{X}_{2[j,\cdot]})\}.\label{eq:e-perm-def}
\end{equation}

Above, we have aggregated over all possible combinations of the observed values for $(Y,X_{2})$ and for $X_{1}$, excluding pairings that are actually observed in the original sample. If the summation over all possible pairs (Eq \ref{eq:e-perm-def}) is computationally prohibitive due to sample size, another estimator of $e_{\text{switch}}(f)$ is
\begin{align}
\hat{e}_{\text{divide}}(f): & =\frac{1}{2\lfloor n/2\rfloor}\sum_{i=1}^{\lfloor n/2\rfloor}\left[L\{f,(\mathbf{y}_{[i]},\mathbf{X}_{1[i+\lfloor n/2\rfloor,\cdot]},\mathbf{X}_{2[i,\cdot]})\}\right.\label{eq:e-divide-def-part1}\\
 & \hspace{2.5cm}\left.+L\{f,(\mathbf{y}_{[i+\lfloor n/2\rfloor]},\mathbf{X}_{1[i,\cdot]},\mathbf{X}_{2[i+\lfloor n/2\rfloor,\cdot]})\}\right].\label{eq:e-divide-def}
\end{align}
Here, rather than summing over all pairs, we divide the sample in half. We then match the first half's values for $(Y,X_{2})$ with the second half's values for $X_{1}$ (Line \ref{eq:e-divide-def-part1}), and vice versa (Line \ref{eq:e-divide-def}). All three of the above estimators (Eqs \ref{eq:e-orig-hat}, \ref{eq:e-perm-def} \& \ref{eq:e-divide-def}) are unbiased for their respective estimands, as we discuss in more detail shortly.

Finally, we can estimate $MR(f)$ with the plug-in estimator 
\begin{equation}
\widehat{MR}(f):=\frac{\hat{e}_{\text{switch}}(f)}{\hat{e}_{\text{orig}}(f)},\label{eq:MR-switch-empirical}
\end{equation}
which we define as the \emph{empirical model reliance} of $f$ on $X_{1}$. In this way, we formalize the empirical MR definition in Eq \ref{eq:informal-MR-hat}.

Again, our definition of empirical MR is very similar to the permutation-based variable importance approach of \citet{breiman2001random_forests}, where \citeauthor{breiman2001random_forests} uses a single random permutation and we consider all possible pairs. To compare these two approaches more precisely, let $\{\bm{\pi}_{1},\dots,\bm{\pi}_{n!}\}$ be a set of $n$-length vectors, each containing a different permutation of the set $\{1,\dots,n\}$. The approach of \citet{breiman2001random_forests} is analogous to computing the loss ${\sum_{i=1}^{n}L\{f,(\mathbf{y}_{[i]},\mathbf{X}_{1[\bm{\pi}_{l[i]},\cdot]},\mathbf{X}_{2[i,\cdot]})\}}$ for a randomly chosen permutation vector $\bm{\pi}_{l}\in\{\bm{\pi}_{1},\dots,\bm{\pi}_{n!}\}$. Similarly, our calculation in Eq \ref{eq:e-perm-def} is proportional to the sum of losses over all possible $(n!)$ permutations, excluding the $n$ unique combinations of the rows of $\mathbf{X}_{1}$ and the rows of $\left[\begin{array}{cc}
\mathbf{X}_{2} & \mathbf{y}\end{array}\right]$ that appear in the original sample (see Appendix \ref{subsec:all-permutations}). Excluding these observations is necessary to preserve the (finite-sample) unbiasedness of $\hat{e}_{\text{switch}}(f)$.

The estimators $\hat{e}_{\text{orig}}(f)$, $\hat{e}_{\text{switch}}(f)$ and $\hat{e}_{\text{divide}}(f)$ all belong to the well-studied class of U-statistics. Thus, under fairly minor conditions, these estimators are unbiased, asymptotically normal, and have finite-sample probabilistic bounds (\citealp{hoeffding1948class_u_stats,hoeffding1963_bounded_sums_inequalities,serfling1980approximation}; see also \citealp{delong1988comparing} for an early use of U-statistics in machine learning, as well as caveats in \citealp{demler2012misuse}). To our knowledge, connections between permutation-based importance and U-statistics have not been previously established. 

While the above results from U-statistics depend on the model $f$ being fixed a priori, we can also leverage these results to create \emph{uniform} bounds on the MR estimation error for all models in a sufficiently regularized class $\mathcal{F}$. We formally present this bound in Section \ref{subsec:Model-class-reliance} (Theorem \ref{thm:MR-unif}), after introducing required conditions on model class complexity. The existence of this uniform bound implies that it is feasible to train a model and to evaluate its importance using the \emph{same data}. This differs from the classical VI approach of Random Forests \citep{breiman2001random_forests}, which avoids in-sample importance estimation. There, each tree in the ensemble is fit on a random subset of data, and VI for the tree is estimated using the held-out data. The tree-specific VI estimates are then aggregated to obtain a VI estimate for the overall ensemble. Although sample-splitting approaches such as this are helpful in many cases, the uniform bound for MR suggests that they are not strictly necessary, depending on the sample size and the complexity of $\mathcal{F}$. 

\subsection{Limitations of Existing Variable Importance Methods\label{sec:Related-Work}}

Several common approaches for variable selection, or for describing relationships between variables, do not necessarily capture a variable's importance. Null hypothesis testing methods may identify a relationship, but do not describe the relationship's strength. Similarly, checking whether a variable is included by a sparse model-fitting algorithm, such as the Lasso \citep{hastie2009elements}, does not describe the extent to which the variable is relied on. Partial dependence plots \citep{breiman2001statistical,hastie2009elements} can be difficult to interpret if multiple variables are of interest, or if the prediction model contains interaction effects.

Another common VI procedure is to run a model-fitting algorithm twice, first on all of the data, and then again after removing $X_{1}$ from the data set. The losses for the two resulting models are then compared to determine the importance, or ``necessity,'' of $X_{1}$ \citep{gevrey2003review}. Because this measure is a function of two prediction models rather than one, it does not measure how much either individual model relies on $X_{1}$. We refer to this approach as measuring empirical \emph{Algorithm Reliance} (AR) on $X_{1}$, as the model-fitting algorithm is the common attribute between the two models. Related procedures were proposed by \citet{breiman2001statistical,breiman2001random_forests}, which measure the sufficiency of $X_{1}$.

As we discuss in Section \ref{subsec:Estimating-MR-u-stats}, the permutation-based VI measure from RFs \citep{breiman2001random_forests,breiman2001statistical} forms the inspiration for our definition of MR. This RF VI measure has been the topic of empirical studies \citep{archer2008empirical,calle2010letter_VI_stability,wang2016experimental}, and several variations of the measure have been proposed \citep{strobl2007bias,strobl2008conditional_VI,altmann2010permutation,hapfelmeier2014new_core_paper}. \citet{mentch2016random_forest_bagg_ustats} use U-statistics to study predictions of ensemble models fit to subsamples, similar to the bootstrap aggregation used in RFs. Procedures related to ``Mean Difference Impurity,'' another VI measure derived for RFs, have been studied theoretically by \citet{louppe2013understanding,kazemitabar2017_VI_stump}. All of this literature focuses on VI measures for RFs, for ensembles, or for individual trees. Our estimator for model reliance differs from the traditional RF VI measure \citep{breiman2001random_forests} in that we permute inputs to the overall model, rather than permuting the inputs to each individual ensemble member. Thus, our approach can be used generally, and is not limited to trees or ensemble models.

Outside of the context of RF VI, \citet{zhu2015reinforcement_trees} propose an estimand similar to our definition of model reliance, and \citet{gregorutti2015grouped_VI,gregorutti2017correlation_LM_VI_RF} propose an estimand analogous to $e_{\text{switch}}(f)-e_{\text{orig}}(f)$. These recent works focus on the model reliance of $f$ on $X_{1}$ specifically when $f$ is equal to the conditional expectation function of $Y$ (that is, $f(x_{1},x_{2})=\mathbb{E}[Y|X_{1}=x_{1},X_{2}=x_{2}]$). In contrast, we consider model reliance for arbitrary prediction models $f$. \citet{datta2016algorithmic_qii} study the extent to which a model's predictions are expected to change when a subset of variables is permuted, regardless of whether the permutation affects a loss function $L$. These VI approaches are specific to a single prediction model, as is MR. In the next section, we consider a more general conception of importance: how much \emph{any} model in a particular set may rely on the variable of interest.

\section{Model Class Reliance\label{subsec:Model-class-reliance}}

Like many statistical procedures, our MR measure (Section \ref{subsec:model-reliance}) produces a description of a \emph{single} predictive model. Given a model with high predictive accuracy, MR describes how much the model's performance hinges on covariates of interest ($X_{1}$). However, there will often be many other models that perform similarly well, and that rely on $X_{1}$ to different degrees. With this notion in mind, we now study how much \emph{any} well-performing model from a prespecified class $\mathcal{F}$ may rely on covariates of interest.

Recall from Section \ref{subsec:Summary-of-Rashomon} that, in order to define a population $\epsilon$-Rashomon set of near-optimal models, we must choose a ``reference'' model $f_{\text{ref}}$ to serve as a performance benchmark. In order to discuss this choice, we now introduce more explicit notation for the population $\epsilon$-Rashomon set, written as 
\begin{equation}
\mathcal{R}(\epsilon,f_{\text{ref}},\mathcal{F}):=\left\{ f\in\mathcal{F}\,:\,e_{\text{orig}}(f)\leq e_{\text{orig}}(f_{\text{ref}})+\epsilon\right\} .\label{eq:explicit-rashomon-def}
\end{equation}
Note that we write $\mathcal{R}(\epsilon,f_{\text{ref}},\mathcal{F})$ and $\mathcal{R}(\epsilon)$ interchangeably when $f_{\text{ref}}$ and $\mathcal{F}$ are clear from context. Similarly, we occasionally write empirical $\epsilon$-Rashomon sets using the more explicit notation $\hat{\mathcal{R}}(\epsilon,f_{\text{ref}},\mathcal{F}):=\left\{ f\in\mathcal{F}\,:\,\hat{e}_{\text{orig}}(f)\leq\hat{e}_{\text{orig}}(f_{\text{ref}})+\epsilon\right\} $, but typically abbreviate these sets as $\hat{\mathcal{R}}(\epsilon)$.

While $f_{\text{ref}}$ could be selected by minimizing the in-sample loss, the theoretical study of $\mathcal{R}(\epsilon,f_{\text{ref}},\mathcal{F})$ is simplified under the assumption that $f_{\text{ref}}$ is prespecified. For example, $f_{\text{ref}}$ may come from a flowchart used to predict injury severity in a hospital's emergency room, or from another quantitative decision rule that is currently implemented in practice. The model $f_{\text{ref}}$ can also be selected using sample splitting. In some cases it may be desirable to fix $f_{\text{ref}}$ equal to the best-in-class model $f^{\star}:=\argmin_{f\in\mathcal{F}}e_{\text{orig}}(f)$, but this is generally infeasible because $f^{\star}$ is unknown. Still, for any $f_{\text{ref}}\in\mathcal{F}$, the Rashomon set $\mathcal{R}(\epsilon,f_{\text{ref}},\mathcal{F})$ defined using $f_{\text{ref}}$ will always be conservative in the sense that it contains the Rashomon set $\mathcal{R}(\epsilon,f^{\star},\mathcal{F})$ defined using $f^{\star}$. 

We can now formalize our definitions of population-level MCR and empirical MCR by simply plugging in our definitions for $MR(f)$ and $\widehat{MR}(f)$ (Section \ref{subsec:model-reliance}) into Eqs \ref{eq:pop-min-max} \& \ref{eq:sample-min-max} respectively. Studying population-level MCR (Eq \ref{eq:pop-min-max}) is the main focus of this paper, as it provides a more comprehensive view of importance than measures from a single model. If $MCR_{+}(\epsilon)$ is low, then \emph{no} well-performing model in $\mathcal{F}$ places high importance on $X_{1}$, and $X_{1}$ can be discarded at low cost regardless of future modeling decisions. If $MCR_{-}(\epsilon)$ is large, then \emph{every} well-performing model in $\mathcal{F}$ must rely substantially on $X_{1}$, and $X_{1}$ should be given careful attention during the modeling process. Here, $\mathcal{F}$ may itself consist of several parametric model forms (for example, all linear models and all decision tree models with less than 6 single-split nodes). We stress that the range $[MCR_{-}(\epsilon),MCR_{+}(\epsilon)]$ does not depend on the fitting algorithm used to select a model $f\in\mathcal{F}$. The range is valid for any algorithm producing models in $\mathcal{F}$, and applies for any $f\in\mathcal{F}$.

In the remainder of this section, we derive finite sample bounds for population-level MCR, from which we argue that empirical MCR provides reasonable estimates of population-level MCR (Section \ref{subsec:Finite-sample-MCR-bounds}). In Appendix \ref{subsec:absolute-rashomon} we consider an alternate formulation of Rashomon sets and MCR where we replace the relative loss threshold in the definition of $\mathcal{R}(\epsilon)$ with an absolute loss threshold. This alternate formulation can be similar in practice, but still requires the specification of a reference function $f_{\text{ref}}$ to ensure that $\mathcal{R}(\epsilon)$ and $\hat{\mathcal{R}}(\epsilon)$ are nonempty.

\subsection{\label{subsec:Finite-sample-MCR-bounds}Motivating Empirical Estimators of MCR by Deriving Finite-sample Bounds}

In this section we derive finite-sample, probabilistic bounds for $MCR_{+}(\epsilon)$ and $MCR_{-}(\epsilon)$. Our results imply that, under minimal assumptions, $\widehat{MCR}_{+}(\epsilon)$ and $\widehat{MCR}_{-}(\epsilon)$ are respectively within a neighborhood of $MCR_{+}(\epsilon)$ and $MCR_{-}(\epsilon)$ with high probability. However, the weakness of our assumptions (which are typical for statistical-learning-theoretic analysis) renders the width of our resulting CIs to be impractically large, and so we use these results only to show conditions under which $\widehat{MCR}_{+}(\epsilon)$ and $\widehat{MCR}_{-}(\epsilon)$ form sensible point estimates. In Sections \ref{sec:Simulations} \& \ref{sec:Data-Analysis}, below, we apply a bootstrap procedure to account for sampling variability.

To derive these results we introduce three bounded loss assumptions, each of which can be assessed empirically. Let $b_{\text{orig}},B_{\text{ind}},B_{\text{ref}},B_{\text{switch}}\in\mathbb{R}$ be known constants.
\begin{assumption}
\label{assu:bnd-ind} (Bounded individual loss) For a given model $f\in\mathcal{F}$, assume that $0\leq L(f,(y,x_{1},x_{2}))\leq B_{\text{ind}}$ for any $(y,x_{1},x_{2})\in(\mathcal{Y}\times\mathcal{X}_{1}\times\mathcal{X}_{2})$.
\end{assumption}

\begin{assumption}
\label{assu:bnd-ref} (Bounded relative loss) For a given model $f\in\mathcal{F}$, assume that $|L(f,(y,x_{1},x_{2}))-L(f_{\text{ref}},(y,x_{1},x_{2}))|\leq B_{\text{ref}}$ for any $(y,x_{1},x_{2})\in\mathcal{Z}$.
\end{assumption}

\begin{assumption}
\label{assu:bnd-avg} (Bounded aggregate loss) For a given model $f\in\mathcal{F}$, assume that $\mathbb{P}\{0<b_{\text{orig}}\leq\hat{e}_{\text{orig}}(f)\}=\mathbb{P}\{\hat{e}_{\text{switch}}(f)\leq B_{\text{switch}}\}=1$.
\end{assumption}

Each assumption is a property of a specific model $f\in\mathcal{F}$. The notation $B_{\text{ind}}$ and $B_{\text{ref}}$ refer to bounds for any individual observation, and the notation $b_{\text{orig}}$ and $B_{\text{switch}}$ refer to bounds on the aggregated loss $L$ in a sample. These boundedness assumptions are central to our finite sample guarantees, shown below. 

Crucially, loss functions $L$ that are unbounded in general may be used so long as $L(f,(y,x_{1},x_{2}))$ is bounded on a particular domain. For example, the squared-error loss can be used if $\mathcal{Y}$ is contained within a known range, and predictions $f(x_{1},x_{2})$ are contained within the same range for $(x_{1},x_{2})\in\mathcal{X}\times\mathcal{X}_{2}$. We give example methods of determining $B_{\text{ind}}$ in Sections \ref{subsec:Upper-bounding-loss-reglm} \& \ref{subsec:Bf-RKHS-kernel}. For Assumption \ref{assu:bnd-avg}, we can approximate $b_{\text{orig}}$ by training a highly flexible model to the data, and setting $b_{\text{orig}}$ equal to half (or any positive fraction) of the resulting cross-validated loss. To determine $B_{\text{switch}}$ we can simply set $B_{\text{switch}}=B_{\text{ind}}$, although this may be conservative. For example, in the case of binary classification models for non-separable groups (see Section \ref{sec:Simulations}), no linear classifier can misclassify all observations, particularly after a covariate is permuted. Thus, it must hold that $B_{\text{ind}}>B_{\text{switch}}$. Similarly, if $f_{\text{ref}}$ satisfies Assumption \ref{assu:bnd-ind}, then $B_{\text{ref}}$ may be conservatively set equal to $B_{\text{ind}}$. If model reliance is redefined as a difference rather than a ratio, then a similar form of the results in this section will apply without Assumption \ref{assu:bnd-avg} (see Appendix \ref{sec:Discussion---Ratios}).

Based on these assumptions, we can create a finite-sample upper bound for $MCR_{+}(\epsilon)$ and lower bound for $MCR_{-}(\epsilon)$. In other words, we create an ``outer'' bound that contains the interval $[MCR_{-}(\epsilon),MCR_{+}(\epsilon)]$ with high probability.
\begin{theorem}
\label{thm:mcr-conserve-bounds}(``Outer'' MCR Bounds) Given a constant $\epsilon\geq0$, let $f_{+,\epsilon}\in\argmax_{\mathcal{R}(\epsilon)}MR(f)$ and $f_{-,\epsilon}\in\argmin_{\mathcal{R}(\epsilon)}MR(f)$ be prediction models that attain the highest and lowest model reliance among models in $\mathcal{R}(\epsilon)$. If $f_{+,\epsilon}$ and $f_{-,\epsilon}$ satisfy Assumptions \ref{assu:bnd-ind}, \ref{assu:bnd-ref} \& \ref{assu:bnd-avg}, then
\begin{align}
\mathbb{P}\left(MCR_{+}(\epsilon)>\ensuremath{\widehat{MCR}_{+}(\epsilon_{\text{out}})}+\mathcal{Q}_{\text{out}}\right) & \leq\delta,\text{ and}\label{eq:result-for-f+}\\
\mathbb{P}\left(MCR_{-}(\epsilon)<\ensuremath{\widehat{MCR}_{-}(\epsilon_{\text{out}})}-\mathcal{Q}_{\text{out}}\right) & \leq\delta,\label{eq:result_for_f-}
\end{align}

where $\epsilon_{\text{out}}:=\epsilon+2B_{\text{ref}}\sqrt{\frac{\log(3\delta^{-1})}{2n}}$, and $\mathcal{Q}_{\text{out}}:=\frac{B_{\text{switch}}}{b_{\text{orig}}}-\frac{B_{\text{switch}}-B_{\text{ind}}\sqrt{\frac{\log(6\delta^{-1})}{n}}}{b_{\text{orig}}+B_{\text{ind}}\sqrt{\frac{\log(6\delta^{-1})}{2n}}}$.
\end{theorem}

Eq \ref{eq:result-for-f+} states that, with high probability, $MCR_{+}(\epsilon)$ is no higher than $\ensuremath{\widehat{MCR}_{+}(\epsilon_{\text{out}})}$ added to an error term $\mathcal{Q}_{\text{out}}$. As $n$ increases, $\epsilon_{\text{out}}$ approaches $\epsilon$ and $\mathcal{Q}_{\text{out}}$ approaches zero. One practical implication is that, roughly speaking, if $\widehat{MCR}_{+}(\epsilon)\approx\widehat{MCR}_{+}(\epsilon_{\text{out}})$, then the empirical estimator $\widehat{MCR}_{+}(\epsilon)$ is unlikely to substantially underestimate $MCR_{+}(\epsilon)$. By similar reasoning, we can conclude from Eq \ref{eq:result_for_f-} that if $\widehat{MCR}_{-}(\epsilon)\approx\widehat{MCR}_{-}(\epsilon_{\text{out}})$, then $\widehat{MCR}_{-}(\epsilon)$ is unlikely to substantially overestimate $MCR_{-}(\epsilon)$. By setting $\epsilon=0$, Theorem \ref{thm:mcr-conserve-bounds} can also be used to create a finite-sample bound for the reliance of the unique (unknown) best-in-class model on $X_{1}$ (see Corollary \ref{cor:best-in-class} in Appendix \ref{subsec:Bound-for-MR-f-star}), although describing individual models is not the main focus of this paper.

\begin{figure}
\begin{centering}
\textcolor{black}{\includegraphics{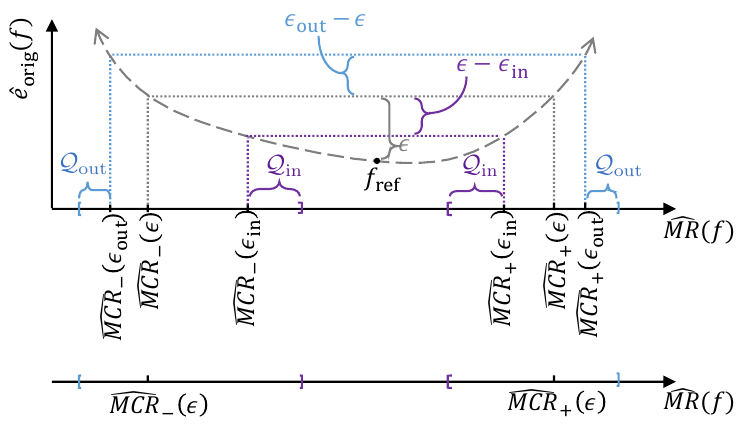}}
\par\end{centering}
\raggedright{}\textcolor{black}{\caption{\textcolor{black}{\label{fig:Illustration-of-terms}Illustration of
terms in }Theorems \ref{thm:mcr-conserve-bounds} and \ref{thm:mcr-consistent}
\textendash{} Above we show the relation between empirical MR (x-axis)
and empirical loss (y-axis) for models $f$ in a hypothetical model
class $\mathcal{F}$. We mark $f_{\text{ref}}$ by the black point.
For each possible model reliance value $r\geq0$, the curved, dashed
line shows the lowest possible empirical loss for a function in $f\in\mathcal{F}$
satisfying $\widehat{MR}(f)=r$. The set $\hat{\mathcal{R}}(\epsilon)$
contains all models in $\mathcal{F}$ within the dotted gray lines.
To create the bounds from Theorem \ref{thm:mcr-conserve-bounds},
we expand the empirical $\epsilon$-Rashomon set by increasing $\epsilon$
to $\epsilon_{\text{out}}$, such that $f_{+,\epsilon}$ (or $f_{-,\epsilon}$)
is contained in $\mathcal{\hat{R}}(\epsilon_{\text{out}})$ with high
probability. We then add (or subtract) $\mathcal{Q}_{\text{out}}$
to account for estimation error of $\widehat{MR}(f_{+,\epsilon})$
(or $\widehat{MR}(f_{-,\epsilon})$). These steps are illustrated
above in blue, with the final bounds shown by the blue bracket symbols
along the x-axis. \textcolor{black}{To create the bounds }for $MCR_{+}(\epsilon)$
(and $MCR_{-}(\epsilon)$)\textcolor{black}{{} in }Theorem \ref{thm:mcr-consistent},
we constrict the empirical $\epsilon$-Rashomon set by decreasing
$\epsilon$ to $\epsilon_{\text{in}}$, such that all models with
high expected loss are simultaneously excluded from $\hat{\mathcal{R}}(\epsilon_{\text{in}})$
with high probability. We then subtract (or add) $\mathcal{Q}_{\text{in}}$
to simultaneously account for MR estimation error for models in $\mathcal{\hat{R}}(\epsilon_{\text{in}})$.
These steps are illustrated above in purple, with the final bounds
shown by the purple bracket symbols along the x-axis. For emphasis,
below this figure we show a copy of the x-axis with selected annotations,
from which it is clear that $\widehat{MCR}_{-}(\epsilon)$ and $\widehat{MCR}_{+}(\epsilon)$
are always within the bounds produced by Theorems \ref{thm:mcr-conserve-bounds}
and \ref{thm:mcr-consistent}. With high probability, $\widehat{MCR}_{-}(\epsilon)$
and $\widehat{MCR}_{+}(\epsilon)$ are within a neighborhood of $MCR_{-}(\epsilon)$
and $MCR_{+}(\epsilon)$ respectively.}
}
\end{figure}

We provide a visual illustration of Theorem \ref{thm:mcr-conserve-bounds} in Figure \ref{fig:Illustration-of-terms}. A brief sketch of the proof is as follows. First, we enlarge the empirical $\epsilon$-Rashomon set by increasing $\epsilon$ to $\epsilon_{\text{out}}$, such that, by Hoeffding's inequality, $f_{+,\epsilon}\in\hat{\mathcal{R}}(\epsilon_{\text{out}})$ with high probability. When $f_{+,\epsilon}\in\hat{\mathcal{R}}(\epsilon_{\text{out}})$, we know that $\ensuremath{\widehat{MR}(f_{+,\epsilon})}\leq\widehat{MCR}_{+}(\epsilon_{\text{out}})$ by the definition of $\widehat{MCR}_{+}(\epsilon_{\text{out}})$. Next, the term $\mathcal{Q}_{\text{out}}$ leverages finite-sample results for U-statistics to account for estimation error of $MR(f_{+,\epsilon})=MCR_{+}(\epsilon)$ when using the estimator $\widehat{MR}(f_{+,\epsilon})$. Thus, we can relate $\ensuremath{\widehat{MR}(f_{+,\epsilon})}$ to both $\widehat{MCR}_{+}(\epsilon_{\text{out}})$ and $\ensuremath{MCR_{+}(\epsilon)}$ in order to obtain Eq \ref{eq:result-for-f+}. Similar steps can be applied to obtain Eq \ref{eq:result_for_f-}.

The bounds in Theorem \ref{thm:mcr-conserve-bounds} naturally account for potential overfitting without an explicit limit on model class complexity (such as a covering number, Rademacher complexity, or VC dimension). Instead, these bounds depend on being able to fully optimize MR across sets in the form of $\hat{\mathcal{R}}(\epsilon)$. If we allow our model class $\mathcal{F}$ to become more flexible, then the size of $\hat{\mathcal{R}}(\epsilon)$ will also increase. Because the bounds in Theorem \ref{thm:mcr-conserve-bounds} result from optimizing over $\hat{\mathcal{R}}(\epsilon)$, increasing the size of $\hat{\mathcal{R}}(\epsilon)$ results in wider, more conservative bounds. In this way, Eqs \ref{eq:result-for-f+} and \ref{eq:result_for_f-} implicitly capture model class complexity.

So far, Theorem \ref{thm:mcr-conserve-bounds} lets us bound the range of MR values corresponding to models that predict well, but it does not tell us whether these bounds are actually attained. Similarly, we can conclude from Theorem \ref{thm:mcr-conserve-bounds} that $[MCR_{-}(\epsilon),MCR_{+}(\epsilon)]$ is unlikely to exceed the estimated range $[\widehat{MCR}_{-}(\epsilon),\widehat{MCR}_{+}(\epsilon)]$ by a substantial margin, but we cannot determine whether this estimated range is unnecessarily wide. For example, consider the models that drive the $\widehat{MCR}_{+}(\epsilon)$ estimator: the models with strong in-sample accuracy, and high empirical reliance on $X_{1}$. These models' in-sample performance could merely be the result of overfitting, in which case they do not tell us direct information about $\mathcal{R}(\epsilon)$. Alternatively, even if all of these models truly do perform well on expectation (that is, even if they are contained in $\mathcal{R}(\epsilon)$), the model with the highest empirical reliance on $X_{1}$ may merely be the model for which our empirical MR estimate contains the most error. Either of these scenarios can cause $\widehat{MCR}_{+}(\epsilon)$ to be unnecessarily high, relative to $MCR_{+}(\epsilon)$.

Fortunately, both problematic scenarios are solved by requiring a limit on the complexity of $\mathcal{F}$. We propose a complexity measure in the form of a covering number, which allows us control a worst case scenario of either overfitting or MR estimation error. Specifically, we define the set of functions $\mathcal{G}_{r}$ as an \emph{$r$-margin-expectation-cover }if for any $f\in\mathcal{F}$ and any distribution $D$, there exists $g\in\mathcal{G}_{r}$ such that
\begin{equation}
\mathbb{E}_{Z\sim D}\left|L\left(f,Z\right)-L\left(g,Z\right)\right|\leq r.\label{eq:cover-number-def}
\end{equation}
We define the \emph{covering number} $\mathcal{N}(\mathcal{F},r)$ to be the size of the smallest $r$-margin-expectation-cover for $\mathcal{F}$. In general, we use $\mathbb{P}_{V\sim D}$ and $\mathbb{E}_{V\sim D}$ to denote probabilities and expectations with respect to a random variable $V$ following the distribution $D$. We abbreviate these quantities accordingly when $V$ or $D$ are clear from context, for example, as $\mathbb{P}_{D}$, $\mathbb{P}_{V}$, or simply $\mathbb{P}$. Unless otherwise stated, all expectations and probabilities are taken with respect to the (unknown) population distribution.

We first show that this complexity measure allows us to control the worst case MR estimation error, that is, the covering number $\mathcal{N}(\mathcal{F},r)$ provides a uniform bound on the error of $\widehat{MR}(f)$ for all $f\in\mathcal{F}$.
\begin{theorem}
\label{thm:MR-unif} (Uniform bound for $\widehat{MR}$) Given $r>0$, if Assumptions \ref{assu:bnd-ind} and \ref{assu:bnd-avg} hold for all $f\in\mathcal{F}$, then 
\[
\mathbb{P}\left[\sup_{f\in\mathcal{F}}\left|\widehat{MR}(f)-MR(f)\right|>q(\delta,r,n)\right]\leq\delta,
\]
 where
\begin{equation}
q(\delta,r,n):=\frac{B_{\text{switch}}}{b_{\text{orig}}}-\frac{B_{\text{switch}}-\left\{ B_{\text{ind}}\sqrt{\frac{\log(4\delta^{-1}\mathcal{N}\left(\mathcal{F},r\sqrt{2}\right))}{n}}+2r\sqrt{2}\right\} }{b_{\text{orig}}+\left\{ B_{\text{ind}}\sqrt{\frac{\log(4\delta^{-1}\mathcal{N}\left(\mathcal{F},r\right))}{2n}}+2r\right\} }.\label{eq:q-func-def}
\end{equation}
\end{theorem}

Theorem \ref{thm:MR-unif} states that, with high probability, the largest possible estimation error for $MR(f)$ across all models in $\mathcal{F}$ is bounded by $q(\delta,r,n$), which can be made arbitrarily small by increasing $n$ and decreasing $r$. As we noted in Section \ref{subsec:Estimating-MR-u-stats}, this means that it is possible to train a model and estimate its reliance on variables without using sample-splitting.

The covering number $\mathcal{N}(\mathcal{F},r)$ can also be used to limit the extent of overfitting (see Appendix \ref{subsec:Proof-of-other-cover-limits}). As a result, it is possible to set an in-sample performance threshold low enough so that it will only be met by models with strong expected performance (that is, by models truly within $\mathcal{R}(\epsilon)$). To implement this idea of a stricter performance threshold, we contract the empirical $\epsilon$-Rashomon set by subtracting a buffer term from $\epsilon$. This requires that we generalize the definition of an empirical $\epsilon$-Rashomon set to $\hat{\mathcal{R}}(\epsilon,f_{\text{ref}},\mathcal{F}):=\left\{ f_{\text{ref}}\right\} \cup\left\{ f\in\mathcal{F}\,:\,\hat{e}_{\text{orig}}(f)\leq\hat{e}_{\text{orig}}(f_{\text{ref}})+\epsilon\right\} $ for $\epsilon\in\mathbb{R}$, where the explicit inclusion of $f_{\text{ref}}$ now ensures that $\hat{\mathcal{R}}(\epsilon,f_{\text{ref}},\mathcal{F})$ is nonempty, even for $\epsilon<0$. As before, we typically omit the notation $f_{\text{ref}}$ and $\mathcal{F}$, writing $\hat{\mathcal{R}}(\epsilon)$ instead.

We are now prepared to answer the questions of whether the bounds from Theorem \ref{thm:mcr-conserve-bounds} are actually attained, and of whether the estimated range $[\widehat{MCR}_{-}(\epsilon),\widehat{MCR}_{+}(\epsilon)]$ is unnecessarily wide. Our answer comes in the form of an upper bound on $MCR_{-}(\epsilon)$, and a lower bound on $MCR_{+}(\epsilon)$. 
\begin{theorem}
\label{thm:mcr-consistent}(``Inner'' MCR Bounds) Given constants $\epsilon\geq0$ and $r>0$, if Assumptions \ref{assu:bnd-ind}, \ref{assu:bnd-ref} and \ref{assu:bnd-avg} hold for all $f\in\mathcal{F}$, and then
\begin{align}
\mathbb{P}\left(MCR_{+}(\epsilon)<\widehat{MCR}_{+}\left(\epsilon_{\text{in}}\right)-\mathcal{Q}_{\text{in}}\right) & \leq\delta,\text{ and}\label{eq:result-inner-for-f+}\\
\mathbb{P}\left(MCR_{-}(\epsilon)>\ensuremath{\widehat{MCR}_{-}\left(\epsilon_{\text{in}}\right)}+\mathcal{Q}_{\text{in}}\right) & \leq\delta,\label{eq:result_inner_for_f-}
\end{align}

where $\epsilon_{\text{in}}:=\epsilon-2B_{\text{ref}}\sqrt{\frac{\log(4\delta^{-1}\mathcal{N}\left(\mathcal{F},r\right))}{2n}}-2r$, and $\mathcal{Q}_{\text{in}}=q\left(\frac{\delta}{2},r,n\right)$, as defined in Eq \ref{eq:q-func-def}.
\end{theorem}

Theorem \ref{thm:mcr-consistent} can allow us to infer an ``inner'' bound that is contained within the interval $[MCR_{-}(\epsilon),MCR_{+}(\epsilon)]$ with high probability. In Figure \ref{fig:Illustration-of-terms}, we illustrate the result of Theorem \ref{thm:mcr-consistent}, and give a sketch of the proof. This proof follows a similar structure to that of Theorem \ref{thm:mcr-conserve-bounds}, but incorporates Theorem \ref{thm:MR-unif}'s uniform bound on MR estimation error ($\mathcal{Q}_{\text{in}}$ term), as well as an additional uniform bound on the probability that any model has in-sample loss too far from its expected loss ($\epsilon_{\text{in}}$ term).

A practical implication of Theorem \ref{thm:mcr-consistent} is that, roughly speaking, if $\widehat{MCR}_{+}(\epsilon_{\text{in}})\approx\widehat{MCR}_{+}(\epsilon)$, then it is unlikely for the empirical estimator $\widehat{MCR}_{+}(\epsilon)$ to substantially underestimate $MCR_{+}(\epsilon)$. Taken together with Theorem \ref{thm:mcr-conserve-bounds}, we can conclude that, if $\widehat{MCR}_{+}(\epsilon_{\text{in}})\approx\widehat{MCR}_{+}(\epsilon_{\text{out}})$, then the estimator $\widehat{MCR}_{+}(\epsilon)$ is unlikely either to overestimate or to underestimate $MCR_{+}(\epsilon)$ by very much. In large samples, it may be plausible to expect the condition $\widehat{MCR}_{+}(\epsilon_{\text{in}})\approx\widehat{MCR}_{+}(\epsilon_{\text{out}})$ to hold, since $\epsilon_{\text{in}}$ and $\epsilon_{\text{out}}$ both approach $\epsilon$ as $n$ increases. In the same way, if $\widehat{MCR}_{-}(\epsilon_{\text{in}})\approx\widehat{MCR}_{-}(\epsilon_{\text{out}})$, we can conclude from Eqs \ref{eq:result_for_f-} \& \ref{eq:result_inner_for_f-} that the empirical estimator $\widehat{MCR}_{-}(\epsilon)$ is unlikely to either overestimate or underestimate $MCR_{-}(\epsilon)$ by very much. For this reason, we argue that $\widehat{MCR}_{-}(\epsilon)$ and $\widehat{MCR}_{+}(\epsilon)$ form sensible estimates of population-level MCR \textendash{} each is contained within a neighborhood of its respective estimand, with high probability. The secondary x-axis of Figure \ref{fig:Illustration-of-terms} gives an illustration of this argument.

\section{Extensions of Rashomon Sets Beyond Variable Importance\label{sec:Connection-confidence-CIs}}

In this section we generalize the Rashomon set approach beyond the study of MR. In Section \ref{subsec:Finite-sample-CIs-general}, we create finite-sample CIs for other summary characterizations of near-optimal, or best-in-class models. The generalization also helps to illustrate a core aspect of the argument underlying Theorem \ref{thm:mcr-conserve-bounds}: models with near-optimal performance in the population tend to have relatively good performance in random samples.

In Section \ref{subsec:Related-Literature-on-Rashomon}, we review existing literature on near-optimal models.

\subsection{Finite-sample Confidence Intervals from Rashomon Sets\label{subsec:Finite-sample-CIs-general}}

Rather than describing how much a model relies on $X_{1}$, here we assume the analyst is interested in an arbitrary characteristic of a model. We denote this characteristic of interest as $\phi:\mathcal{F}\rightarrow\mathbb{R}$. For example, if $f_{\bm{\beta}}$ is the linear model $f_{\bm{\beta}}(x)=x'\bm{\beta}$, then $\phi$ may be defined as the norm of the associated coefficient vector (that is, $\phi(f_{\bm{\beta}})=\|\bm{\beta}\|_{2}^{2}$) or the prediction $f_{\bm{\beta}}$ would assign given a specific covariate profile $x_{\text{new}}$ (that is, $\phi(f_{\bm{\beta}})=f_{\bm{\beta}}(x_{\text{new}})$).

Given a descriptor $\phi$, we now show a general result that allows creation of finite-sample CIs for the best performing models $\mathcal{R}(\epsilon)$. The resulting CIs are themselves based on empirical Rashomon sets. 
\begin{proposition}
\label{rem:simplified-thm3-general}(Finite sample CIs from Rashomon sets) Let $\epsilon':=\epsilon+2B_{\text{ref}}\sqrt{\frac{\log(2\delta^{-1})}{2n}}$, let $\hat{\phi}_{-}(\epsilon'):=\min_{f\in\hat{\mathcal{R}}\left(\epsilon'\right)}\phi(f)$ and let $\hat{\phi}_{+}(\epsilon'):=\max_{f\in\hat{\mathcal{R}}\left(\epsilon'\right)}\phi(f)$.

If Assumption \ref{assu:bnd-ref} holds for all $f\in\mathcal{R}(\epsilon)$, then 

\[
\mathbb{P}\left[\left\{ \phi(f):f\in\mathcal{R}(\epsilon)\right\} \subseteq\left[\hat{\phi}_{-}(\epsilon'),\hat{\phi}_{+}(\epsilon')\right]\right]\geq1-\delta.
\]
 
\end{proposition}

Proposition \ref{rem:simplified-thm3-general} generates a finite-sample CI for the range of values $\phi(f)$ corresponding to well-performing models, $\left\{ \phi(f):f\in\mathcal{R}(\epsilon)\right\} $. This CI, denoted by $\left[\hat{\phi}_{-}(\epsilon'),\hat{\phi}_{+}(\epsilon')\right]$, can itself be interpreted as the range of values $\phi(f)$ corresponding to models $f$ with empirical loss not substantially above that of $f_{\text{ref}}$. Thus, the interval has both a rigorous coverage rate and a coherent in-sample interpretation. The proof of Proposition \ref{rem:simplified-thm3-general} uses Hoeffding's inequality to show that models in $\mathcal{F}$ are contained in $\hat{\mathcal{R}}\left(\epsilon'\right)$ with high probability, that is, that models with good expected performance tend to perform well in random samples.

An immediate corollary of Proposition \ref{rem:simplified-thm3-general} is that we can generate finite-sample CIs for all best-in-class models $f^{\star}\in\arg\min_{f\in\mathcal{F}}\mathbb{E}L(f,Z)$ by setting $\epsilon=0$. This corollary can be further strengthened if a single model $f^{\star}$ is assumed to uniquely minimize $\mathbb{E}L(f,Z)$ over $f\in\mathcal{F}$ (see Appendix \ref{subsec:Proof-of-Propositions-general}).

Note that Proposition \ref{rem:simplified-thm3-general} implicitly assumes that $\phi(f)$ can be determined exactly for any model $f\in\mathcal{F}$, in order for the interval $\left[\hat{\phi}_{-}(\epsilon'),\hat{\phi}_{+}(\epsilon')\right]$ to be precisely determined. This assumption does not hold, for example, if $\phi(f)=MR(f)$, or if $\phi(f)=\text{Var}\{f(X_{1},X_{2})\}$, as these quantities depend on both $f$ and the (unknown) population distribution. In such cases, an additional correction factor must be incorporated to account for estimation error of $\phi(f)$, such as the $\mathcal{Q}_{\text{out}}$ term in Theorem \ref{thm:mcr-conserve-bounds}. 

In concurrent work, \citet{coker2018hacking_interval} show that profile likelihood intervals take the same form as the interval $[\hat{\phi}_{-}(\epsilon'),\hat{\phi}_{+}(\epsilon')]$ in Proposition \ref{rem:simplified-thm3-general}. This means that a profile likelihood interval can also be expressed by minimizing and maximizing over an empirical Rashomon set. More specifically, consider the case where the loss function $L$ is the negative of the \emph{known} log likelihood function, and where $f_{\text{ref}}$ is the maximum likelihood estimate of the ``true model,'' which in this case is $f^{\star}$. If additional minor assumptions are m\textcolor{black}{et (see Appendix \ref{subsec:Rashomon-sets-and-profile} for details)}, then the $(1-\delta)$-level profile likelihood interval for $\phi(f^{\star})$ is equal to $[\hat{\phi}_{-}(\frac{\chi_{1,1-\delta}}{2n}),\hat{\phi}_{+}(\frac{\chi_{1,1-\delta}}{2n})]$, where $\hat{\phi}_{-}$ and $\hat{\phi}_{+}$ are defined as in Proposition \ref{rem:simplified-thm3-general}, and $\chi_{1,1-\delta}$ is the $1-\delta$ percentile of a chi-square distribution with 1 degree of freedom. 

Relative to a profile likelihood approach, the advantage of Proposition \ref{rem:simplified-thm3-general} is that it does not require asymptotics, it does not require that the likelihood be known up to a parametric form, and it can be extended to study the \emph{set} of near-optimal prediction models $\mathcal{R}(\epsilon)$, rather than a single, potentially misspecified prediction model $f^{\star}$. This is especially useful when different near-optimal models accurately describe different aspects of the underlying data generating process, but none capture it completely. The disadvantage of Proposition \ref{rem:simplified-thm3-general} is that the required performance threshold of $\epsilon'=\epsilon+2B_{\text{ref}}\sqrt{\frac{\log(2\delta^{-1})}{2n}}$ decreases more slowly than the performance threshold of $\frac{\chi_{1,1-\delta}}{2n}$ required in a profile likelihood interval. Because our results from Section \ref{subsec:Finite-sample-MCR-bounds} carry a similar disadvantage, we use these results primarily to motivate point estimates describing the Rashomon set $\mathcal{R}(\epsilon)$.

Still, it is worth emphasizing the generality of Proposition \ref{rem:simplified-thm3-general}. Through this result, Rashomon sets allow us to reframe a wide set of finite-sample inference problems as in-sample optimization problems. The implied CIs are not necessarily in closed form, but the approach still opens an exciting pathway for deriving non-asymptotic results. For example, they imply that existing methods for profile likelihood intervals might be able to be reapplied to achieve finite-sample results. For highly complex model classes where profile likelihoods are difficult to compute, such as neural networks or random forests, approximate inference is sometimes achieved via approximate optimization procedures (for example, Markov chain Monte Carlo for Bayesian additive regression trees, in \citealp{chipman2010bart}). Proposition \ref{rem:simplified-thm3-general} shows that similar approximate optimization methods \emph{could be repurposed to establish approximate, finite-sample inferences for the same model classes.}

\subsection{Related Literature on the Rashomon Effect\label{subsec:Related-Literature-on-Rashomon}}

\citet{breiman2001statistical} introduced the ``Rashomon effect'' of statistics as a problem of ambiguity: if many models fit the data well, it is unclear which model we should try to interpret. Breiman suggests that the ensembling many well-performing models together can resolve this ambiguity, as the new ensemble model may perform better than any of its individual members. However, this approach may only push the problem from the member level to the ensemble level, as there may also be many different ensemble models that fit the data well.

The Rashomon effect has also been considered in several subject areas outside of VI, including those in non-statistical academic disciplines \citep{heider1988rashomon,roth2002rashomon}. \citet{tulabandhula_rudin2014robust_opt} optimize a decision rule to perform well under the predicted range of outcomes from any well-performing model. \citet{statnikov2013multiple_markov_boundaries} propose an algorithm to discover multiple Markov boundaries, that is, minimal sets of covariates such that conditioning on any one set induces independence between the outcome and the remaining covariates. \citet{nevo2015minimal_class} report interpretations corresponding to a set of well-fitting, sparse linear models. \citet{meinshausen2010stability_selection} estimate structural aspects of an underlying model (such as the variables included in that model) based on how stable those aspects are across a set of well-fitting models. This set of well-fitting models is identified by repeating an estimation procedure in a series of perturbed samples, using varying levels of regularization (see also \citealp{azen2001criticality}). \citet{letham2016dynamical_systems} search for a pair of well-fitting dynamical systems models that give maximally different predictions.

\section{Calculating Empirical Estimates of Model Class Reliance \label{sec:Calculating-MCR}}

In this section, we propose a binary search procedure to bound the values of $\widehat{MCR}_{-}(\epsilon)$ and $\widehat{MCR}_{+}(\epsilon)$ (see Eq \ref{eq:sample-min-max}), which respectively serve as estimates of $MCR_{-}(\epsilon)$ and $MCR_{+}(\epsilon)$ (see Section \ref{subsec:Finite-sample-MCR-bounds}). Each step of this search consists of minimizing a linear combination of $\hat{e}_{\text{orig}}(f)$ and $\hat{e}_{\text{switch}}(f)$ across $f\in\mathcal{F}$. Our approach is related to the fractional programming approach of \citet{dinkelbach1967nonlinear_FP}, but accounts for the fact that the problem is constrained by the value of the denominator, $\hat{e}_{\text{orig}}(f)$. We additionally show that, for many model classes, computing $\widehat{MCR}_{-}(\epsilon)$ only requires that we minimize convex combinations of $\hat{e}_{\text{orig}}(f)$ and $\hat{e}_{\text{switch}}(f)$, which is no more difficult than minimizing the average loss over an expanded and reweighted sample (See Eq \ref{eq:h-expand-weighted} \& Proposition \ref{rem:(Convexity-for-}).

\textcolor{black}{Computing $\widehat{MCR}_{+}(\epsilon)$ however will require that we are able to minimize arbitrary linear combinations of $\hat{e}_{\text{orig}}(f)$ and $\hat{e}_{\text{switch}}(f)$. In Section \ref{subsec:Convex-models}, we outline how this can be done for convex model classes \textendash{} classes for which the loss function is convex in the model parameter. Later, in Section \ref{subsec:linear-additive-interpretation-computation}, we give more specific computational procedures}\textcolor{red}{{} }for when $\mathcal{F}$ is the class of linear models, regularized linear models, or linear models in a reproducing kernel Hilbert space (RKHS). We summarize the tractability of computing empirical MCR for different model classes in Table \ref{tab:Tractability}.

\begin{table}
\begin{tabular}{|>{\raggedright}p{0.4\columnwidth}|>{\centering}p{0.25\columnwidth}|>{\centering}p{0.25\columnwidth}|}
\hline 
\textbf{Model class and loss function (}$\mathcal{F}$\textbf{ \& $L$)} & \textbf{Computing $\widehat{MCR}_{-}$} & \textbf{Computing} $\widehat{MCR}_{+}$\tabularnewline
\hline 
(L2 Regularized) Linear models, with the squared error loss & Highly tractable (QP1QC, see Sections \ref{subsec:LM-get-MCR} \& \ref{subsec:Regularized-Linear-Models}) & Highly tractable (QP1QC, see Sections \ref{subsec:LM-get-MCR} \& \ref{subsec:Regularized-Linear-Models})\tabularnewline
\hline 
Linear models in a reproducing kernel Hilbert space, with the squared error loss & Moderately tractable (QP1QC, see Section \ref{subsec:Calculating-MCR-kernel}) & Moderately tractable (QP1QC, see Section \ref{subsec:Calculating-MCR-kernel})\tabularnewline
\hline 
Cases where irrelevant covariates do not improve predictions & Moderately tractable (Convex optimization problems, see Proposition \ref{rem:(Convexity-for-}) & Potentially intractable\tabularnewline
\hline 
Cases where minimizing the empirical loss is a convex optimization problem & Potentially intractable (DC programs, see Section \ref{subsec:Convex-models}) & Potentially intractable (DC programs, see Section \ref{subsec:Convex-models})\tabularnewline
\hline 
\end{tabular}

\caption{\label{tab:Tractability}Tractability of empirical MCR computation for different model classes \textendash{} For each case, we describe the tractability of computing \textbf{$\widehat{MCR}_{-}$} and \textbf{$\widehat{MCR}_{+}$} using our proposed approaches. Computing empirical MCR can be reduced to a sequence of optimization problems, the form of which are noted in parentheses within the above table. }
\end{table}

To simplify notation associated with the reference model $f_{\text{ref}}$, we present our computational results in terms of bounds on empirical MR subject to performance thresholds on the \emph{absolute} scale. More specifically, we present bound functions $b_{-}$ and $b_{+}$ satisfying $b_{-}(\epsilon_{\text{abs}})\leq\widehat{MR}(f)\leq b_{+}(\epsilon_{\text{abs}})$ simultaneously for all $\{f,\epsilon_{\text{abs}}\,:\,\hat{e}_{\text{orig}}(f)\leq\epsilon_{\text{abs}},f\in\mathcal{F},\epsilon_{\text{abs}}>0\}$ (Figures \ref{fig:A-Rashomon-set-calc} \& \ref{fig:Broward} show examples of these bounds). The binary search procedures we propose can be used to tighten these boundaries at a particular value $\epsilon_{\text{abs}}$ of interest.

We briefly note that as an alternative to the global optimization procedures we discuss below, \textcolor{black}{heuristic optimization procedures such as simulated annealing can also prove useful in bounding empirical MCR. By definition, the empirical MR for any model in $\hat{\mathcal{R}}(\epsilon)$ forms a lower bound for $\widehat{MCR}_{+}(\epsilon)$, and an upper bound for $\widehat{MCR}_{-}(\epsilon)$. Heuristic maximization and minimization of empirical MR can be used to tighten these boundaries.}

Throughout this section, we assume that $0<\min_{f\in\mathcal{F}}\hat{e}_{\text{orig}}(f)$, to ensure that MR is finite. 

\subsection{Binary Search for Empirical MR Lower Bound \label{subsec:Binary-Search-MCR-}}

Before describing our binary search procedure, we introduce additional notation used in this section. Given a constant $\gamma\in\mathbb{R}$ and prediction model $f\in\mathcal{F}$, we define the linear combination $\hat{h}_{-,\gamma}$, and its minimizers (for example, $\hat{g}_{-,\gamma,\mathcal{F}}$), as
\begin{align*}
\hat{h}_{-,\gamma}(f) & :=\gamma\hat{e}_{\text{orig}}(f)+\hat{e}_{\text{switch}}(f),\hspace{1em}\text{and}\hspace{1em}\hat{g}_{-,\gamma,\mathcal{F}}\in\argmin_{f\in\mathcal{F}}\hat{h}_{-,\gamma}(f).
\end{align*}
We do not require that $\hat{h}_{-,\gamma}$ is uniquely minimized, and we frequently use the abbreviated notation $\hat{g}_{-,\gamma}$ when $\mathcal{F}$ is clear from context.

Our goal in this section is to derive a lower bound on $\widehat{MR}$ for subsets of $\mathcal{F}$ in the form of $\{f\in\mathcal{F}:\hat{e}_{\text{orig}}(f)\leq\epsilon_{\text{abs}}\}$. We achieve this by minimizing a series of linear objective functions in the form of $\hat{h}_{-,\gamma}$, using a similar method to that of \citet{dinkelbach1967nonlinear_FP}. Often, minimizing the linear combination $\hat{h}_{-,\gamma}(f)$ is more tractable than minimizing the MR ratio directly. 

Almost all of the results shown in this section, and those in Section \ref{subsec:Binary-Search-MCR+}, also hold if we replace $\hat{e}_{\text{switch}}$ with $\hat{e}_{\text{divide}}$ throughout (see Eq \ref{eq:e-divide-def}), including in the definition of $\widehat{MR}$ and $\hat{h}_{-,\gamma}(f)$. The exception is Proposition \ref{rem:(Convexity-for-}, below, which we may still expect to approximately hold if we replace $\hat{e}_{\text{switch}}$ with $\hat{e}_{\text{divide}}$.

Given an observed sample, we define the following condition for a pair of values $\{\gamma,\epsilon_{\text{abs}}\}\in\mathbb{R}\times\mathbb{R}_{>0}$, and argmin function $\hat{g}_{-,\gamma}$:
\begin{condition}
\label{cond:mcr--bound} \emph{(Criteria to continue search for $\widehat{MR}$ lower bound) } $\hat{h}_{-,\gamma}\left(\hat{g}_{-,\gamma}\right)\geq0$ and $\hat{e}_{\text{orig}}(\hat{g}_{-,\gamma})\leq\epsilon_{\text{abs}}$.
\end{condition}

We are now equipped to determine conditions under which we can tractably create a lower bound for empirical MR.
\begin{lemma}
(Lower bound for $\widehat{MR}$)\label{lem:mcr--bound-compute} If $\gamma\in\mathbb{R}$ satisfies $\hat{h}_{-,\gamma}(\hat{g}_{-,\gamma})\geq0$, then 
\begin{equation}
\frac{\hat{h}_{-,\gamma}(\hat{g}_{-,\gamma})}{\epsilon_{\text{abs}}}-\gamma\leq\widehat{MR}(f)\label{eq:lower-bnd-mcr--thm}
\end{equation}
for all $f\in\mathcal{F}$ satisfying $\hat{e}_{\text{orig}}(f)\le\epsilon_{\text{abs}}$. It also follows that 
\[
-\gamma\leq\widehat{MR}(f)\text{\hspace{1cm} for all }f\in\mathcal{F}.
\]

Additionally, if $f=\hat{g}_{-,\gamma}$ and at least one of the inequalities in Condition \ref{cond:mcr--bound} holds with equality, then Eq \ref{eq:lower-bnd-mcr--thm} holds with equality. 
\end{lemma}

Lemma \ref{lem:mcr--bound-compute} reduces the challenge of lower-bounding $\widehat{MR}(f)$ to the task of minimizing the linear combination $\hat{h}_{-,\gamma}(f)$. The result of Lemma \ref{lem:mcr--bound-compute} is not only a single boundary for a particular value of $\epsilon_{\text{abs}}$, but a boundary \emph{function} that holds all values of $\epsilon_{\text{abs}}>0$, with lower values of $\epsilon_{\text{abs}}$ leading to more restrictive lower bounds on $\widehat{MR}(f)$.

\begin{figure}
\includegraphics[scale=0.35]{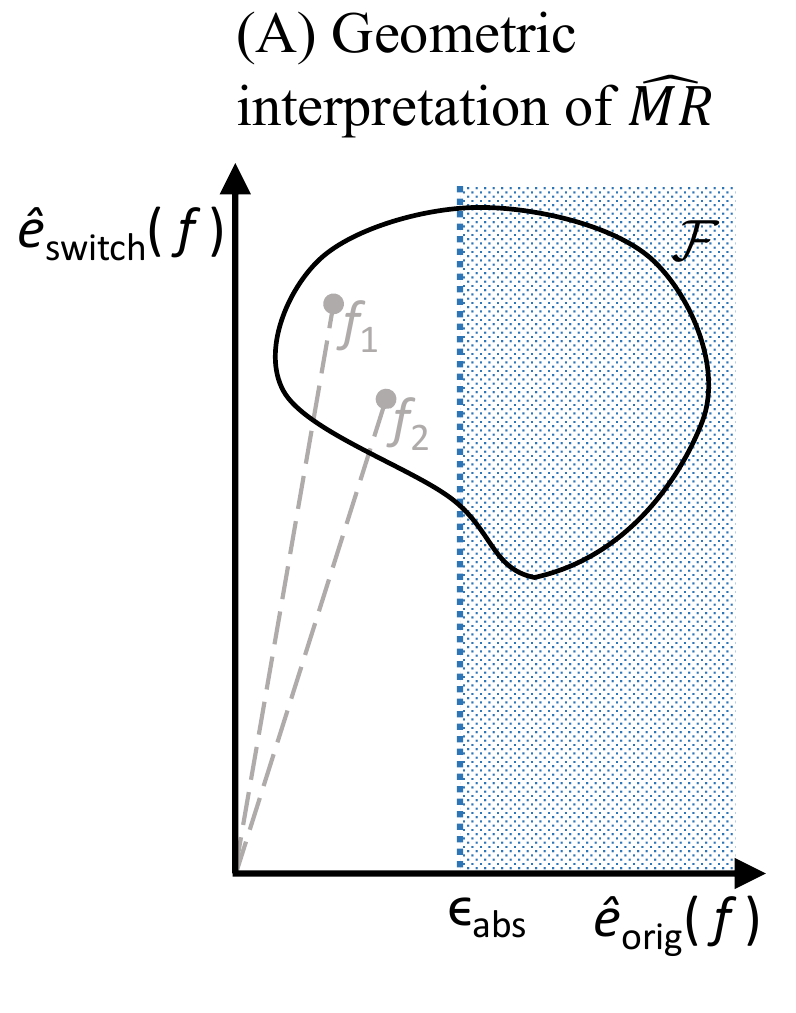}
\includegraphics[scale=0.35]{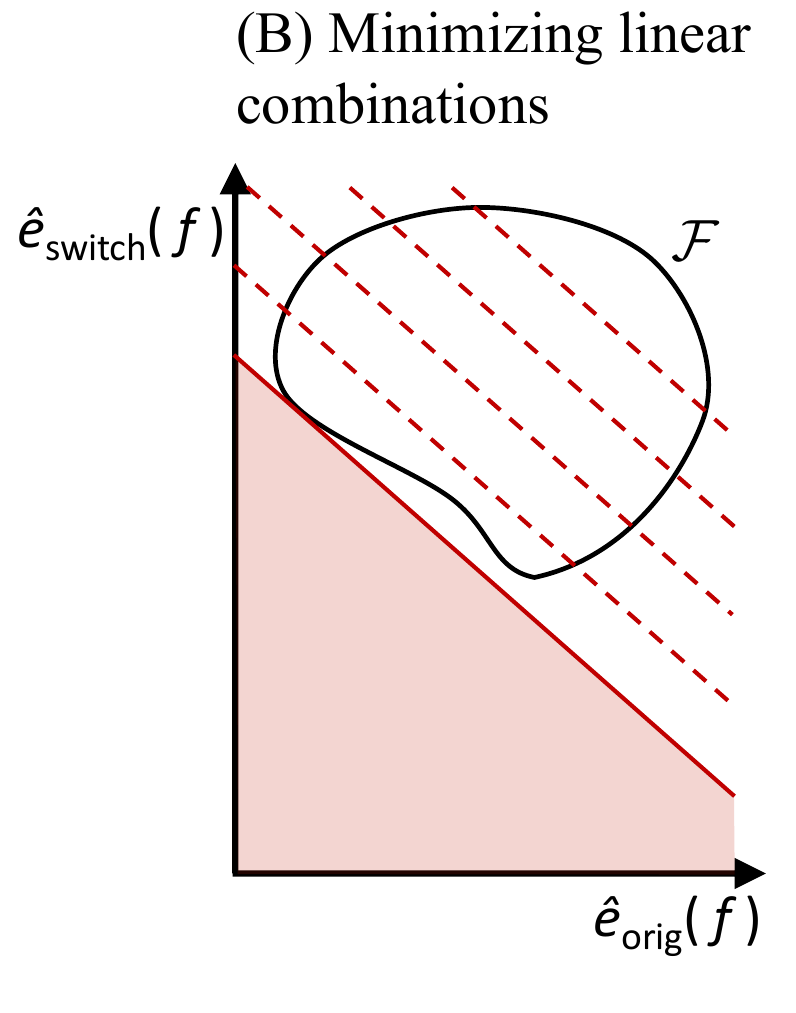}
\includegraphics[scale=0.35]{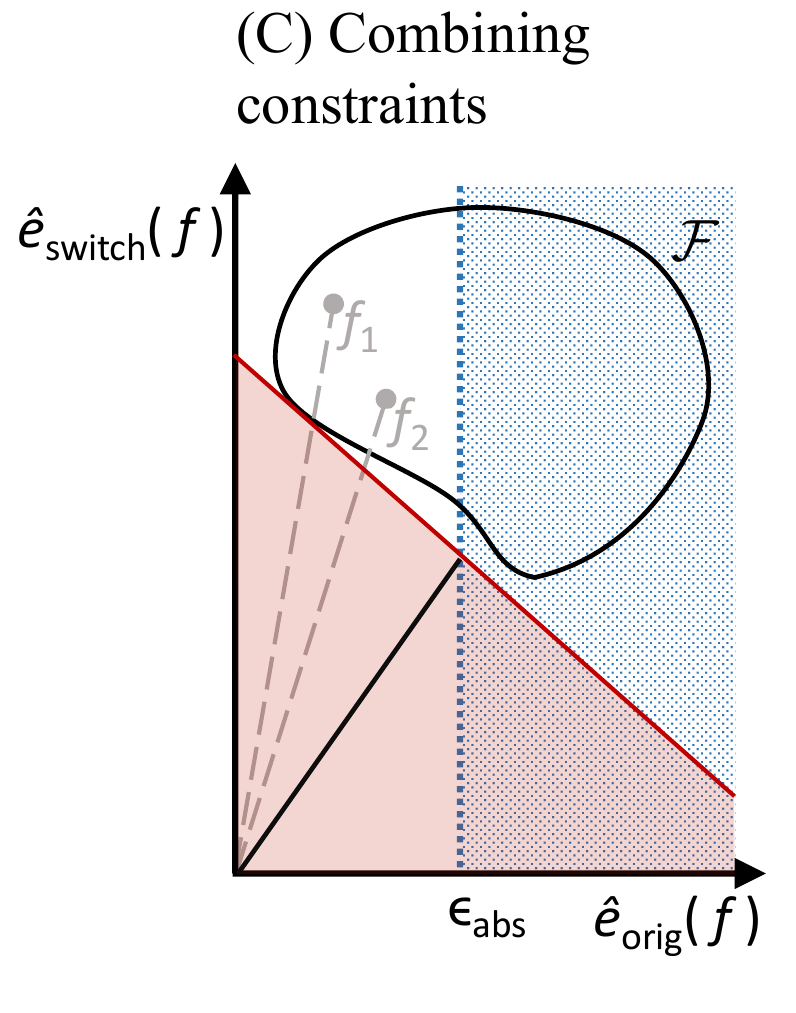}\caption{
Above, we illustrate the geometric intuition for Lemma \ref{lem:mcr--bound-compute}. In Panel (A), we show an example of a hypothetical model class $\mathcal{F}$, marked by the enclosed region. For each model $f\in\mathcal{F}$, the x-axis shows $\hat{e}_{\text{orig}}(f)$ and the y-axis shows $\hat{e}_{\text{switch}}(f)$. Here, we can see that the condition ${\min_{f\in\mathcal{F}}\hat{e}_{\text{orig}}(f)>0}$ holds. The blue dotted region marks models with higher empirical loss. We mark two example models within $\mathcal{F}$ as $f_{1}$ and $f_{2}$. The slopes of the lines connecting the origin to $f_{1}$ and $f_{2}$ are equal to $\widehat{MR}(f_{1})$ and $\widehat{MR}(f_{2})$ respectively. Our goal is to lower-bound the slope corresponding to $\widehat{MR}$ for any model $f$ satisfying $\hat{e}_{\text{orig}}(f)\leq\epsilon_{\text{abs}}$. 
In Panel (B), we consider the linear combination $\hat{h}_{-,\gamma}(f)=\gamma\hat{e}_{\text{orig}}(f)+\hat{e}_{\text{switch}}(f)$ for $\gamma=1$. Above, contour lines of $\hat{h}_{-,\gamma}$ are shown in red. The solid red line indicates the smallest possible value of $\hat{h}_{-,\gamma}$ across $f\in\mathcal{F}$. Specifically, its y-intercept equals $\min_{f\in\mathcal{F}}\hat{h}_{-,\gamma}(f)$. If we can determine this minimum, we can determine a linear border constraint on $\mathcal{F}$, that is, we will know that no points corresponding to models $f\in\mathcal{F}$ may lie in the shaded region above. Additionally, if ${\min_{f\in\mathcal{F}}\hat{h}_{-,\gamma}(f)\geq0}$ (see Lemma \ref{lem:mcr--bound-compute}), then we know that the origin is either excluded by this linear constraint, or is on the boundary.
In Panel (C), we combine the two constraints from Panels (A) \& (B) to see that models $f\in\mathcal{F}$ satisfying $\hat{e}_{\text{orig}}(f)\le\epsilon_{\text{abs}}$ must correspond to points in the white, unshaded region above. Thus, as long as the unshaded region does not contain the origin, any line connecting the origin to the a model $f$ satisfying $\hat{e}_{\text{orig}}(f)\le\epsilon_{\text{abs}}$ (for example, here, $f_{1}$,$f_{2}$) must have a slope at least as high as that of the solid black line above. It can be shown algebraically that the black line has slope equal to the left-hand side of Eq \ref{eq:lower-bnd-mcr--thm}. Thus the left-hand side of Eq \ref{eq:lower-bnd-mcr--thm} is a lower bound for $\widehat{MR}(f)$ for all $\left.\{f\in\mathcal{F}:\hat{e}_{\text{orig}}(f)\le\epsilon_{\text{abs}}\}\right.$.\label{fig:-bin-search}}
\end{figure}

In addition to the formal proof for Lemma \ref{lem:mcr--bound-compute}, we provide a heuristic illustration of the result in Figure \ref{fig:-bin-search}, to aid intuition.

It remains to determine which value of $\gamma$ should be used in Eq \ref{eq:lower-bnd-mcr--thm}. The following lemma implies that this value can be determined by a binary search, given a particular value of interest for $\epsilon_{\text{abs}}$.
\begin{lemma}
(Monotonicity for $\widehat{MR}$ lower bound binary search)\label{lem:mcr--compute-mono}  The following monotonicity results hold:
\begin{enumerate}
\item \label{enu:thm-h-mono}$\hat{h}_{-,\gamma}(\hat{g}_{-,\gamma})$ is monotonically increasing in $\gamma$.
\item \label{enu:thm-e0--1}$\hat{e}_{\text{orig}}(\hat{g}_{-,\gamma})$ is monotonically decreasing in $\gamma$.
\item \label{enu:thm-lowerbnd-mono}Given $\epsilon_{\text{abs}}$, the lower bound from Eq \ref{eq:lower-bnd-mcr--thm}, $\left\{ \frac{\hat{h}_{-,\gamma}(\hat{g}_{-,\gamma})}{\epsilon_{\text{abs}}}-\gamma\right\} $, is monotonically decreasing in $\gamma$ in the range where $\hat{e}_{\text{orig}}(\hat{g}_{-,\gamma})\leq\epsilon_{\text{abs}}$, and increasing otherwise.
\end{enumerate}
\end{lemma}

\begin{figure}[t]
\centering{}\includegraphics[scale=0.35]{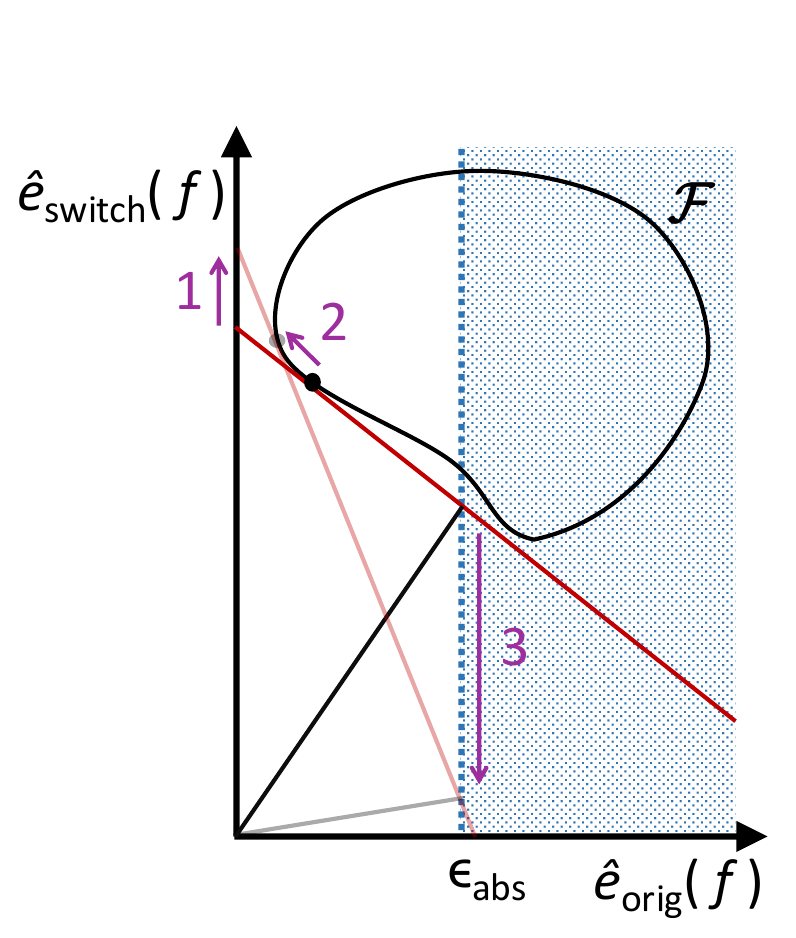}\caption{Monotonicity for binary search. Above we show a version of Figure \ref{fig:-bin-search}-C for two alternative values of $\gamma$. This figure is meant to add intuition for the monotonicity results in Lemma \ref{lem:mcr--compute-mono}, in addition to the formal proof. Increasing $\gamma$ is equivalent to \emph{decreasing} the slope of the red line in Figure \ref{fig:-bin-search}-C. We define two values $\gamma_{1}<\gamma_{2}$, where $\gamma_{1}$ corresponds to the solid red line, above, and $\gamma_{2}$ corresponds to the semi-transparent red line. The y-intercept values of these lines are equal to $\hat{h}_{-,\gamma_{1}}(\hat{g}_{-,\gamma_{1}})$ and $\hat{h}_{-,\gamma_{2}}(\hat{g}_{-,\gamma_{2}})$ respectively (see Figure \ref{fig:-bin-search}-C caption). The solid and semi-transparent black dots mark $\hat{g}_{-,\gamma_{1}}$ and $\hat{g}_{-,\gamma_{2}}$ respectively. Plugging $\gamma_{1}$ and $\gamma_{2}$ into Eq \ref{eq:lower-bnd-mcr--thm} yields two lower bounds for $\widehat{MR}$, marked by the slopes of the solid and semi-transparent black lines respectively (see Figure \ref{fig:-bin-search}-C caption). We see that (1) $\hat{h}_{-,\gamma_{1}}(\hat{g}_{-,\gamma_{1}})\leq\hat{h}_{-,\gamma_{2}}(\hat{g}_{-,\gamma_{2}})$, that (2) $\hat{e}_{\text{orig}}(\hat{g}_{-,\gamma_{1}})\geq\hat{e}_{\text{orig}}(\hat{g}_{-,\gamma_{2}})$, and that (3) the left-hand side of Eq \ref{eq:lower-bnd-mcr--thm} is decreasing in $\gamma$ when $\hat{e}_{\text{orig}}(\hat{g}_{-,\gamma})\leq\epsilon_{\text{abs}}$. These three conclusions are marked by arrows in the above figure, with numbering matching the enumerated list in Lemma \ref{lem:mcr--compute-mono}. \label{fig:-bin-search-DD}}
\end{figure}

Given a particular performance level of interest, $\epsilon_{\text{abs}}$, Point \ref{enu:thm-lowerbnd-mono} of Lemma \ref{lem:mcr--compute-mono} tells us that the value of $\gamma$ resulting in the tightest lower bound from Eq \ref{eq:lower-bnd-mcr--thm} occurs when $\gamma$ is as low as possible while still satisfying Condition \ref{cond:mcr--bound}. Points \ref{enu:thm-h-mono} and \ref{enu:thm-e0--1} show that if $\gamma_{0}$ satisfies Condition \ref{cond:mcr--bound}, and one of the equalities in Condition \ref{cond:mcr--bound} holds with equality, then Condition \ref{cond:mcr--bound} holds for all $\gamma\geq\gamma_{0}$. Together, these results imply that we can use a binary search to determine the value of $\gamma$ to be used in Lemma \ref{lem:mcr--bound-compute}, reducing this value until Condition \ref{cond:mcr--bound} is no longer met. In addition to the formal proof for Lemma \ref{lem:mcr--compute-mono}, we provide an illustration of the result in Figure \ref{fig:-bin-search-DD} to aid intuition.

Next we present simple conditions under which the binary search for values of $\gamma$ can be restricted to the nonnegative real line. This result substantially extends the computational tractability of our approach, as minimizing $\hat{h}_{-,\gamma}$ for $\gamma\geq0$ is equivalent to minimizing a reweighted empirical loss over an expanded sample of size $n^{2}$:
\begin{align}
\hat{h}_{-,\gamma}(f) & =\gamma\hat{e}_{\text{orig}}(f)+\hat{e}_{\text{switch}}(f)=\sum_{i=1}^{n}\sum_{j=1}^{n}w_{\gamma}(i,j)L\{f,(\mathbf{y}_{[i]},\mathbf{X}_{1[j,\cdot]},\mathbf{X}_{2[i,\cdot]})\},\label{eq:h-expand-weighted}
\end{align}
where $w_{\gamma}(i,j)=\frac{\gamma1(i=j)}{n}+\frac{1(i\neq j)}{n(n-1)}\geq0$. 
\begin{proposition}
\label{rem:(Convexity-for-}(Nonnegative weights for $\widehat{MR}$ lower bound binary search) Assume that $L$ and $\mathcal{F}$ satisfy the following conditions.
\begin{enumerate}
\item \label{cond:(Predictions-are-sufficient}(Predictions are sufficient for computing the loss) The loss $L\{f,(Y,X_{1},X_{2})\}$ depends on the covariates $(X_{1},X_{2})$ only via the prediction function $f$, that is, $L\{f,(y,x_{1}^{(a)},x_{2}^{(a)})\}=L\{f,(y,x_{1}^{(b)},x_{2}^{(b)})\}$ whenever $f(x_{1}^{(a)},x_{2}^{(a)})=f(x_{1}^{(b)},x_{2}^{(b)})$. 
\item \label{cond:independence}(Irrelevant information does not improve predictions) For any distribution $D$ satisfying $X_{1}\perp_{D}(X_{2},Y)$, there exists a function $f_{D}$ satisfying
\[
\mathbb{E}_{D}L\{f_{D},(Y,X_{1},X_{2})\}=\min_{f\in\mathcal{F}}\mathbb{E}_{D}L\{f,(Y,X_{1},X_{2})\},
\]
and
\begin{equation}
f_{D}(x_{1}^{(a)},x_{2})=f_{D}(x_{1}^{(b)},x_{2})\text{ for any }x_{1}^{(a)},x_{1}^{(b)}\in\mathcal{X}_{1}\text{ and }x_{2}\in\mathcal{X}_{2}.\label{eq:fD-unaffected-by-X1}
\end{equation}
\end{enumerate}
Let $\gamma=0$. Under the above assumptions, it follows that either (i) there exists a function $\hat{g}_{-,0}$ minimizing $\hat{h}_{-,0}$ that does not satisfy Condition \ref{cond:mcr--bound}, or (ii) $\hat{e}_{\text{orig}}(\hat{g}_{-,0})\leq\epsilon_{\text{abs}}$ and $\widehat{MR}(g_{-,0})\leq1$ for any function $\hat{g}_{-,0}$ minimizing $\hat{h}_{-,0}$.
\end{proposition}

The implication of Proposition \ref{rem:(Convexity-for-} is that, when the conditions of Proposition \ref{rem:(Convexity-for-} are met, the search region for $\gamma$ can be limited to the nonnegative real line, and minimizing $\hat{h}_{-,\gamma}$ will be no harder than minimizing a reweighted empirical loss over an expanded sample (Eq \ref{eq:h-expand-weighted}). To see this, recall that for a fixed value of $\epsilon_{\text{abs}}$ we can tighten the boundary in Lemma \ref{lem:mcr--bound-compute} by conducting a binary search for the smallest value of $\gamma$ that satisfies Condition \ref{cond:mcr--bound}. If setting $\gamma$ equal to $0$ does not satisfy Condition \ref{cond:mcr--bound}, and the search for $\gamma$ can be restricted to the nonnegative real line, where minimizing $\hat{h}_{-,0}$ is more tractable (see Eq \ref{eq:h-expand-weighted}). Alternatively, if $\hat{e}_{\text{orig}}(g_{-,0})\leq\epsilon_{\text{abs}}$ and $\widehat{MR}(g_{-,0})\leq1$, then we have identified a well-performing model $g_{-,0}$ with empirical MR no greater than 1. For $\epsilon_{\text{abs}}=\hat{e}_{\text{orig}}(f_{\text{ref}})+\epsilon$, this implies that $\widehat{MCR}_{-}(\epsilon)\leq1$, which is a sufficiently precise conclusion for most interpretational purposes (see Appendix \ref*{subsec:Model-reliance-leq-1} ).

Because of the fixed pairing structure used in $\hat{e}_{\text{divide}}$, Proposition \ref{rem:(Convexity-for-} will not necessarily hold if we replace $\hat{e}_{\text{switch}}$ with $\hat{e}_{\text{divide}}$ throughout (see Appendix \ref*{subsec:Proof-of-Remark-convexity}). However, since $\hat{e}_{\text{divide}}$ approximates $\hat{e}_{\text{switch}}$, we can expect Proposition \ref{rem:(Convexity-for-} to hold approximately. The bound from Eq \ref{eq:lower-bnd-mcr--thm} still remains valid if we replace $\hat{e}_{\text{switch}}$ with $\hat{e}_{\text{divide}}$ and limit $\gamma$ to the nonnegative reals, although in some cases it may not be as tight.

\subsection{Binary Search for Empirical MR Upper Bound\label{subsec:Binary-Search-MCR+}}

We now briefly present a binary search procedure to upper bound $\widehat{MR}$, which mirrors the procedure from Section \ref{subsec:Binary-Search-MCR-}. Given a constant $\gamma\in\mathbb{R}$ and prediction model $f\in\mathcal{F}$, we define the linear combination $\hat{h}_{+,\gamma}$, and its minimizers (for example, $\hat{g}_{+,\gamma,\mathcal{F}}$), as
\begin{align*}
\hat{h}_{+,\gamma}(f):= & \hat{e}_{\text{orig}}(f)+\gamma\hat{e}_{\text{switch}}(f),\hspace{1em}\text{and}\hspace{1em}\hat{g}_{+,\gamma,\mathcal{F}}\in\argmin_{f\in\mathcal{F}}\hat{h}_{+,\gamma}(f).
\end{align*}
As in Section \ref{subsec:Binary-Search-MCR-}, $\hat{h}_{+,\gamma}$ need not be uniquely minimized, and we generally abbreviate $\hat{g}_{+,\gamma,\mathcal{F}}$ as $\hat{g}_{+,\gamma}$ when $\mathcal{F}$ is clear from context. 

Given an observed sample, we define the following condition for a pair of values $\{\gamma,\epsilon_{\text{abs}}\}\in\mathbb{R}_{\leq0}\times\mathbb{R}_{>0}$, and argmin function $\hat{g}_{+,\gamma}$:
\begin{condition}
\label{cond:mcr+-bound}\emph{ (Criteria to continue search for $\widehat{MR}$ upper bound)} $\hat{h}_{+,\gamma}\left(\hat{g}_{+,\gamma}\right)\geq0$ and $\hat{e}_{\text{orig}}(\hat{g}_{+,\gamma})\leq\epsilon_{\text{abs}}$.
\end{condition}

We can now develop a procedure to upper bound $\widehat{MR}$, as shown in the next lemma.
\begin{lemma}
(Upper bound for $\widehat{MR}$) \label{lem:mcr+-compute-bound} If $\gamma\in\mathbb{R}$ satisfies $\gamma\leq0$ and $\hat{h}_{+,\gamma}(\hat{g}_{+,\gamma})\geq0$, then
\begin{equation}
\widehat{MR}(f)\leq\left\{ \frac{\hat{h}_{+,\gamma}(\hat{g}_{+,\gamma})}{\epsilon_{\text{abs}}}-1\right\} \gamma^{-1}\label{eq:thm-bnd-mcr+}
\end{equation}
for all $f\in\mathcal{F}$ satisfying $\hat{e}_{\text{orig}}(f)\leq\epsilon_{\text{abs}}$. It also follows that 
\begin{equation}
\widehat{MR}(f)\leq|\gamma^{-1}|\hspace{1cm}\text{for all }f\in\mathcal{F}.\label{eq:global-upper-bnd}
\end{equation}

Additionally, if $f=\hat{g}_{+,\gamma}$ and at least one of the inequalities in Condition \ref{cond:mcr+-bound} holds with equality, then Eq \ref{eq:thm-bnd-mcr+} holds with equality.
\end{lemma}

As in Section \ref{subsec:Binary-Search-MCR-}, it remains to determine the value of $\gamma$ to use in Lemma \ref{lem:mcr+-compute-bound}, given a value of interest for $\epsilon_{\text{abs}}\geq\min_{f\in\mathcal{F}}\hat{e}_{\text{orig}}(f)$. The next lemma tells us that the boundary from Lemma \ref{lem:mcr+-compute-bound} is tightest when $\gamma$ is as low as possible while still satisfying Condition \ref{cond:mcr+-bound}.
\begin{lemma}
(Monotonicity for $\widehat{MR}$ upper bound binary search)\label{lem:mcr+-compute-mono} The following monotonicity results hold:
\begin{enumerate}
\item \label{enu:thm-h+-mono}$\hat{h}_{+,\gamma}(\hat{g}_{+,\gamma})$ is monotonically increasing in $\gamma$.
\item \label{enu:thm-e0-+1}$\hat{e}_{\text{orig}}(\hat{g}_{+,\gamma})$ is monotonically decreasing in $\gamma$ for $\gamma\leq0$, and Condition \ref{cond:mcr+-bound} holds for $\gamma=0$ and $\epsilon_{\text{abs}}\geq\min_{f\in\mathcal{F}}\hat{e}_{\text{orig}}(f)$.
\item \label{enu:thm-upperbnd-mono}Given $\epsilon_{\text{abs}}$, the upper boundary $\left\{ \frac{\hat{h}_{+,\gamma}(\hat{g}_{+,\gamma})}{\epsilon_{\text{abs}}}-1\right\} \gamma^{-1}$ is monotonically increasing in $\gamma$ in the range where $\hat{e}_{\text{orig}}(\hat{g}_{+,\gamma})\leq\epsilon_{\text{abs}}$ and $\gamma<0$, and decreasing in the range where $\hat{e}_{\text{orig}}(\hat{g}_{+,\gamma})>\epsilon_{\text{abs}}$ and $\gamma<0$.
\end{enumerate}
\end{lemma}

Together, the results from Lemma \ref{lem:mcr+-compute-mono} imply that we can use a binary search across $\gamma\in\mathbb{R}$ to tighten the boundary on $\widehat{MR}$ from Lemma \ref{lem:mcr+-compute-bound}.

\subsection{Convex Models\label{subsec:Convex-models}}

In this section we show that empirical MCR can be conservatively computed when the loss function is convex in the model parameters \textendash{} that is, when the models $f_{\theta}\in\mathcal{F}$ are indexed by a $d$-dimensional parameter $\theta\in\Theta\subseteq\mathbb{R}^{d}$, and when the loss function $L(f_{\theta},(y,x_{1},x_{2}))$ is convex in $\theta$ for all $(x_{1},x_{2},y)\in\mathcal{X}_{1}\times(\mathcal{X}_{2},\mathcal{Y})$. 

Fortunately, neither Lemma \ref{lem:mcr--bound-compute} nor Lemma \ref{lem:mcr+-compute-bound} require an exact minimum for $\hat{h}_{-,\gamma}$ or $\hat{h}_{+,\gamma}$. For Lemma \ref{lem:mcr--bound-compute}, any lower bound on $\hat{h}_{-,\gamma}$ is sufficient to determine a lower bound on $MR(f)$. Likewise, for Lemma \ref{lem:mcr+-compute-bound}, any lower bound on $\hat{h}_{+,\gamma}$ is sufficient to determine an upper bound on $MR(f)$. 

To find these lower bounds, we note that for ``convex'' model classes (defined above) the optimization problems in Sections \ref{subsec:Binary-Search-MCR-} \& \ref{subsec:Binary-Search-MCR+} can be written either as convex optimization problems, or as difference convex function (DC) programs. A DC program is one that can be written as 
\[
\min_{\{\theta:c_{\text{DC}}(\theta)\leq k,\theta\in\Theta\}}g_{\text{DC}}(\theta)-h_{\text{DC}}(\theta),
\]
where $c_{\text{DC}}$ is a constraint function, $k\in\mathbb{R}^{1}$, and $g_{\text{DC}}$, $h_{\text{DC}}$, and $c_{\text{DC}}$ are convex. Although precise solutions to DC problems are not always tractable, lower bounds can be attained by branch-and-bound (B\&B) methods \citep{horst1999dc}. A simple B\&B approach is to partition $\Theta$ into a set of simplexes. Within the $j^{th}$ simplex, a lower bound on $g_{\text{DC}}(\theta)-h_{\text{DC}}(\theta)$ can be determined by replacing $h_{\text{DC}}$ with the hyperplane function $h_{j}$ satisfying $h_{j}(v)=h_{\text{DC}}(v)$ at each vertex $v$ of the $j^{th}$ simplex. Within this partition, $g_{\text{DC}}(\theta)-h_{\text{DC}}(\theta)$ is lower bounded by $l_{j}:=\min_{\theta}g_{\text{DC}}(\theta)-h_{j}(\theta)$, which can be computed as the solution to a convex optimization problem. Any partition for which $l_{j}$ is found to be too high is disregarded. Once a bound $l_{j}$ is computed for each partition, the partition with the lowest value $l_{j}$ is selected to be subdivided further, and additional lower bounds are recomputed for each new, resulting partition. This procedure continues until a sufficiently tight lower bound is attained (for more detailed procedures, see \citealp{horst1999dc}).

This approach allows us to conservatively approximate bounds on $\widehat{MR}(f)$ in the form of Eq \ref{eq:lower-bnd-mcr--thm} \& \ref{eq:thm-bnd-mcr+} by replacing $\hat{h}_{-,\gamma}(\hat{g}_{-,\gamma})$ and $\hat{h}_{+,\gamma}(\hat{g}_{+,\gamma})$ with lower bounds from the B\&B procedure. Although it will always yield valid bounds, the procedure may converge slowly when the dimension of $\Theta$ is large, giving highly conservative results. For some special cases of model classes however, even high dimensional DC problems simplify greatly. We discuss these cases in the next section.

\section{MR \& MCR for Linear Models, Additive Models, and Regression Models in a Reproducing Kernel Hilbert Space\label{subsec:linear-additive-interpretation-computation}}

For linear or additive models, many simplifications can be made to our approaches for MR and MCR. To simplify the interpretation of MR, we show below that population-level MR for a linear model can be expressed in terms of the model's coefficients (Section \ref{subsec:Interpreting-and-computing-MR-linear}). To simplify computation, we show that the cost of computing empirical MR for a linear model grows only linearly in $n$ (Section \ref{subsec:Interpreting-and-computing-MR-linear}), even though the number of terms in the definition of empirical MR grows quadratically (see Eqs \ref{eq:e-perm-def} \& \ref{eq:MR-switch-empirical}).

Moving on from MR, we show how empirical MCR can be computed for the class of linear models (Section \ref{subsec:LM-get-MCR}), for regularized linear models (Section \ref{subsec:Regularized-Linear-Models}), and for regression models in a reproducing kernel Hilbert space (RKHS, Section \ref{subsec:Kernel-Least-Squares}). To do this, we build on the approach in Section \ref{sec:Calculating-MCR} by giving approaches for minimizing arbitrary combinations of $\hat{e}_{\text{switch}}(f)$ and $\hat{e}_{\text{orig}}(f)$ across $f\in\mathcal{F}$. Even when the associated objective functions are non-convex, we can tractably obtain global minima for these model classes. We also discuss procedures to determine an upper bound $B_{\text{ind}}$ on the loss for any observation when using these model classes (see Assumption \ref{assu:bnd-ind}).

Throughout this section, we assume that $\mathcal{X}\subset\mathbb{R}^{p}$ for $p\in\mathbb{Z}^{+}$, that $\mathcal{Y}\subset\mathbb{R}^{1}$, and that $L$ is the squared error loss function $L(f,(y,x_{1},x_{2})=(y-f(x_{1},x_{2}))^{2}$. As in Section \ref{sec:Calculating-MCR}, we also assume that $0<\min_{f\in\mathcal{F}}\hat{e}_{\text{orig}}(f)$, to ensure that empirical MR is finite. 

\subsection{Interpreting and Computing MR for Linear or Additive Models\label{subsec:Interpreting-and-computing-MR-linear}}

We begin by considering MR for linear models evaluated with the squared error loss. For this setting, we can show both an interpretable definition of MR, as well as a computationally efficient formula for $\hat{e}_{\text{switch}}(f)$.
\begin{proposition}
\label{thm:Linear-models} (Interpreting MR, and computing empirical MR for linear models) For any prediction model $f$, let $e_{\text{orig}}(f)$, $e_{\text{switch}}(f)$, \textup{$\hat{e}_{\text{orig}}(f)$}, and $\hat{e}_{\text{switch}}(f)$ be defined based on the squared error loss $L(f,(y,x_{1},x_{2})):=(y-f(x_{1},x_{2}))^{2}$ for $y\in\mathbb{R}$, $x_{1}\in\mathbb{R}^{p_{1}}$, and $x_{2}\in\mathbb{R}^{p_{2}}$, where $p_{1}$ and $p_{2}$ are positive integers. Let $\beta=(\beta_{1},\beta_{2})$ and $f_{\beta}$ satisfy $\beta_{1}\in\mathbb{R}^{p_{1}}$, $\beta_{2}\in\mathbb{R}^{p_{2}}$, and $f_{\beta}(x)=x'\beta=x_{1}'\beta_{1}+x_{2}'\beta_{2}$. Then 
\begin{equation}
MR(f_{\beta})=1+\frac{2}{e_{\text{orig}}(f_\beta)}\left\{ \text{Cov}(Y,X_{1})\beta_{1}-\beta_{2}'\text{Cov}(X_{2},X_{1})\beta_{1}\right\} ,\label{eq:linear-models-pop}
\end{equation}

and, for finite samples,
\begin{equation}
\hat{e}_{\text{switch}}(f_{\beta})=\frac{1}{n}\left\{ \mathbf{y}'\mathbf{y}-2\left[\begin{array}{c}
\mathbf{X}_{1}'\mathbf{W}\mathbf{y}\\
\mathbf{X}_{2}'\mathbf{y}
\end{array}\right]^{'}\beta+\beta'\left[\begin{array}{cc}
\mathbf{X}_{1}'\mathbf{X}_{1} & \mathbf{X}_{1}'\mathbf{W}\mathbf{X}_{2}\\
\mathbf{X}_{2}'\mathbf{W}\mathbf{X}_{1} & \mathbf{X}_{2}'\mathbf{X}_{2}
\end{array}\right]\beta\right\} ,\label{eq:linear-models-finite}
\end{equation}

where $\mathbf{W}:=\frac{1}{n-1}(\mathbf{1}_{n}\mathbf{1}_{n}'-\mathbf{I}_{n})$, $\mathbf{1}_{n}$ is the $n$-length vector of ones, and $\mathbf{I}_{n}$ is the $n\times n$ identity matrix. 
\end{proposition}

Eq \ref{eq:linear-models-pop} shows that model reliance for linear models can be interpreted in terms of the population covariances, the model coefficients, and the model's accuracy. \citet{gregorutti2017correlation_LM_VI_RF} show an equivalent formulation of Eq \ref{eq:linear-models-pop} under the stronger assumptions that $f_{\beta}$ is equal to the conditional expectation function of $Y$ (that is, $f_{\beta}(x)=\mathbb{E}(Y|X=x)$), and the covariates $X_{1}$ and $X_{2}$ are centered.

Eq \ref{eq:linear-models-finite} shows that, although the number of terms in the definition of $\hat{e}_{\text{switch}}$ grows quadratically in $n$ (see Eq \ref{eq:e-perm-def}), the computational complexity of $\hat{e}_{\text{switch}}(f_{\text{\ensuremath{\beta}}})$ for a linear model $f_{\text{\ensuremath{\beta\ }}}$ grows\emph{ only linearly} in $n$. Specifically, the terms $\mathbf{X}_{1}'\mathbf{W}\mathbf{y}$ and $\mathbf{X}_{1}'\mathbf{W}\mathbf{X}_{2}$ in Eq \ref{eq:linear-models-finite} can be computed as $\frac{1}{n-1}\left\{ (\mathbf{X}_{1}'\mathbf{1}_{n})(\mathbf{1}_{n}'\mathbf{y})-(\mathbf{X}_{1}'\mathbf{y})\right\} $ and $\frac{1}{n-1}\left\{ (\mathbf{X}_{1}'\mathbf{1}_{n})(\mathbf{1}_{n}'\mathbf{X}_{2})-(\mathbf{X}_{1}'\mathbf{X}_{2})\right\} $ respectively, where the computational complexity of each term in parentheses grows linearly in $n$.

As in \citet{gregorutti2017correlation_LM_VI_RF}, both results in Proposition \ref{thm:Linear-models} readily generalize to additive models of the form $f_{g_{1},g_{2}}(X_{1},X_{2}):=g_{1}(X_{1})+g_{2}(X_{2})$, since permuting $X_{1}$ is equivalent to permuting $g_{1}(X_{1})$. 

\subsection{Computing Empirical MCR for Linear Models\label{subsec:LM-get-MCR}}

Building on the computational result from the previous section, we now consider empirical MCR computation for linear model classes of the form
\[
\mathcal{F}_{\text{lm}}:=\left\{ f_{\beta}\text{ : }f_{\beta}(x)=x'\beta,\hspace{1em}\beta\in\mathbb{R}^{p}\right\} .
\]
In order to implement the computational procedure from Sections \ref{subsec:Binary-Search-MCR-} and \ref{subsec:Binary-Search-MCR+}, we must be able to minimize arbitrary linear combinations of $\hat{e}_{\text{orig}}(f_{\beta})$ and $\hat{e}_{\text{switch}}(f_{\beta})$. Fortunately, for linear models, this minimization reduces to a quadratic program, as we show in the next remark.
\begin{remark}
\label{rem:For-any-LM-comp} (Tractability of empirical MCR for linear model classes) For any $f_{\beta}\in\mathcal{F}_{\text{lm}}$ and any fixed coefficients $\xi_{\text{orig}},\xi_{\text{switch}}\in\mathbb{R}$, the linear combination 
\begin{equation}
\xi_{\text{orig}}\hat{e}_{\text{orig}}(f_{\beta})+\xi_{\text{switch}}\hat{e}_{\text{switch}}(f_{\beta})\label{eq:linear-obj}
\end{equation}
 is proportional in $\beta$ to the quadratic function $-2\mathbf{q}'\beta+\beta'\mathbf{Q}\beta,$ where 

\[
\mathbf{Q}:=\xi_{\text{orig}}\mathbf{X}'\mathbf{X}+\xi_{\text{switch}}\left[\begin{array}{cc}
\mathbf{X}_{1}'\mathbf{X}_{1} & \mathbf{X}_{1}'\mathbf{W}\mathbf{X}_{2}\\
\mathbf{X}_{2}'\mathbf{W}\mathbf{X}_{1} & \mathbf{X}_{2}'\mathbf{X}_{2}
\end{array}\right],\text{\hspace{1em}}\hspace{1em}\mathbf{q}:=\left(\xi_{\text{orig}}\mathbf{y}'\mathbf{X}+\xi_{\text{switch}}\left[\begin{array}{c}
\mathbf{X}_{1}'\mathbf{W}\mathbf{y}\\
\mathbf{X}_{2}'\mathbf{y}
\end{array}\right]^{'}\right)^{'},
\]
and $\mathbf{W}:=\frac{1}{n-1}(\mathbf{1}_{n}\mathbf{1}_{n}'-\mathbf{I}_{n})$. Thus, minimizing $\xi_{\text{orig}}\hat{e}_{\text{orig}}(f_{\beta})+\xi_{\text{switch}}\hat{e}_{\text{switch}}(f_{\beta})$ is equivalent to an unconstrained (possibly non-convex) quadratic program.
\end{remark}

Because our empirical MCR computation procedure from Sections \ref{subsec:Binary-Search-MCR-} and \ref{subsec:Binary-Search-MCR+} consists of minimizing a sequence of objective functions in the form of Eq \ref{eq:linear-obj}, Remark \ref{rem:For-any-LM-comp} shows us that this procedure is tractable for the class of unconstrained linear models.

\subsection{Regularized Linear Models\label{subsec:Regularized-Linear-Models}}

Next, we continue to build on the results from Section \ref{subsec:LM-get-MCR} to calculate boundaries on $\widehat{MR}$ for \emph{regularized} linear models. We consider model classes formed by quadratically constrained subsets of $\mathcal{F}_{\text{lm}}$, defined as
\begin{equation}
\mathcal{F}_{\text{lm},r_{\text{lm}}}:=\left\{ f_{\beta}\text{ : }f_{\beta}(x)=x'\beta,\hspace{1em}\beta\in\mathbb{R}^{p},\text{ }\hspace{1em}\beta'\mathbf{M}_{\text{lm}}\beta\le r_{\text{lm}}\right\} ,\label{eq:F-reg-def}
\end{equation}
where $\mathbf{M}_{\text{lm}}$ and $r_{\text{lm}}$ are pre-specified. Again, this class describes linear models with a quadratic constraint on the coefficient vector.

\subsubsection{Calculating MCR}

As in Section \ref{subsec:LM-get-MCR}, calculating bounds on $\widehat{MR}$ via Lemmas \ref{lem:mcr--bound-compute} \& \ref{lem:mcr+-compute-bound} requires that are able to minimizing linear combinations $\xi_{\text{orig}}\hat{e}_{\text{orig}}(f_{\beta})+\xi_{\text{switch}}\hat{e}_{\text{switch}}(f_{\beta})$ across $f_{\beta}\in\mathcal{F}_{\text{lm},r_{\text{lm}}}$ for arbitrary $\xi_{\text{orig}},\xi_{\text{switch}}\in\mathbb{R}$. Applying Remark \ref{rem:For-any-LM-comp}, we can again equivalently minimize $-2\mathbf{q}'\beta+\beta'\mathbf{Q}\beta$ subject to the constraint in Eq \ref{eq:F-reg-def}:
\begin{align}
\text{minimize}\text{\hspace{1em}} & -2\mathbf{q}'\beta+\beta'\mathbf{Q}\beta\nonumber \\
\text{subject to\hspace{1em}} & \beta'\mathbf{M}_{\text{lm}}\beta\le r_{\text{lm}}.\label{eq:lm-constr}
\end{align}

The resulting optimization problem is a (possibly non-convex) quadratic program with one quadratic constraint (QP1QC). This problem is well-studied, and is related to the trust region problem \citep{boyd2004convex,polik2007survey,park2017general_qp1qc}. Thus, the bounds on MCR presented in Sections \ref{subsec:Binary-Search-MCR-} and \ref{subsec:Binary-Search-MCR+} again become computationally tractable for the class of quadratically constrained linear models.

\subsubsection{Upper Bounding the Loss\label{subsec:Upper-bounding-loss-reglm}}

One benefit of constraining the coefficient vector ($\beta'\mathbf{M}_{\text{lm}}\beta\le r_{\text{lm}}$) is that it facilitates determining an upper bound $B_{\text{ind}}$ on the loss function $L(f_{\beta},(y,x))=(y-x'\beta)^{2}$, which automatically satisfies Assumption \ref{assu:bnd-ind} for all $f\in\mathcal{F}_{\text{lm},r_{\text{lm}}}$. The following lemma gives sufficient conditions to determine $B_{\text{ind}}$.
\begin{lemma}
\label{lem:-lm-get-BB} (Loss upper bound for linear models) If $\mathbf{M}_{\text{lm}}$ is positive definite, $Y$ is bounded within a known range, and there exists a known constant $r_{\mathcal{X}}$ such that $x'\mathbf{M}_{\text{lm}}^{-1}x\le r_{\mathcal{X}}$ for all $x\in(\mathcal{X}_{1}\times\mathcal{X}_{2})$, then Assumption \ref{assu:bnd-ind} holds for the model class $\mathcal{F}_{\text{lm},r_{\text{lm}}}$, the squared error loss function, and the constant
\[
B_{\text{ind}}=\max\left[\left\{ \min_{y\in\mathcal{Y}}\left(y\right)-\sqrt{r_{\mathcal{X}}r_{\text{lm}}}\right\} ^{2},\left\{ \max_{y\in\mathcal{Y}}\left(y\right)+\sqrt{r_{\mathcal{X}}r_{\text{lm}}}\right\} ^{2}\right].
\]
\end{lemma}

In practice, the constant $r_{\mathcal{X}}$ can be approximated by the empirical distribution of $X$ and $Y$. The motivation behind the restriction $x'\mathbf{M}_{\text{lm}}^{-1}x\le r_{\mathcal{X}}$ in Lemma \ref{lem:-lm-get-BB} is to create complementary constraints on $X$ and $\beta$. For example, if $\mathbf{M}_{\text{lm}}$ is diagonal, then the smallest elements of $\mathbf{M}_{\text{lm}}$ correspond to directions along which $\beta$ is least restricted by $\beta'\mathbf{M}_{\text{lm}}\beta\le r_{\text{lm}}$ (Eq \ref{eq:lm-constr}), as well as the directions along which $x$ is most restricted by $x'\mathbf{M}_{\text{lm}}^{-1}x\le r_{\mathcal{X}}$ (Lemma \ref{lem:-lm-get-BB}).

\subsection{Regression Models in a Reproducing Kernel Hilbert Space (RKHS)\label{subsec:Kernel-Least-Squares}}

We now expand our scope of model classes by considering regression models in a reproducing kernel Hilbert space (RKHS), which allow for nonlinear and nonadditive features of the covariates. We show that, as in Section \ref{subsec:Regularized-Linear-Models}, minimizing a linear combination of $\hat{e}_{\text{orig}}(f)$ and $\hat{e}_{\text{switch}}(f)$ across models $f$ in this class can be expressed as a QP1QC, which allows us to implement the binary search procedure of Sections \ref{subsec:Binary-Search-MCR-} \& \ref{subsec:Binary-Search-MCR+}.

First we introduce notation required to describe regression in a RKHS. Let $\mathbf{D}$ be a $(R\times p)$ matrix representing a pre-specified dictionary of $R$ reference points, such that each row of $\mathbf{D}$ is contained in $\mathcal{X}=\mathbb{R}^{p}$. Let $k$ be a pre-specified positive definite kernel function, and let $\mu$ be a prespecified estimate of $\mathbb{E}Y$. Let $\mathbf{K}_{\mathbf{D}}$ be the $R\times R$ matrix with $\mathbf{K}_{\mathbf{D}[i,j]}=k(\mathbf{D}{}_{[i,\cdot]},\mathbf{D}_{[j,\cdot]})$. We consider prediction models of the following form, where the distance to each reference point is used as a regression feature:
\begin{equation}
\mathcal{F}_{\mathbf{D},r_{k}}=\left\{ f_{\alpha}\text{ : }f_{\alpha}(x)=\mu+\sum_{i=1}^{R}k\left(x,\mathbf{D}_{[i,\cdot]}\right)\alpha_{[i]},\hspace{1em}\|f_{\alpha}\|_{k}\leq r_{k},\hspace{1em}\alpha\in\mathbb{R}^{R}\right\} .\label{eq:RKHS-def}
\end{equation}
Above, the norm $\|f_{\alpha}\|_{k}$ is defined as
\begin{align}
\|f_{\alpha}\|_{k}:=\sum_{i=1}^{R}\sum_{j=1}^{R}\alpha_{[i]}\alpha_{[j]}k(\mathbf{D}_{[i,\cdot]},\mathbf{D}_{[j,\cdot]}) & =\alpha'\mathbf{K}_{\mathbf{D}}\alpha.\label{eq:f_k_norm_def}
\end{align}

In the next two sections, we show that bounds on empirical MCR can again be tractably computed for this class, and that the loss for models in this class can be feasibly upper bounded.

\subsubsection{Calculating MCR\label{subsec:Calculating-MCR-kernel}}

Again, calculating bounds on $\widehat{MR}$ from Lemmas \ref{lem:mcr--bound-compute} \& \ref{lem:mcr+-compute-bound} requires us to be able to minimize arbitrary linear combinations of $\hat{e}_{\text{orig}}(f_{\alpha})$ and $\hat{e}_{\text{switch}}(f_{\alpha})$.

Given a size-$n$ sample of test observations $\mathbf{Z}=\left[\begin{array}{cc}
\mathbf{y} & \mathbf{X}\end{array}\right]$, let $\mathbf{K}_{\text{orig}}$ be the $n\times R$ matrix with elements $\mathbf{K}_{\text{orig}[i,j]}=k\left(\mathbf{X}_{[i,\cdot]},\mathbf{D}_{[j,\cdot]}\right)$. Let $\mathbf{Z}_{\text{switch}}=\left[\begin{array}{cc}
\mathbf{y}_{\text{switch}} & \mathbf{X}_{\text{switch}}\end{array}\right]$ be the $(n(n-1))\times(1+p)$ matrix with rows that contain the set $\{(\mathbf{y}_{[i]},\mathbf{X}_{1[j,\cdot]},\mathbf{X}_{2[i,\cdot]}):i,j\in\{1,\dots,n\}\text{ and }i\neq j\}$. Finally, let $\mathbf{K}_{\text{switch}}$ be the $n(n-1)\times R$ matrix with $\mathbf{K}_{\text{switch}[i,j]}=k\left(\mathbf{X}_{\text{switch}[i,\cdot]},\mathbf{D}_{[j,\cdot]}\right)$. 

For any two constants $\xi_{\text{orig}},\xi_{\text{switch}}\in\mathbb{R}$, we can show that minimizing the linear combination $\xi_{\text{orig}}\hat{e}_{\text{orig}}(f_{\alpha})+\xi_{\text{switch}}\hat{e}_{\text{switch}}(f_{\alpha})$ over $\mathcal{F}_{\mathbf{D},r_{k}}$ is equivalent to the minimization problem
\begin{align}
\text{minimize }\hspace{1em} & \frac{\xi_{\text{orig}}}{n}\|\mathbf{y}-\mu-\mathbf{K}_{\text{orig}}\alpha\|_{2}^{2}+\frac{\xi_{\text{switch}}}{n(n-1)}\|\mathbf{y}_{\text{switch}}-\mu-\mathbf{K}_{\text{switch}}\alpha\|_{2}^{2}\label{eq:kernel-min-h-start}\\
\text{subject to}\hspace{1em} & \alpha'\mathbf{K}_{\mathbf{D}}\alpha<r_{k}.\label{eq:kernel-min-h-end}
\end{align}

Like Problem \ref{eq:lm-constr}, Problem \ref{eq:kernel-min-h-start}-\ref{eq:kernel-min-h-end} is a QP1QC. To show Eqs \ref{eq:kernel-min-h-start}-\ref{eq:kernel-min-h-end}, we first write $\hat{e}_{\text{orig}}(f_{\alpha})$ as 
\begin{align}
\hat{e}_{\text{orig}}(f_{\alpha}) & =\frac{1}{n}\sum_{i=1}^{n}\left\{ \mathbf{y}_{[i]}-f_{\alpha}(\mathbf{X}_{[i,\cdot]})\right\} ^{2}\label{eq:start-kernel-sum}\\
 & =\frac{1}{n}\sum_{i=1}^{n}\left\{ \mathbf{y}_{[i]}-\mu-\sum_{j=1}^{R}k(\mathbf{X}_{[i,\cdot]},\mathbf{D}_{[j,\cdot]})\alpha_{[j]}\right\} ^{2}\nonumber \\
 & =\frac{1}{n}\sum_{i=1}^{n}\left\{ \mathbf{y}_{[i]}-\mu-\mathbf{K}_{\text{orig}[i,\cdot]}'\alpha\right\} ^{2}\nonumber \\
 & =\frac{1}{n}\|\mathbf{y}-\mu-\mathbf{K}_{\text{orig}}\alpha\|_{2}^{2}.\label{eq:end-kernel-sum}
\end{align}

Following similar steps, we can obtain
\begin{align*}
\hat{e}_{\text{switch}}(f_{\alpha}) & =\frac{1}{n(n-1)}\|\mathbf{y}_{\text{switch}}-\mu-\mathbf{K}_{\text{switch}}\alpha\|_{2}^{2}.
\end{align*}

Thus, for any two constants $\xi_{\text{orig}},\xi_{\text{switch}}\in\mathbb{R}$, we can see that $\xi_{\text{orig}}\hat{e}_{\text{orig}}(f_{\alpha})+\xi_{\text{switch}}\hat{e}_{\text{switch}}(f_{\alpha})$ is quadratic in $\alpha$. This means that we can tractably compute bounds on empirical MCR for this class as well.

\subsubsection{Upper Bounding the Loss\label{subsec:Bf-RKHS-kernel}}

Using similar steps as in Section \ref{subsec:Upper-bounding-loss-reglm}, the following lemma gives sufficient conditions to determine $B_{\text{ind}}$ for the case of regression in a RKHS.
\begin{lemma}
\label{lem:BB-kernel} (Loss upper bound for regression in a RKHS) Assume that $Y$ is bounded within a known range, and there exists a known constant $r_{\mathbf{D}}$ such that $v(x)'\mathbf{K}_{\mathbf{D}}^{-1}v(x)\leq r_{\mathbf{D}}$ for all $x\in(\mathcal{X}_{1}\times\mathcal{X}_{2})$, where $v:\mathbb{R}^{p}\rightarrow\mathbb{R}^{R}$ is the function satisfying $v(x)_{[i]}=k(x,\mathbf{D}_{[i,\cdot]})$. Under these conditions, Assumption \ref{assu:bnd-ind} holds for the model class $\mathcal{F}_{\mathbf{D},r_{k}}$, the squared error loss function, and the constant
\[
B_{\text{ind}}=\max\left[\left\{ \min_{y\in\mathcal{Y}}\left(y\right)-\left(\mu+\sqrt{r_{\mathcal{\mathbf{D}}}r_{k}}\right)\right\} ^{2},\left\{ \max_{y\in\mathcal{Y}}\left(y\right)+\left(\mu+\sqrt{r_{\mathcal{\mathbf{D}}}r_{k}}\right)\right\} ^{2}\right].
\]
\end{lemma}

Thus, for regression models in a RKHS, we can satisfy Assumption \ref{assu:bnd-ind} for all models in the class.

\section{Connections Between MR and Causality\label{sec:Connections-between-MR-causality}}

Our MR approach can be fundamentally described as studying how a model's behavior changes under an intervention on the underlying data. We aim to study the causal effect of this intervention on the \emph{model's} performance. This goal mirror's the conventional causal inference goal of studying how an intervention on variables will change outcomes generated by a process in \emph{nature}. 

This section explores this connection to causal inference further. Section \ref{subsec:causal-MR} shows that when the prediction model in question is the conditional expectation function from nature itself, MR reduces to commonly studied quantities in the causal literature. Section \ref{subsec:mcr-impute} proposes an alternative to MR that focuses on interventions, or data perturbations, that are likely to occur in the underlying data generating process.

\subsection{Model Reliance and Causal Effects\label{subsec:causal-MR}}

In this section, we show a connection between population-level model reliance and the conditional average causal effect. For consistency with the causal inference literature, we temporarily rename the random variables $(Y,X_{1},X_{2})$ as $(Y,T,C)$, with realizations $(y,t,c)$. Here, $T:=X_{1}$ represents a binary treatment indicator, $C:=X_{2}$ represents a set of baseline covariates (``C'' is for ``covariates''), and $Y$ represents an outcome of interest. Under this notation, $e_{\text{orig}}(f)$ represents the expected loss of a prediction function $f$, and $e_{\text{switch}}(f)$ denotes the expected loss in a pair of observations in which the treatment has been switched. Let $f_{0}(t,c):=\mathbb{E}(Y|C=c,T=t)$ be the (unknown) conditional expectation function for $Y$, where we place no restrictions on the functional form of $f_{0}$.

Let $Y_{1}$ and $Y_{0}$ be potential outcomes under treatment and control respectively, such that $Y=Y_{0}(1-T)+Y_{1}T$. The treatment effect for an individual is defined as $Y_{1}-Y_{0}$, and the average treatment effect is defined as $\mathbb{E}(Y_{1}-Y_{0})$. Let $\text{CATE}(c):=\mathbb{E}(Y_{1}-Y_{0}|C=c)$ be the (unknown) conditional average treatment effect of $T$ for all patients with $C=c$. Causal inference methods typically assume $(Y_{1},Y_{0})\perp T|C$ (conditional ignorability), and $0<\mathbb{P}(T=1|C=c)<1$ for all values of $c$ (positivity), in order for $f_{0}$ and $\text{CATE}$ to be well defined and identifiable.

The next proposition quantifies the relation between the conditional average treatment effect function ($\text{CATE}$) and the model reliance of $f_{0}$ on $X_{1}$.
\begin{proposition}
(Causal interpretations of MR) For any prediction model $f$, let $e_{\text{orig}}(f)$ and $e_{\text{switch}}(f)$ be defined based on the squared error loss $L(f,(y,t,c)):=(y-f(t,c))^{2}$. 

If $(Y_{1},Y_{0})\perp T|C$ (conditional ignorability) and $0<\mathbb{P}(T=1|C=c)<1$ for all values of $c$ (positivity), then $MR(f_{0}) $ is equal to \label{thm:VI_TRT}
\begin{eqnarray}
1+\frac{\text{Var}(T)}{\mathbb{E}_{T,C}\text{Var}(Y|T,C)}\sum_{t\in\{0,1\}}\left\{ \mathbb{E}(Y_{1}-Y_{0}|T=t)^{2}+\text{Var}\left(\text{CATE}(C)|T=t\right)\right\} ,\label{eq:TE-result}
\end{eqnarray}
where $\text{Var}(T)$ is the marginal variance of the treatment assignment.
\end{proposition}

We see above that model reliance decomposes into several terms that are each individually important in causal inference: the treatment prevalence (via $\text{Var}(T)$); the variability in $Y$ that is not explained by $C$ or $T$; the magnitude of the average treatment effect, conditional on $T$; and the variance of the conditional average treatment effect across subgroups. For example, if all patients are treated, then scrambling the treatment in a random pair of observations has no effect on the loss. In this case we see that $\text{Var}(T)=0$ and $MR(f_{0})=1$, indicating no reliance. When $\text{Var}(T)>0$, a higher average treatment effect magnitude ($\mathbb{E}(Y_{1}-Y_{0}|T=t)^{2}$) corresponds to $f_{0}$ requiring $T$ more heavily to predict $Y$, all else equal. Similarly, if there is a high degree of treatment effect heterogeneity across subgroups (that is, when $\text{Var}\left(\text{CATE}(C)|T=t\right)$ is large),  the model $f_{0}$ will again use $T$ more heavily when predicting $Y$. For example, a treatment may be important for predicting $Y$ even if the average treatment effect is zero, so long as the treatment helps some subgroups more than others.

\subsection[Conditional Importance: Adjusting for Dependence Between Covariates]{Conditional Importance: Adjusting for Dependence Between $X_{1}$ and $X_{2}$\label{subsec:mcr-impute}}

One common scenario where multiple models achieve low loss is when the sets of predictors $X_{1}$ and $X_{2}$ are highly correlated, or contain redundant information. Models may predict well either through reliance on $X_{1}$, or through reliance on $X_{2}$, and so MCR will correctly identify a wide range of potential reliances on $X_{1}$. However, we may specifically be interested how much models rely on the information in $X_{1}$ that cannot alternatively be gleaned from $X_{2}$.

For example, age and accumulated wealth may be correlated, and both may be predictive of future promotion. We may wish to know the how much a model for predicting promotion relies on information that is uniquely available from wealth measurements.

To formalize this notion, we define an alternative to $e_{\text{switch}}$ where noise is added to $X_{1}$ in a way that accounts for the dependence between $X_{1}$ and $X_{2}$. Given a fixed prediction model $f$, we ask: how well would the model $f$ perform if the values of $X_{1}$ were scrambled\emph{ across observations with the same value for $X_{2}$.} Specifically, let $Z^{(a)}=(Y^{(a)},X_{1}^{(a)},X_{2}^{(a)})$ and $Z^{(b)}=(Y^{(b)},X_{1}^{(b)},X_{2}^{(b)})$ denote a pair of independent random vectors following the same distribution as $Z=(Y,X_{1},X_{2})$, as in Section \ref{subsec:model-reliance}, and let
\begin{equation}
e_{\text{cond}}(f):=\mathbb{E}_{X_{2}}\mathbb{E}_{(Y^{(b)},X_{1}^{(a)},X_{2}^{(b)})}\left[L\{f,(Y^{(b)},X_{1}^{(a)},X_{2}^{(b)})\}|X_{2}^{(a)}=X_{2}^{(b)}=X_{2}\right].\label{eq:e-cond-def}
\end{equation}
In words, $e_{\text{cond}}(f)$ is the expected loss of a given model $f$ across pairs of observations $(Z^{(a)},Z^{(b)})$ in which the values of $X_{1}^{(a)}$ and $X_{1}^{(b)}$ have been switched, given that these pairs match on $X_{2}$. This quantity can also be interpreted as the expected loss of $f$ if noise were added to $X_{1}$ in such a way that $X_{1}$ was no longer informative of $Y$, given $X_{2}$, but that the joint distribution of the covariates $(X_{1},X_{2})$ was maintained. 

We then define conditional model reliance, or ``core'' model reliance (CMR) for a fixed function $f$ as 
\[
CMR(f)=\frac{e_{\text{cond}}(f)}{e_{\text{orig}}(f)}.
\]
That is, CMR is the factor by which the model's performance degrades when the information unique to $X_{1}$ is removed. If $X_{1}\perp X_{2}$, then $X_{1}$ contains no redundant information, and CMR and MR are equivalent. Otherwise, all else equal, CMR will decrease as $X_{2}$ becomes more predictive of $X_{1}$. Analogous to MCR, we define conditional MCR (CMCR) in the same way as in Eq \ref{eq:pop-min-max}, but with MR replaced with CMR. In comparison with MCR, CMCR will generally result in a range that is closer to 1 (null reliance).

An advantage of CMR is that it restricts the ``noise-corrupted'' inputs to be within the domain $\mathcal{X}$, rather than the expanded domain $\mathcal{X}_{1}\times\mathcal{X}_{2}$ considered by MR. This means that CMR will not be influenced by impossible combinations of $x_{1}$ and $x_{2}$, while MR may be influenced by them. \citet{hooker2007generalized} discuss a similar issue, arguing that evaluations of a prediction model's behavior in different circumstances should be weighted by, for example, how likely those circumstances are to occur.

A challenge facing the CMR approach is that matched pairs such as those in Eq \ref{eq:e-cond-def} may occur rarely, making it difficult to estimate CMR nonparametrically. We explore this estimation issue next.

\subsubsection{Estimation of CMR by Weighting, Matching, or Imputation}

If the covariate space is discrete and low dimensional, nonparametric methods based on weighting or matching can be effective means of estimating CMR. Specifically, we can weight each pair of sample points $i,j$ according to how likely the covariate combination $(\mathbf{X}_{1[i,\cdot]},\mathbf{X}_{2[j,\cdot]})$ is to occur, as in

\[
\hat{e}_{\text{weight}}(f):=\frac{1}{n(n-1)}\sum_{i=1}^{n}\sum_{j\neq i}w(\mathbf{X}_{1[i,\cdot]},\mathbf{X}_{2[j,\cdot]})\times L\{f,(\mathbf{y}_{[j]},\mathbf{X}_{1[i,\cdot]},\mathbf{X}_{2[j,\cdot]})\},
\]
where $w(x_{1},x_{2}):=\frac{\mathbb{P}(X_{1}=x_{1}|X_{2}=x_{2})}{\mathbb{P}(X_{1}=x_{1})}$ is an importance weight (see also \citealp{hooker2007generalized}). Here, pairs of observations corresponding to unlikely or impossible combinations of covariates are down-weighted or discarded, respectively. If the probabilities $\mathbb{P}(X_{1}=x_{1}|X_{2}=x_{2})$ and $\mathbb{P}(X_{1}=x_{1})$ are known, then $\hat{e}_{\text{weight}}(f)$ is unbiased for $e_{\text{cond}}(f)$ (see Appendix \ref{subsec:Unbiased-CMR}).

Alternatively, if $X_{2}$ is discrete and low dimensional, we can restrict estimates of $e_{\text{cond}}(f)$ to only consider pairs of sample observations in which $X_{2}$ is constant, or ``matched,'' as in
\begin{equation}
\hat{e}_{\text{match}}(f):=\frac{1}{n(n-1)}\sum_{i=1}^{n}\sum_{j\neq i}\frac{1(\mathbf{X}_{2[j,\cdot]}=\mathbf{X}_{2[i,\cdot]})}{\mathbb{P}(X_{2}=\mathbf{X}_{2[i,\cdot]})}\times L\{f,(\mathbf{y}_{[j]},\mathbf{X}_{1[i,\cdot]},\mathbf{X}_{2[j,\cdot]})\}.\label{eq:e-match}
\end{equation}
This approach allows estimation of CMR without knowledge of the conditional distribution $\mathbb{P}(X_{1}=x_{1}|X_{2}=x_{2})$. If the inverse probability weight $\mathbb{P}(X_{2}=\mathbf{X}_{2[i,\cdot]})^{-1}$ is known, then $\hat{e}_{\text{match}}(f)$ is unbiased for $e_{\text{cond}}(f)$ (see Appendix \ref{subsec:Unbiased-CMR}). The weight $\mathbb{P}(X_{2}=\mathbf{X}_{2[i,\cdot]})^{-1}$ accounts for the fact that, for any given value $x_{2}$, the proportion of observations of $X_{2}$ taking the value $x_{2}$ will generally not be the same as the proportion of matched pairs $(X_{2}^{(a)},X_{2}^{(b)})$ taking value the $x_{2}$, and so simply summing over all matched pairs would lead to bias. In practice, the proportion $\mathbb{P}(X_{2}=\mathbf{X}_{2[i,\cdot]})$ can be approximated as $\frac{1}{n-1}\sum_{j'\neq i}1(\mathbf{X}_{2[i,\cdot]}=\mathbf{X}_{2[j',\cdot]})$, with minor adjustments to Eq \ref{eq:e-match} to avoid dividing by zero. The resulting estimate is analogous to exact matching procedures commonly used in causal inference, which are known to work best when the covariates are discrete and low dimensional, in order for exact matches to be common \citep{stuart2010matching}.

However, when the covariate space is continuous or high dimensional, we typically cannot estimate CMR nonparametrically. For such cases, we propose to estimate CMR under an assumption of homogeneous residuals. Specifically, we define $\mu_{1}$ to be the conditional expectation function $\mu_{1}(x_{2})=\mathbb{E}(X_{1}|X_{2}=x_{2})$, and assume that the random residual $X-\mu_{1}(X_{2})$ is independent of $X_{2}$. Under this assumption, it can be shown that
\begin{align*}
e_{\text{cond}}(f) & =\mathbb{E}L\left[f,(Y^{(b)},\left\{ X_{1}^{(a)}-\mu_{1}(X_{2}^{(a)})\right\} +\mu_{1}(X_{2}^{(b)}),X_{2}^{(b)})\right].
\end{align*}
That is, $e_{\text{cond}}(f)$ is equal to the expected loss of $f$ across random pairs of observations $(Z^{(a)},Z^{(b)})$ in which the value \emph{of the residual terms} (in curly braces) have been switched. Because of the independence assumption, no matching or weighting is required. If $\mu_{1}$ is known, then we can again produce an unbiased estimate using the U-statistic
\[
\hat{e}_{\text{impute}}(f):=\frac{1}{n(n-1)}\sum_{i=1}^{n}\sum_{j\neq i}L\left[f,(\mathbf{y}_{[j]},\left\{ \mathbf{X}_{1[i,\cdot]}-\mu_{1}(\mathbf{X}_{2[i,\cdot]})\right\} +\mu_{1}(\mathbf{X}_{2[j,\cdot]}),\mathbf{X}_{2[j,\cdot]})\right].
\]
This estimator aggregates over all pairs in our sample, switching the values of the residual terms (in curly braces) within each pair. In practice, when $\mu_{1}$ is not known, an estimate of $\mu_{1}$ can be achieved via regression or related machine learning techniques, and plugged in to the above equation. In this way, the assumption that $X-\mu_{1}(X_{2})\perp X_{2}$ allows us to estimate CMR without explicitly modeling the joint distribution of $X_{1}$ and $X_{2}$.

In the existing literature, \citet{strobl2008conditional_VI} introduce a similar procedure for estimating conditional variable importance. However, a formal comparison to \citeauthor{strobl2008conditional_VI} is complicated by the fact that the authors do not define a specific estimand, and that their approach is limited to tree-based regression models. Other existing approaches conditional importance approaches include methods for redefining $X_{1}$ and $X_{2}$ to induce approximate independence, before computing an importance measure analogous to MR. This can be done by reducing the total number of covariates used, and hence reducing how well any one variable can be predicted by the others (as in \citealp{gregorutti2017correlation_LM_VI_RF}). Alternatively, variables in $X_{2}$ that are predictive of $X_{1}$ can be regrouped directly into $X_{1}$ (as in \citealp{tolocsi2011classification}; see also the discussion from Kirk, Lewin and Stumpf, in \citealt{meinshausen2010stability_selection}).

In summary, CMR allows us to see how much a model relies on the information uniquely available in $X_{1}$. While CMR is more difficult to estimate than MR, several tractable approaches exist when $X_{2}$ is discrete, or when a homogenous residual assumption can be applied. One may also consider extending CMR by conditioning only on a subset of $X_{2}$. For example, we may consider conditioning only on elements of $X_{2}$ that are believed to causally effect $X_{1}$, by changing the outer expectation in Eq \ref{eq:e-cond-def}. For simplicity, we focus on the base case of estimating MR in this paper. Similar results could potentially be carried over for CMR as well.

\section{Simulations\label{sec:Simulations-all}}

In this section, we first present a toy example to illustrate the concepts of MR, MCR, and AR. We then present a Monte Carlo simulation studying the effectiveness of bootstrap CIs for MCR.

\subsection{Illustrative Toy Example with Simulated Data\label{sec:Simulations}}

To illustrate the concepts of MR, MCR, and AR (see Section \ref{sec:Related-Work}), we consider a toy example where $X=(X_{1},X_{2})\in\mathbb{R}^{2}$, and $Y\in\{-1,1\}$ is a binary group label. Our primary goal in this section is to build intuition for the differences between these three importance measures, and so we demonstrate them here only in a single sample. We focus on the empirical versions of our importance metrics ($\widehat{MR}$, $\widehat{MCR}_{-}$ and $\widehat{MCR}_{+}$), and compare them against AR, which is typically interpreted as an in-sample measure \citep{breiman2001random_forests}, or as an intermediate step to estimate an alternate importance measure in terms of variable rankings \citep{gevrey2003review,olden2004accurate_neural}.

We simulate $X|Y=-1$ from an independent, bivariate normal distribution with means $\mathbb{E}(X_{1}|Y=-1)=\mathbb{E}(X_{2}|Y=-1)=0$ and variances $\text{Var}(X_{1}|Y=-1)=\text{Var}(X_{2}|Y=-1)=\frac{1}{9}$. We simulate $X|Y=1$ by drawing from the same bivariate normal distribution, and then adding the value of a random vector $(C_{1},C_{2}):=(\text{cos}(U),\text{sin}(U))$, where $U$ is a random variable uniformly distributed on the interval $[-\pi,\pi]$. Thus, $(C_{1},C_{2})$ is uniformly distributed across the unit circle.

Given a prediction model $f:\mathcal{X}\rightarrow\mathbb{R}$, we use the sign of $f(X_{1},X_{2})$ as our prediction of $Y$. For our loss function, we use the hinge loss $L(f,(y,x_{1},x_{2}))=(1-yf(x_{1},x_{2}))_{+}$, where $(a)_{+}=a$ if $a\geq0$ and $(a)_{+}=0$ otherwise. The hinge loss function is commonly used as a convex approximation to the zero-one loss ${L(f,(y,x_{1},x_{2}))=1[y\neq\text{sign}\{f(x_{1},x_{2})\}]}$.

We simulate two samples of size 300 from the data generating process described above, one to be used for training, and one to be used for testing. Then, for the class of models used to predict $Y$, we consider the set of degree-3 polynomial classifiers
\begin{align*}
\mathcal{F}_{d3} & =\left\{ f_{\bm{\theta}}:f_{\bm{\theta}}(x_{1},x_{2})=\bm{\theta}_{[1]}+\bm{\theta}_{[2]}x_{1}+\bm{\theta}_{[3]}x_{2}\right.\\
 & \hspace{1cm}\hspace{1cm}\hspace{1cm}+\bm{\theta}_{[4]}x_{1}^{2}+\bm{\theta}_{[5]}x_{2}^{2}+\bm{\theta}_{[6]}x_{1}x_{2}\\
 & \hspace{1cm}\hspace{1cm}\hspace{1cm}\left.+\bm{\theta}_{[7]}x_{1}^{3}++\bm{\theta}_{[8]}x_{2}^{3}+\bm{\theta}_{[9]}x_{1}^{2}x_{2}+\bm{\theta}_{[10]}x_{1}x_{2}^{2};\,\ \|\bm{\theta}_{[-1]}\|_{2}^{2}\leq r_{d3}\right\} ,
\end{align*}
where $\bm{\theta}_{[-1]}$ denotes all elements of $\bm{\theta}$ except $\bm{\theta}_{[1]}$, and where we set $r_{d3}$ to the value that minimizes the 10-fold cross-validated loss in the training data. Let $\mathcal{A}_{d3}$ be the algorithm that minimizes the hinge loss over the (convex) feasible region $\{f_{\theta}:\|\bm{\theta}_{[-1]}\|_{2}^{2}\leq r_{d3}\}$. We apply $\mathcal{A}_{d3}$ to the training data to determine a reference model $f_{\text{ref}}$. Also using the training data, we set $\epsilon$ equal to 0.10 multiplied by the cross-validated loss of $\mathcal{A}_{d3}$, such that $\mathcal{R}(\epsilon,f_{\text{ref}},\mathcal{F}_{d3})$ contains all models in $\mathcal{F}_{d3}$ that exceed the loss of $f_{\text{ref}}$ by no more than approximately 10\% (see Eq \ref{eq:explicit-rashomon-def}). We then calculate empirical AR, MR, and MCR using the test observations.

\begin{figure}[t]
\begin{centering}
\includegraphics[width=1\columnwidth]{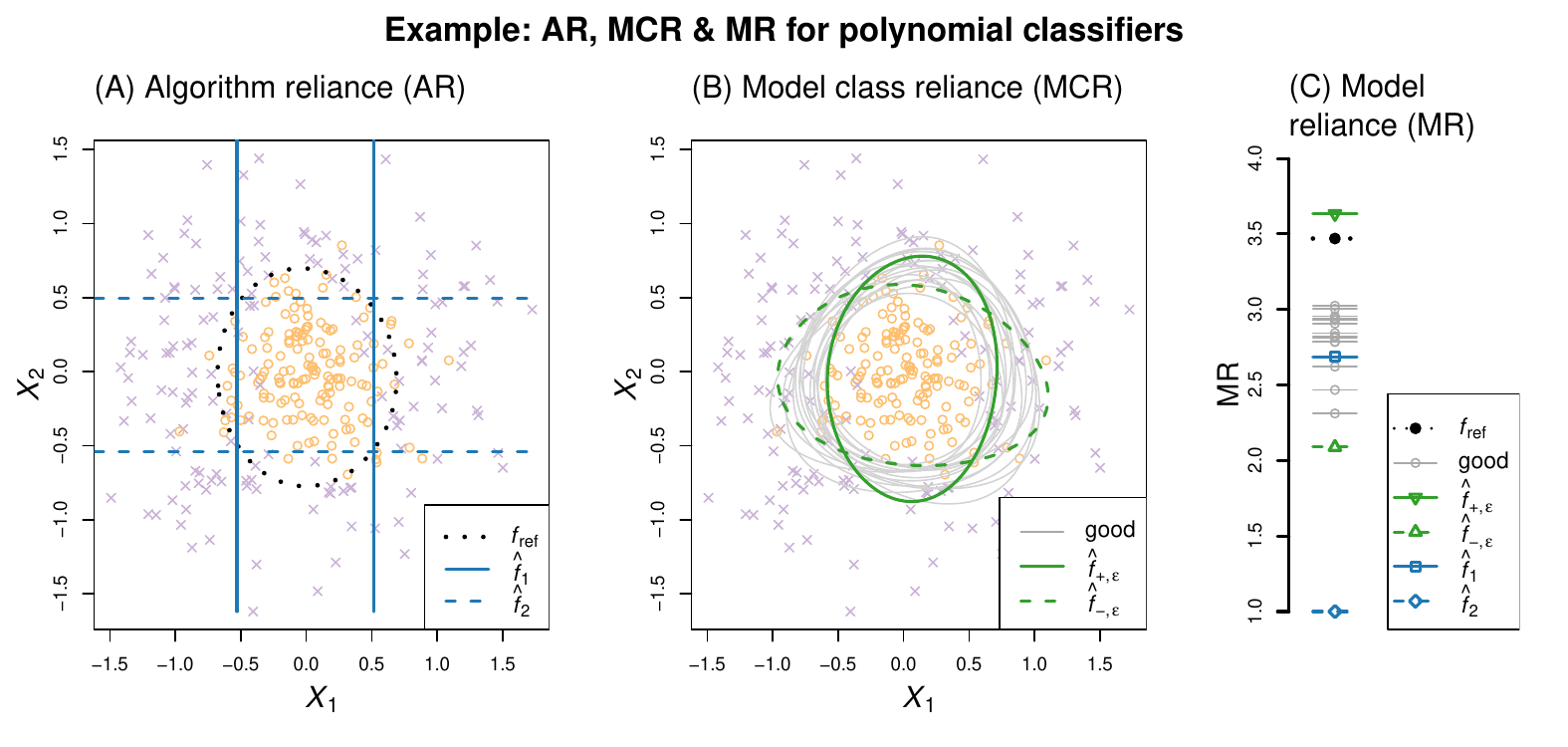}
\par\end{centering}
\caption{Example of AR, MCR \& MR for polynomial classifiers. Panels (A) \& (B) show the same 300 draws from a simulated data set, with the classification of each data point marked by ``x'' for $Y=1$, and ``o'' for $Y=-1$. In Panel (A), for AR, we show single-feature models formed by dropping a covariate. Because these models take only a single input, we represent their classification boundaries as straight lines. In Panel (B), for MCR, we show the classification boundaries for several (two-feature) models with low in-sample loss. Of these models, the model with minimal dependence on $X_{1}$ is shown by the dashed green oval, and the model with maximal dependence on $X_{1}$ is shown by the solid green oval. Panel (C) shows the empirical model reliance on $X_{1}$ for each of the models in Panels (A) \& (B). We see in Panel (C) that, as expected, no well-performing model relies (empirically) on $X_{1}$ more than $\hat{f}_{+,\epsilon}$ does, or relies (empirically) on $X_{1}$ less than $\hat{f}_{-,\epsilon}$ does. That is, no well-performing model has an empirical MR value greater than $\widehat{MCR}_{+}(\epsilon)$, or less than $\widehat{MCR}_{-}(\epsilon)$. \label{fig:Example}}
\end{figure}

We begin by considering the AR of $\mathcal{A}_{d3}$ on $X_{1}$. Calculating AR requires us to fit two separate models, first using all of the variables to fit a model on the training data, and then again using only $X_{2}$. In this case, the first model is equivalent to $f_{\text{ref}}$. We denote the second model as $\hat{f}_{2}$. To compute AR, we evaluate $f_{\text{ref}}$ and $\hat{f}_{2}$ in the test observations. We illustrate this AR computation in Figure \ref{fig:Example}-A, marking the classification boundaries for $f_{\text{ref}}$ and $\hat{f}_{2}$ by the black dotted line and the blue dashed lines respectively, and marking the test observations by labelled points (``x'' for $Y=1$, and ``o'' for $Y=-1$). Comparing the loss associated with these two models gives one form of AR\textendash an estimate of the necessity of $X_{1}$ for the algorithm $\mathcal{A}_{d3}$. Alternatively, to estimate the \emph{sufficiency} of $X_{1}$, we can compare the reference model $f_{\text{ref}}$ against the model resulting from retraining algorithm $\mathcal{A}_{d3}$ only using $X_{1}$. We refer to this third model as $\hat{f}_{1}$, and mark its classification boundary by the solid blue lines in Figure \ref{fig:Example}-A. 

Each of the classifiers in Figure \ref{fig:Example}-A can also be evaluated for its reliance on $X_{1}$, as shown in Figure \ref{fig:Example}-C. Here, we use $\hat{e}_{\text{divide}}$ in our calculation of $\widehat{MR}$ (see Eq \ref{eq:e-divide-def}). Unsurprisingly, the classifier fit without using $X_{1}$ (blue dashed line) has a model reliance of $\widehat{MR}(\hat{f}_{2})=1$. The reference model $f_{\text{ref}}$ (dotted black line) has a model reliance of $\widehat{MR}(f_{\text{ref}})=3.47$. Each $\widehat{MR}$ value has an interpretation contained to a single model. That is, $\widehat{MR}$ compares a \emph{single model's} behavior under different data distributions, rather than the AR approach of comparing \emph{different models'} behavior on marginal distributions from a single joint distribution.

We illustrate MCR in Figure \ref{fig:Example}-B. In contrast to AR, MCR is only ever a function of well-performing prediction models. Here, we consider the empirical $\epsilon$-Rashomon set $\hat{\mathcal{R}}(\epsilon,f_{\text{ref}},\mathcal{F}_{d3})$, the subset of models in $\mathcal{F}_{d3}$ with test loss no more than $\epsilon$ above that of $f_{\text{ref}}$. We show the classification boundary associated with 15 well-performing models contained in $\hat{\mathcal{R}}(\epsilon,f_{\text{ref}},\mathcal{F}_{d3})$ by the gray solid lines. We also show two of the models in $\hat{\mathcal{R}}(\epsilon,f_{\text{ref}},\mathcal{F}_{d3})$ that approximately maximize and minimize empirical reliance on $X_{1}$ among models in $\hat{\mathcal{R}}(\epsilon,f_{\text{ref}},\mathcal{F}_{d3})$. We denote these models as $\hat{f}_{+,\epsilon}$ and $\hat{f}_{-,\epsilon}$, and mark them by the solid green and dashed green lines respectively. For every model shown in Figure \ref{fig:Example}-B, we also mark its model reliance in Figure \ref{fig:Example}-C. We can then see from Figure \ref{fig:Example}-C that $\widehat{MR}$ for each model in $\hat{\mathcal{R}}(\epsilon,f_{\text{ref}},\mathcal{F}_{d3})$ is contained between $\widehat{MR}(\hat{f}_{-,\epsilon})$ and $\widehat{MR}(\hat{f}_{+,\epsilon})$, up to a small approximation error.

In summary, unlike AR, MCR is only a function of models that fit the data well.

\subsection{Simulations of Bootstrap Confidence Intervals\label{subsec:Simulations-of-bootstrap}}

In this section we study the performance of MCR under model class misspecification. Our goal will be to estimate how much the conditional expectation function $f_{0}(x)=\mathbb{E}(Y|X=x)$ relies on subsets of covariates. Given a reference model $f_{\text{ref}}$ and model class $\mathcal{F}$, our ability to describe $MR(f_{0})$ will hinge on two conditions:
\begin{condition}
\emph{(Nearly correct model class)} The class $\mathcal{F}$ contains a well-performing model $\tilde{f}\in\mathcal{R}(\epsilon,f_{\text{ref}},\mathcal{F})$ satisfying $MR(\tilde{f})=MR(f_{0})$ (see Eq \ref{eq:explicit-rashomon-def}).\label{cond:pop-mcr-contains-truth}
\end{condition}

\begin{condition}
\emph{(Bootstrap coverage)} Bootstrap CIs for empirical MCR give appropriate coverage of population-level MCR. \label{cond:bootstrap-works}
\end{condition}

Condition \ref{cond:pop-mcr-contains-truth} ensures that the interval $[MCR_{-}(\epsilon),MCR_{+}(\epsilon)]$ contains $MR(f_{0})$, and Condition \ref{cond:bootstrap-works} ensures that this interval can be estimated in finite samples. Condition \ref{cond:pop-mcr-contains-truth} can also be interpreted as saying that the model reliance value of $MR(f_{c})$ is ``well supported'' by the class $\mathcal{F}$, even if $\mathcal{F}$ does not contain $f_{0}$. Our primary goal is to assess whether CIs derived from MCR can give appropriate coverage of $MR(f_{0})$, which depends on both conditions. As a secondary goal, we also would like to be able to assess Conditions \ref{cond:pop-mcr-contains-truth} \& \ref{cond:bootstrap-works} individually.

Verifying the above conditions requires that we are able to calculate population-level MCR. To this end, we draw samples with replacement from a finite population of 20,000 observations, in which MCR can also be calculated directly. To derive a CI based on MCR, we divide each simulated sample $\mathcal{Z}_{s}$ into a training subset and analysis subset. We use the training subset to fit a reference model $f_{\text{ref},s}$, which is required for our definition of population-level MCR. We calculate a bootstrap CI by drawing 500 bootstrap samples from the analysis subset, and computing $\widehat{MCR}_{-}(\epsilon)$ and $\widehat{MCR}_{+}(\epsilon)$ in each bootstrap sample by optimizing over $\hat{\mathcal{R}}(\epsilon,f_{\text{ref},s},\mathcal{F})$. We then take the 2.5\% percentile of $\widehat{MCR}_{-}(\epsilon)$ values across bootstrap samples, and the 97.5\% percentile of $\widehat{MCR}_{+}(\epsilon)$ values across bootstrap samples, as the lower and upper endpoints of our CI, respectively. We repeat this procedure for both $X_{1}$ and $X_{2}$.

We generate data according to a model with increasing amounts of nonlinearity. For $\gamma\in\{0,0.1,0.2,0.3,0.4,0.5\}$, we simulate continuous outcomes as $Y=f_{0}(X)+E$, where $f_{0}$ is the function $f_{0}(\mathbf{x})=\sum_{j=1}^{p}j\mathbf{x}_{[j]}-\gamma\mathbf{x}_{[j]}^{2}$; the covariate dimension $p$ is equal to 2, with $X_{1}$ and $X_{2}$ defined as the first and second elements of $X$; the covariates $X$ are drawn from a multivariate normal distribution with $\mathbb{E}(X_{1})=\mathbb{E}(X_{2})=0$, $\text{Var}(X_{1})=\text{Var}(X_{2})=0$, and $\text{Cov}(X_{1},X_{2})=1/4$; and $E$ is a normally distributed noise variable with mean zero and variance equal to $\sigma_{E}^{2}:=\text{Var}(f_{0}(X))$. We consider sample sizes of $n=400$ and $800$, of which $n_{tr}=200$ or $300$ observations are assigned to the training subset respectively. 

To implement our approach, we use the model class $\mathcal{F}_{\text{lm}}=\{f_{\bm{\beta}}:f_{\bm{\beta}}(\mathbf{x})=\bm{\beta}_{[1]}+\sum_{j=1}^{2}\mathbf{x}_{[j]}\bm{\beta}_{[j+1]},\bm{\beta}\in\mathbb{R}^{3}\}$. We set the performance threshold $\epsilon$ equal to $0.1\times\sigma_{E}^{2}$. We refer to this MCR implementation with $\mathcal{F}_{\text{lm}}$ as ``MCR-Linear.''

As a comparator method, we consider a simpler bootstrap approach, which we refer to as ``Standard-Linear.'' Here, we take 500 bootstrap samples from the simulated data $\mathcal{Z}_{s}$. In each bootstrap sample, indexed by $b$, we set aside $n_{tr}$ training points to train a model $f_{b}\in\mathcal{F}_{\text{lm}}$, and calculate $\widehat{MR}(f_{b})$ from the remaining data points. We then create a 95\% bootstrap percentile CI for $MR(f_{0})$ by taking the $2.5\%$ and $97.5\%$ percentiles of $\widehat{MR}(f_{b})$ across $b=1,\dots,500$. 

\subsubsection{Results\label{subsec:sim-Results}}

Overall, we find that MCR provides more robust and conservative intervals for the reliance of $f_{0}$ on $X_{1}$ and $X_{2}$, relative to standard bootstrap approaches. We also find that higher sample size generally exacerbates coverage errors due to misspecification, as methods become more certain of biased results. 

\begin{figure}
\begin{centering}
\includegraphics[width=1\columnwidth]{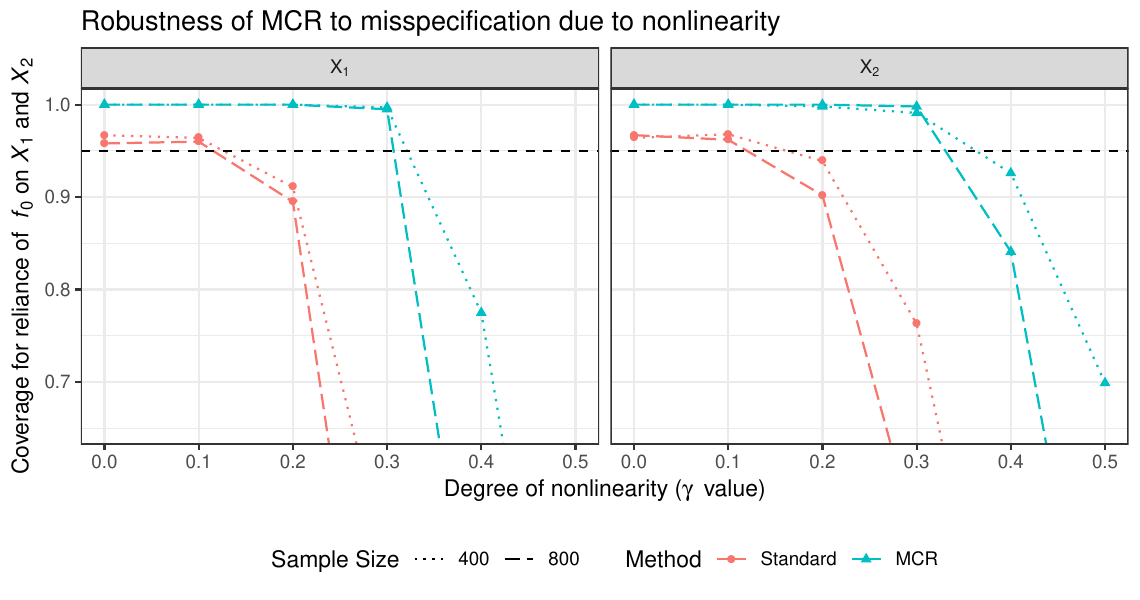}
\par\end{centering}
\caption{MR Coverage - The y-axis shows coverage rate for the reliance of $f_{0}$ on either $X_{1}$ (left column) or $X_{2}$ (right column), where $X_{2}$ is simulated to be more influential that $X_{1}$. The x-axis shows increasing levels of misspecification ($\gamma$). All methods aim to have at least 95\% coverage for each scenario (dashed horizontal line).\label{fig:lm-sim}}
\end{figure}

MCR-Linear gave proper coverage for up to moderate levels of misspecification ($\gamma=0.3$), where Standard-Linear began to break down (Figure \ref{fig:lm-sim}). For larger levels of misspecification ($\gamma\geq0.4$), both MCR-Linear and Standard-Linear failed to give appropriate coverage.

The increased robustness of MCR comes at the cost of wider CIs. Intervals for MCR-Linear were typically larger than intervals for Standard-Linear by a factor of approximately 2-4. This is partly due to the fact that CIs for MCR are meant to cover the range of values $[MCR_{-}(\epsilon),MCR_{+}(\epsilon)]$ (defined using $f_{\text{ref},s}$), rather than to cover a single point.

When investigating Conditions \ref{cond:pop-mcr-contains-truth} \& \ref{cond:bootstrap-works} individually, we find that the coverage errors for MCR-Linear were largely attributable to violations of Condition \ref{cond:pop-mcr-contains-truth}. Condition \ref{cond:bootstrap-works} appears to hold conservatively for all scenarios studied\textendash within each scenario, at least $95.9\%$ of bootstrap CIs contained population-level MCR.

These simulation results highlight an aspect of MCR that is both a strength and a weakness: MCR is generic. MCR does not assume a particular means by which misspecification may occur, and is less powerful than sensitivity analyses which make that assumption correctly. Nonetheless, MCR still appears to add robustness. For sufficiently strong signals, an informative interval may still be returned. In our applied data analysis, below, we see that this is indeed the case.

\section{Data Analysis: Reliance of Criminal Recidivism Prediction Models on Race and Sex\label{sec:Data-Analysis}}

Evidence suggests that bias exists among judges and prosecutors in the criminal justice system \citep{spohn2000thirty_years_reform,blair2004Afrocentric_facial_features,paternoster2008reassessing}\uline{}. In an aim to counter this bias, machine learning models trained to predict recidivism are increasingly being used to inform judges' decisions on pretrial release, sentencing, and parole \citep{monahan2016risk,courtInnovation}. Ideally, prediction models can avoid human bias and provide judges with empirically tested tools. But prediction models can also mirror the biases of the society that generates their training data, and perpetuate the same bias at scale. In the case of recidivism, if arrest rates across demographic groups are not representative of underlying crime rate \citep{beckett2006understanding_seattle,ramchand2006racial_marijuana,usdoj2016_Baltimore_PD_investigation}, then bias can be created in both (1) the outcome variable, future crime, which is measured imperfectly via arrests or convictions, and (2) the covariates, which include the number of prior convictions on a defendant's record \citep{washington_post2016_computer_bias_not_clear,lum2016predict_and_serve}. Further, when a prediction model's behavior and mechanisms are an opaque black box, the model can evade scrutiny, and fail to offer recourse or explanations to individuals rated as ``high risk.''

We focus here on the issue of transparency, which takes an important role in the recent debate about the proprietary recidivism prediction tool COMPAS \citep{propublica2016_how_we_analyzed_COMPAS,washington_post2016_computer_bias_not_clear}. While COMPAS is known to not rely explicitly on race, there is concern that it may rely implicitly on race via proxies\textendash variables statistically dependent with race (see further discussion in Section \ref{sec:Concluding-discussion}).

Our goal is to identify bounds for how much COMPAS relies on different covariate subsets, either implicitly or explicitly, under certain assumptions (defined below). We analyze a public data set of defendants from Broward County, Florida, in which COMPAS scores have been recorded \citep{propublica2016_how_we_analyzed_COMPAS}. Within this data set, we only included defendants measured as African-American or Caucasian (3,373 in total) due to sparseness in the remaining categories. The outcome of interest ($Y$) is the COMPAS violent recidivism score. Of the available covariates, we consider three variables which we refer to as ``admissible'': an individual's age, their number of priors, and an indicator of whether the current charge is a felony. We also consider two variables which we refer to as ``inadmissible'': an individual's race and sex. Our labels of ``admissible'' and ``inadmissible'' are not intended to be legally precise\textendash indeed, the boundary between these types of labels is not always clear (see Section \ref{subsec:Discussion-=000026-limitations}). We compute empirical MCR and AR for each variable group, as well as bootstrap CIs for MCR (see Section \ref{subsec:Simulations-of-bootstrap}).

To compute empirical MCR and AR, we consider a flexible class of linear models in a RKHS to predict the COMPAS score (described in more detail below). Given this class, the MCR range (See Eq \ref{eq:pop-min-max}) captures the highest and lowest degree to which any model in the class may rely on each covariate subset. We assume that our class contains at least one model that relies on ``inadmissible variables'' to the same extent that COMPAS relies either on ``inadmissible variables'' or on proxies that are unmeasured in our sample (analogous to Condition \ref{cond:pop-mcr-contains-truth}). We make the same assumption for ``admissible variables.'' These assumptions can be interpreted as saying that the reliance values of COMPAS are relatively ``well supported'' by our chosen model class, and allows us to identify bounds on the MR values for COMPAS. We also consider the more conventional, but less robust approach of AR (Section \ref{sec:Related-Work}), that is, how much would the accuracy suffer for a model-fitting algorithm trained on COMPAS score if a variable subset was removed?

These computations require that we predefine our loss function, model class, and performance threshold. We define MR, MCR, and AR in terms of the squared error loss $L(f,(y,x_{1},x_{2}))=\{y-f(x_{1},x_{2})\}^{2}$. We define our model class $\mathcal{F}_{\mathbf{D},r_{k}}$ in the form of Eq \ref{eq:RKHS-def}, where we determine $\mathbf{D}$, $\mu$, $k$, and $r_{k}$ based on a subset $\mathcal{S}$ of 500 training observations. We set $\mathbf{D}$ equal to the matrix of covariates from $\mathcal{S}$; we set $\mu$ equal to the mean of $Y$ in $\mathcal{S}$; we set $k$ equal to the radial basis function $k_{\sigma_{s}}(\mathbf{x},\tilde{\mathbf{x}})=\text{exp}\left(-\frac{\|\mathbf{x}-\tilde{\mathbf{x}}\|^{2}}{2\sigma_{s}}\right)$, where we choose $\sigma_{s}$ to minimize the cross-validated loss of a Nadaraya-Watson kernel regression \citep{hastie2009elements} fit to $\mathcal{S}$; and we select the parameters $r_{k}$ by cross-validation on $\mathcal{S}$. We set $\epsilon$ equal to 0.1 times the cross-validated loss on $\mathcal{S}$. Also using $\mathcal{S}$, we train a reference model $f_{\text{ref}}\in\mathcal{F}_{\mathbf{D},r_{k}}$. Using the held-out 2,873 observations, we then estimate $MR(f_{\text{ref}})$ and MCR for $\mathcal{F}_{\mathbf{D},r_{k}}$. To calculate AR, we train models from $\mathcal{F}_{\mathbf{D},r_{k}}$ using $\mathcal{S}$, and evaluate their performance in the held-out observations. 

\subsection{Results\label{subsec:Results-data}}

\begin{figure}[ph]
\begin{centering}
\includegraphics[width=1\columnwidth]{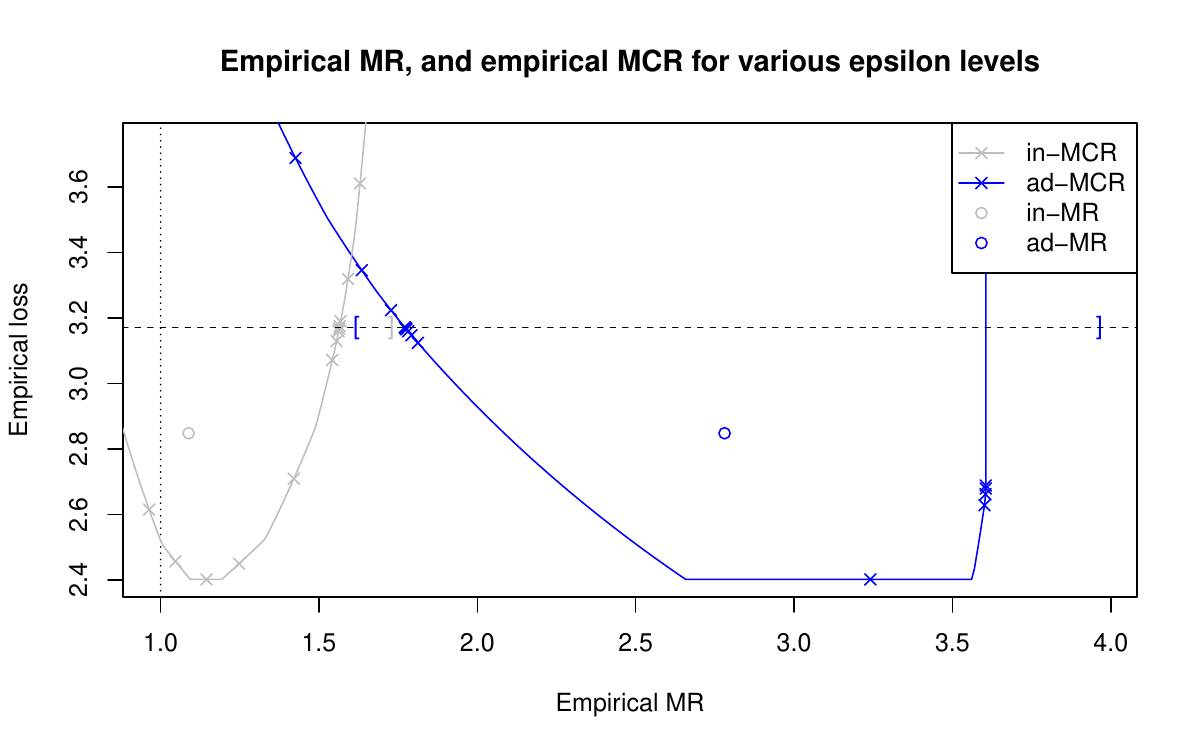}
\par\end{centering}
\caption{\label{fig:Broward}Empirical MR and MCR for Broward County criminal records data set - For any prediction model $f$, the y-axis shows empirical loss ($\hat{e}_{\text{stnd}}(f)$) and the x-axis shows empirical reliance ($\widehat{MR}(f)$) on each covariate subset. Null reliance (MR equal to 1.0) is marked by the vertical dotted line. Reliances on different covariate subsets are marked by color (``admissible'' = blue; ``inadmissible'' = gray). For example, model reliance values for $f_{\text{ref}}$ are shown by the two circular points, one for ``admissible'' variables and one for ``inadmissible'' variables. MCR for different values of $\epsilon$ can be represented as boundaries on this coordinate space. To this end, for each covariate subset, we compute conservative boundary functions (shown as solid lines, or ``bowls'') guaranteed to contain \emph{all models in the class} (see Section \ref{sec:Calculating-MCR}). Specifically, all models in $f\in\mathcal{F}_{\mathbf{D},r_{k}}$ are guaranteed to have an empirical loss ($\hat{e}_{\text{stnd}}(f)$) and empirical MR value ($\widehat{MR}(f)$) for ``inadmissible variables'' corresponding to a point within the gray bowl. Likewise, all models in $\mathcal{F}_{\mathbf{D},r_{k}}$ are guaranteed to have an empirical loss and empirical MR value for ``admissible variables'' corresponding to a point within the blue bowl. Points shown as ``$\times$'' represent additional models in $\mathcal{F}_{\mathbf{D},r_{k}}$ discovered during our computational procedure, and thus show where the ``bowl'' boundary is tight. The goal of our computation procedure (see Section \ref{sec:Calculating-MCR}) is to tighten the boundary as much as possible near the $\epsilon$ value of interest, shown by the dashed horizontal line above. This dashed line has a y-intercept equal to the loss of the reference model plus the $\epsilon$ value of interest. Bootstrap CIs for $MCR_{-}(\epsilon)$ and $MCR_{+}(\epsilon)$ are marked by brackets.}
\end{figure}

Our results imply that race and sex play somewhere between a null role and a modest role in determining COMPAS score, but that they are less important than ``admissible'' factors (Figure \ref{fig:Broward}). As a benchmark for comparison, the empirical MR of $f_{\text{ref}}$ is equal to 1.09 for ``inadmissible variables,'' and 2.78 for ``admissible variables.'' The AR is equal to 0.94 and 1.87 for ``inadmissible'' and ``admissible'' variables respectively, roughly in agreement with MR. The MCR range for ``inadmissible variables'' is equal to {[}1.00,1.56{]}, indicating that \emph{for any model in} $\mathcal{F}_{\mathbf{D},r_{k}}$ with empirical loss no more than $\epsilon$ above that of $f_{\text{ref}}$, the model's loss can increase by no more than 56\% if race and sex are permuted. Such a statement cannot be made solely based on AR or MR methods, as these methods do not upper bound the reliance values of well-performing models. The bootstrap 95\% CI for MCR on ``inadmissible variables'' is {[}1.00, 1.73{]}. Thus, \emph{under our assumptions, if COMPAS relied on sex, race, or their unmeasured proxies by a factor greater than 1.73}, \emph{then intervals as low as what we observe would occur with probability $<0.05$.}

For ``admissible variables'' the MCR range is equal to {[}1.77,3.61{]}, with a 95\% bootstrap CI of {[}1.62, 3.96{]}. \emph{Under our assumptions, this implies if COMPAS relied on age, number of priors, felony indication, or their unmeasured proxies by a factor lower than 1.77, then intervals as high as what we observe would occur with probability $<0.05$.} This result is consistent with \citet{RudinWaCo19}, who find age to be highly predictive of COMPAS score. 

It is worth noting that the upper limit of 3.61 maximizes empirical MR on ``admissible variables'' not only among well-performing models, but globally across all models in the class (see Figure \ref{fig:Broward}, and Eq \ref{eq:global-upper-bnd}). In other words, it is not possible to find models in $\mathcal{F}_{\mathbf{D},r_{k}}$ that perform arbitrarily poorly on perturbed data, but still perform well on unperturbed data, and so the ratio of $\hat{e}_{\text{switch}}(f)$ to $\hat{e}_{\text{orig}}(f)$ has a finite upper bound. Because the regularization constraints of $\mathcal{F}_{\mathbf{D},r_{k}}$ preclude MR values higher than 3.61, the MR of COMPAS on ``admissible variables'' may be underestimated by empirical MCR. Note also that both MCR intervals are left-truncated at 1, as it is often sufficiently precise to conclude that there exists a well-performing model with no reliance on the variables of interest (that is, MR equal to 1; see Appendix \ref{subsec:Model-reliance-leq-1}). 

\subsection{Discussion \& Limitations\label{subsec:Discussion-=000026-limitations}}

Asking whether a proprietary model relies on sex and race, after adjusting for other covariates, is related to the fairness metric known as conditional statistical parity (CSP). A decision rule satisfies CSP if its decisions are independent of a sensitive variable, conditional on a set of ``legitimate'' covariates $C$ (\citealp{corbett2017algorithmic}; see also \citealp{kamiran2013quantifying}). Roughly speaking, CSP reflects the idea that groups of people with similar covariates $C$ are treated similarly \citep{dwork2012awareness_similar_lipschitz}, regardless of the sensitive variable (for example, race or sex). However, the criteria becomes superficial if too many variables are included in $C$, and care should be taken to avoid including proxies for the sensitive variables. Several other fairness metrics have also been proposed, which often form competing objectives \citep{kleinberg2016inherent_tradeoff,chouldechova2017fair_conflicting_definition,nabi2018fair_inference,corbett2017algorithmic}. Here, if COMPAS was not influenced by race, sex, or variables related to race or sex (conditional on a set of ``legitimate'' variables), it would satisfy CSP.

Unfortunately, it is often difficult to distinguish between ``legitimate'' (or ``admissible'') variables and ``illegitimate'' variables. Some variables function both as part of a reasonable predictor for risk, and, separately, as a proxy for race. Because of disproportional arrest rates, particularly for misdemeanors and drug-related offenses \citep{usdoj2016_Baltimore_PD_investigation,lum2016predict_and_serve}, prior misdemeanor convictions may act as such a proxy \citep{washington_post2016_computer_bias_not_clear,lum2016predict_and_serve}.

Proxy variables for race (defined as being statistically dependent with race) that are unmeasured in our sample are also not the only reason that race could be predictive of COMPAS score. Other inputs to the COMPAS algorithm might be associated with race \emph{only} \emph{conditionally} on variables we categorize as ``admissible.'' However, our result from Section \ref{subsec:Results-data} that race has limited predictive utility for COMPAS score suggests that such conditional relationships are also limited.

\section{Conclusion\label{sec:Concluding-discussion}}

In this article, we propose MCR as the upper and lower limit on how important a set of variables can be to any well-performing model in a class. In this way, MCR provides a more comprehensive and robust measure of importance than traditional importance measures for a single model. We derive bounds on MCR, which motivate our choice of point estimates. We also derive connections between permutation importance, U-statistics, conditional variable importance, and conditional causal effects. We apply MCR in a data set of criminal recidivism, in order to help inform the characteristics of the proprietary model COMPAS. 

Several exciting areas remain open for future research. One research direction closely related to our current work is the development of exact or approximate MCR computation procedures for other model classes and loss functions. We have shown that, for model classes where minimizing the empirical loss is a convex optimization problem, MCR can be conservatively computed via a series of convex optimization problems. Further, we have shown that computing $\widehat{MCR}_{-}$ is often no more challenging that minimizing the empirical loss over a reweighted sample. General computation procedures for MCR are still an open research area.

Another direction is to consider MCR for variable selection. If $MCR_{+}$ is small for a variable, then no well-performing predictive model can heavily depend on that variable, indicating that it can be eliminated.

Our theoretical analysis of Rashomon sets depends on $\mathcal{F}$ and $f_{\text{ref}}$ being prespecified. Above, we have actualized this by splitting our sample into subsets of size $n_{1}$ and $n_{2}$, using the first subset to determine $\mathcal{F}$ and $f_{\text{ref}}$, and conditioning on $\mathcal{F}$ and $f_{\text{ref}}$ when estimating MCR in the second subset. As a result, the boundedness constants in our assumptions ($B_{\text{ind}}$, $B_{\text{ref}},B_{\text{switch}},$ and $b_{\text{orig}}$) depend on $\mathcal{F}$, and hence on $n_{1}$. However, because our results are non-asymptotic, we have not explored how Rashomon sets behave when $n_{1}$ and $n_{2}$ grow at different rates. An exciting future extension of this work is to study sequences of triples $\{\epsilon_{n_{1}},f_{\text{ref},n_{1}},\mathcal{F}_{n_{1}}\}$ that change as $n_{1}$ increases, and the corresponding Rashomon sets $\mathcal{R}(\epsilon_{n_{1}},f_{\text{ref},n_{1}},\mathcal{F}_{n_{1}})$, as this may more thoroughly capture how model classes are determined by analysts.

While we develop Rashomon sets with the goal of studying MR, Rashomon sets can also be useful for finite sample inferences about a wide variety of other attributes of best-in-class models (for example, Section \ref{sec:Connection-confidence-CIs}). Characterizations of a Rashomon set itself may also be of interest. For example, in ongoing work, we are studying the size of a Rashomon set, and its connection to generalization of models and model classes \citep{semenova2019study}. We are additionally developing methods for visualizing Rashomon sets \citep{dong2019variable}.

\acks{}

Support for this work was provided by the National Institutes of Health (grants P01CA134294, R01GM111339, R01ES024332, R35CA197449, R01ES026217, P50MD010428, DP2MD012722, R01MD012769, \& R01ES028033), by the Environmental Protection Agency (grants 83615601 \& 83587201-0), and by the Health Effects Institute (grant 4953-RFA14-3/16-4).

\appendix

\section{Miscellaneous Supplemental Sections}

All labels for items in the following appendices begin with a letter (for example, Section \ref{subsec:Model-reliance-leq-1}), while references to items in the main text contain only numbers (for example, Proposition \ref{thm:VI_TRT}).

\subsection{Code}

R code for our example in Section \ref{sec:Simulations} and analysis in Section \ref{sec:Data-Analysis}  is available at
\url{https://github.com/aaronjfisher/mcr-supplement}.

\subsection{Model Reliance Less than 1\label{subsec:Model-reliance-leq-1}}

While it is counterintuitive, it is possible for the expected loss of a prediction model to \emph{decrease }when the information in $X_{1}$ is removed. Roughly speaking, a ``pathological'' model $f_{\text{silly}}$ may use the information in $X_{1}$ to ``intentionally'' misclassify $Y$, such that $e_{\text{switch}}(f_{\text{silly}})<e_{\text{orig}}(f_{\text{silly}})$ and $MR(f_{\text{silly}})<1$. The model $f_{\text{silly}}$ may even be included in a population $\epsilon$-Rashomon set (see Section \ref{subsec:Model-class-reliance}) if it is still possible to predict $Y$ sufficiently well from the information in $X_{2}$.

However, in these cases there will often exist another model that outperforms $f_{\text{silly}}$, and that has MR equal to 1 (i.e., no reliance on $X_{1}$). To see this, consider the case where $\mathcal{F}=\{f_{\bm{\theta}}:\bm{\theta}\in\mathbb{R}^{d}\}$ is indexed by a parameter $\bm{\theta}$. Let $\bm{\theta}_{\text{silly}}$ and $\bm{\theta}^{\star}$ be parameter values such that $f_{\bm{\theta}_{\text{silly}}}$ is equivalent to $f_{\text{silly}}$, and $f_{\bm{\theta}^{\star}}$ is the best-in-class model. If $f_{\bm{\theta}^{\star}}$ satisfies $MR(f_{\bm{\theta}^{\star}})>1$ and if the model reliance function $MR$ is continuous in $\bm{\theta}$, then there exists a parameter value $\bm{\theta}_{1}$ between $\bm{\theta}_{\text{silly}}$ and $\bm{\theta}^{\star}$ such that $MR(f_{\bm{\theta}_{1}})=1$. Further, if the loss function $L$ is convex in $\bm{\theta}$, then $e_{\text{orig}}(f_{\bm{\theta}^{\star}})\leq e_{\text{orig}}(f_{\bm{\theta}_{1}})\leq e_{\text{orig}}(f_{\text{silly}})$, and any population $\epsilon$-Rashomon set containing $f_{\text{silly}}$ will also contain $f_{\bm{\theta}_{1}}$.

\subsection[Relating the ``Switched'' Loss to All Possible Permutations of the Sample]{Relating $\hat{e}_{\text{switch}}(f)$ to All Possible Permutations of the Sample\label{subsec:all-permutations}}

\textcolor{black}{Following the notation in Section \ref{subsec:model-reliance}, let $\{\bm{\pi}_{1},\dots,\bm{\pi}_{n!}\}$ be a set of $n$-length vectors, each containing a different permutation of the set $\{1,\dots,n\}$. We show in this section that $\hat{e}_{\text{switch}}(f)$ is equal to the product of 
\begin{equation}
{\sum_{l=1}^{n!}\sum_{i=1}^{n}L\{f,(\mathbf{y}_{[i]},\mathbf{X}_{1[\bm{\pi}_{l[i]},\cdot]},\mathbf{X}_{2[i,\cdot]})\}1(\bm{\pi}_{l[i]}\neq i)},\label{eq:all-perm-indicator}
\end{equation}
and a proportionality constant that is only a function of $n$.}

\textcolor{black}{First, consider the sum 
\begin{equation}
\sum_{l=1}^{n!}\sum_{i=1}^{n}L\{f,(\mathbf{y}_{[i]},\mathbf{X}_{1[\bm{\pi}_{l[i]},\cdot]},\mathbf{X}_{2[i,\cdot]})\},\label{eq:all-perm}
\end{equation}
 which omits the indicator function found in Eq \ref{eq:all-perm-indicator}.}

\textcolor{black}{The summation in Eq \ref{eq:all-perm} contains $n(n!)$ terms, each of which is a two-way combination of the form $L\{f,(\mathbf{y}_{[i]},\mathbf{X}_{1[j,\cdot]},\mathbf{X}_{2[i,\cdot]})\}$ for $i,j\in\{1,\dots n\}$. There are only $n^{2}$ unique combinations of this form, and each must occur in at least $(n-1)!$ of the $n(n!)$ terms in Eq \ref{eq:all-perm}. To see this, consider selecting two integer values $\tilde{i},\tilde{j}\in\{1,\dots n\}$, and enumerating all occurrences of the term $L\{f,(\mathbf{y}_{[\tilde{i}]},\mathbf{X}_{1[\tilde{j},\cdot]},\mathbf{X}_{2[\tilde{i},\cdot]})\}$ within the sum in Eq \ref{eq:all-perm}. Of the permutation vectors $\{\bm{\pi}_{1},\dots,\bm{\pi}_{n!}\}$, we know that $(n-1)!$ of them place $\tilde{i}$ in the $\tilde{j}^{th}$ position, i.e., that satisfy $\bm{\pi}_{l[\tilde{i}]}=\tilde{j}$. For each such permutation $\bm{\pi}_{l}$, the inner summation in Eq \ref{eq:all-perm} over all possible values of $i$ must include the term $L\{f,(\mathbf{y}_{[\tilde{i}]},\mathbf{X}_{1[\bm{\pi}_{l[\tilde{i}]},\cdot]},\mathbf{X}_{2[\tilde{i},\cdot]})\}=L\{f,(\mathbf{y}_{[\tilde{i}]},\mathbf{X}_{1[\tilde{j},\cdot]},\mathbf{X}_{2[\tilde{i},\cdot]})\}$. Thus, Eq \ref{eq:all-perm} contains at least $(n-1)!$ occurrences of the term $L\{f,(\mathbf{y}_{[\tilde{i}]},\mathbf{X}_{1[\tilde{j},\cdot]},\mathbf{X}_{2[\tilde{i},\cdot]})\}$. }

\textcolor{black}{So far, we have shown that each unique combination occurs at least $(n-1)!$ times, but it also follows that each unique combination must occur precisely $(n-1)!$ times. This is because each of the $n^{2}$ unique combinations must occur at least $(n-1)!$ times, which accounts for $n^{2}((n-1)!)=n(n!)$ terms in total. As noted above, Eq has \ref{eq:all-perm} has only $n(n)!$ terms, so there can be no additional terms. We can then simplify Eq \ref{eq:all-perm} as
\[
\sum_{l=1}^{n!}\sum_{i=1}^{n}L\{f,(\mathbf{y}_{[i]},\mathbf{X}_{1[\bm{\pi}_{l[i]},\cdot]},\mathbf{X}_{2[i,\cdot]})\}=(n-1)!\sum_{i=1}^{n}\sum_{j=1}^{n}L\{f,(\mathbf{y}_{[i]},\mathbf{X}_{1[j,\cdot]},\mathbf{X}_{2[i,\cdot]})\}.
\]
}

\textcolor{black}{By the same logic, we can simplify Eq \ref{eq:all-perm-indicator} as}

\textcolor{black}{
\begin{align}
 & \sum_{l=1}^{n!}\sum_{i=1}^{n}L\{f,(\mathbf{y}_{[i]},\mathbf{X}_{1[\bm{\pi}_{l[i]},\cdot]},\mathbf{X}_{2[i,\cdot]})\}1(\bm{\pi}_{l[i]}\neq i)\nonumber \\
= & (n-1)!\left\{ \sum_{i=1}^{n}\sum_{j=1}^{n}L\{f,(\mathbf{y}_{[i]},\mathbf{X}_{1[j,\cdot]},\mathbf{X}_{2[i,\cdot]})\}1(j\neq i)\right\} \nonumber \\
= & (n-1)!\sum_{i=1}^{n}\sum_{j\neq i}L\{f,(\mathbf{y}_{[i]},\mathbf{X}_{1[j,\cdot]},\mathbf{X}_{2[i,\cdot]})\},\label{eq:simplified-all-perm}
\end{align}
and Line \ref{eq:simplified-all-perm} is proportional to $\hat{e}_{\text{switch}}(f)$ up to a function of $n$.}

\subsection{Bound for MR of the Best-in-class Prediction Model\label{subsec:Bound-for-MR-f-star}}

Although describing individual models is not the primary focus of this work, a corollary of Theorem \ref{thm:mcr-conserve-bounds} is that we can create a probabilistic bound for the reliance of the (unknown) best-in-class model $f^{\star}$ on $X_{1}$.
\begin{corollary}
\label{cor:best-in-class}(Bound on Best-in-class MR) Let $f^{\star}\in\argmin_{f\in\mathcal{F}}e_{\text{orig}}(f)$ be a prediction model that attains the lowest possible expected loss, and let $f_{+,\epsilon}$ and $f_{-,\epsilon}$ be defined as in Theorem \ref{thm:mcr-conserve-bounds}. If $f_{+,\epsilon}$ and $f_{-,\epsilon}$ satisfy Assumptions \ref{assu:bnd-ind}, \ref{assu:bnd-ref} and \ref{assu:bnd-avg}, then
\begin{align*}
\mathbb{P}\left(MR(f^{\star})\in\left[\widehat{MCR}_{-}\left(\epsilon_{\text{best}}\right)-\mathcal{Q}_{\text{best}},\hspace{1em}\widehat{MCR}_{+}\left(\epsilon_{\text{best}}\right)+\mathcal{Q}_{\text{best}}\right]\right) & \geq1-\delta,
\end{align*}

where $\epsilon_{\text{best}}:=2B_{\text{ref}}\sqrt{\frac{\log(6\delta^{-1})}{2n}}$ , and $\mathcal{Q}_{\text{best}}:=\frac{B_{\text{switch}}}{b_{\text{orig}}}-\frac{B_{\text{switch}}-B_{\text{ind}}\sqrt{\frac{\log(12\delta^{-1})}{n}}}{b_{\text{orig}}+B_{\text{ind}}\sqrt{\frac{\log(12\delta^{-1})}{2n}}}$.
\end{corollary}

\textcolor{black}{The above result does not require that $f^{\star}$ be unique. If several models achieve the minimum possible expected loss, the above boundaries apply simultaneously for each of them. In the special case when the true conditional expectation function $\mathbb{E}(Y|X_{1},X_{2})$ is equal to $f^{\star}$, then we have a boundary for the reliance of the function $\mathbb{E}(Y|X_{1},X_{2})$ on $X_{1}$. This reliance bound can also be translated into a causal statement using Proposition \ref{thm:VI_TRT}.}

\subsection{Ratios versus Differences in MR Definition\label{sec:Discussion---Ratios}}

We choose our ratio-based definition of model reliance, $MR(f)=\frac{e_{\text{switch}}(f)}{e_{\text{orig}}(f)}$, so that the measure can be comparable across problems, regardless of the scale of $Y$. However, several existing works define VI measures in terms of differences \citep{strobl2008conditional_VI,datta2016algorithmic_qii,gregorutti2017correlation_LM_VI_RF}, analogous to 
\begin{equation}
MR_{\text{difference}}(f):=e_{\text{switch}}(f)-e_{\text{orig}}(f).\label{eq:mr-diff}
\end{equation}
While this difference measure is less readily interpretable, it has several computational advantages. The mean, variance, and asymptotic distribution of the estimator $\widehat{MR}_{\text{difference}}(f):=\hat{e}_{\text{switch}}(f)-\hat{e}_{\text{orig}}(f)$ can be easily determined using results for U-statistics, without the use of the delta method (\citealp{dorfman1938note,lehmann2006theory_point_estimation}; see also \citealp{ver2012invented}). Estimates in the form of $\widehat{MR}_{\text{difference}}(f)$ will also be more stable when $\min_{f\in\mathcal{F}}e_{\text{orig}}(f)$ is small, relative to estimates for the ratio-based definition of MR. To improve interpretability, we may also normalize $MR_{\text{difference}}(f)$ by dividing by the variance of $Y$, which can be easily estimated without the use of models, as in \citet{williamson2017nonparametric_VI}.

Under the difference-based definition for MR (Eq \ref{eq:mr-diff}), the results from Theorem \ref{thm:mcr-conserve-bounds}, Theorem \ref{thm:mcr-consistent}, and Corollary \ref{cor:best-in-class} will still hold under the following modified definitions of $\mathcal{Q}_{\text{out}}$, $\mathcal{Q}_{\text{in}}$, and $\mathcal{Q}_{\text{best}}$:
\begin{align*}
\mathcal{Q}_{\text{out},\text{difference}}:= & \left(1+\frac{1}{\sqrt{2}}\right)B_{\text{ind}}\sqrt{\frac{\log(6\delta^{-1})}{n}},\\
\mathcal{Q}_{\text{in},\text{difference}}:= & B_{\text{ind}}\left\{ \sqrt{\frac{\log(8\delta^{-1}\mathcal{N}\left(\mathcal{F},r\sqrt{2}\right))}{n}}+\sqrt{\frac{\log(8\delta^{-1}\mathcal{N}\left(\mathcal{F},r\right))}{2n}}\right\} +2r(\sqrt{2}+1),\text{ and}\\
\mathcal{Q}_{\text{best},\text{difference}}:= & \left(1+\frac{1}{\sqrt{2}}\right)B_{\text{ind}}\sqrt{\frac{\log(12\delta^{-1})}{n}}.
\end{align*}

Respectively replacing $\mathcal{Q}_{\text{out}}$, $\mathcal{Q}_{\text{in}}$, $\mathcal{Q}_{\text{best}}$, $MR$, and $\widehat{MR}$ with $\mathcal{Q}_{\text{out},\text{difference}}$, $\mathcal{Q}_{\text{in},\text{difference}}$, $\mathcal{Q}_{\text{best},\text{difference}}$, $MR_{\text{difference}}$ and $\widehat{MR}{}_{\text{difference}}$ entails only minor changes to the corresponding proofs (see Appendices \ref{subsec:Proof-mcr-conserve-bounds}, \ref{subsec:Proof-of-learning-bnd}, and \ref{subsec:Proof-cor-best-in-class}). The results will also hold without Assumption \textcolor{black}{\ref{assu:bnd-avg}}, as is suggested by the fact that $b_{\text{orig}}$ and $B_{\text{switch}}$ do not appear in $\mathcal{Q}_{\text{out},\text{difference}}$, $\mathcal{Q}_{\text{in},\text{difference}}$, or $\mathcal{Q}_{\text{best},\text{difference}}$.

We also prove an analogous version of Theorem \ref{thm:MR-unif}, on uniform bounds for $\widehat{MR}{}_{\text{difference}}$, in Appendix \ref{subsec:Proof-of-other-cover-limits}.

\subsection{Rashomon Sets and Profile Likelihood Intervals\label{subsec:Rashomon-sets-and-profile}}

We note in Section \ref{subsec:Finite-sample-CIs-general} that, under certain conditions, the CIs returned from Proposition \ref{rem:simplified-thm3-general} take the same form as profile likelihood CIs \citep{coker2018hacking_interval}. For completeness, we briefly review this connection. We assume here that models $f_{\theta}\in\mathcal{F}$ are indexed by a finite dimensional parameter vector $\theta\in\Theta$, where $\theta=(\gamma,\psi)$ contains a 1-dimensional parameter of interest $\gamma\in\mathbb{R}^{1}$, and a nuisance parameter $\psi\in\Psi$. We further assume and that $e_{\text{orig}}(f_{\theta})$ is minimized by a unique parameter value $\theta^{\star}=(\gamma^{\star},\psi^{\star})\in\Theta$, and that our goal is to learn about $\gamma^{\star}$.

If $s_{\theta}:=\int_{\mathcal{Z}}\exp\{-L(f_{\theta},z)\}dz$ is finite for all $\theta\in\Theta$, we can convert $L$ into the likelihood function $\mathcal{L}:(\mathcal{Z}\times\Theta)\rightarrow\mathbb{R}^{1}$ satisfying $\mathcal{L}(z;\theta)=\exp\{-L(f_{\theta},z)\}/s_{\theta}$. As an abbreviation, let $\mathcal{L}(\mathbf{Z};\theta)$ denote $\prod_{i=1}^{n}\mathcal{L}(\mathbf{Z}_{[i,\cdot]};\theta)$. Additionally, let $\hat{\theta}:=\argmin_{\theta\in\Theta}\hat{e}_{\text{orig}}(f_{\theta})$ be the empirical loss minimizer, and hence the maximum likelihood estimator of $\theta^{\star}$. If $\mathcal{L}$ is indeed the correct likelihood function, then $\theta^{\star}=(\gamma^{\star},\psi^{\star})$ corresponds to the true parameter vector. Further, if $\phi(f_{\theta})=\phi(f_{(\gamma,\psi)})=\gamma$ returns the parameter element of interest ($\gamma$), then the $(1-\delta)$-level profile likelihood interval for $\phi(f_{\theta^{\star}})=\gamma^{\star}$ is
\begin{align}
\text{PLI}(\delta):= & \left\{ \gamma\::\:\log\mathcal{L}(\mathbf{Z};\hat{\theta})-\log\mathcal{L}(\mathbf{Z};\hat{\theta}_{\gamma})\leq\frac{\chi_{1,1-\delta}}{2},\:\text{ where }\:\hat{\theta}_{\gamma}=\argmax_{\{\theta\in\Theta\,:\,\phi(f_{\theta})=\gamma\}}\mathcal{L}(\mathbf{Z};\theta)\right\} \nonumber \\
= & \left\{ \gamma\::\:\exists\hat{\theta}_{\gamma}\:\text{ satisfying }\:\phi\left(f_{\hat{\theta}_{\gamma}}\right)=\gamma\:\text{ and }\:\log\mathcal{L}(\mathbf{Z};\hat{\theta})-\log\mathcal{L}(\mathbf{Z};\hat{\theta}_{\gamma})\leq\frac{\chi_{1,1-\delta}}{2}\right\} \nonumber \\
= & \left\{ \gamma\::\:\exists\hat{\theta}_{\gamma}\:\text{ satisfying }\:\phi\left(f_{\hat{\theta}_{\gamma}}\right)=\gamma\:\text{ and }\:\hat{e}_{\text{orig}}\left(f_{\hat{\theta}_{\gamma}}\right)\leq\hat{e}_{\text{orig}}\left(f_{\hat{\theta}}\right)+\frac{\chi_{1,1-\delta}}{2n}\right\} \nonumber \\
= & \left\{ \gamma\::\:\exists f_{\hat{\theta}_{\gamma}}\:\text{ satisfying }\:\phi\left(f_{\hat{\theta}_{\gamma}}\right)=\gamma\:\text{ and }\:f_{\hat{\theta}_{\gamma}}\in\hat{\mathcal{R}}\left(\frac{\chi_{1,1-\delta}}{2n},f_{\hat{\theta}},\mathcal{F}\right)\right\} \label{eq:like-a-rashomon-1}
\end{align}
where $\chi_{1,1-\delta}$ is the $1-\delta$ percentile of a chi-square distribution with 1 degree of freedom. If $\text{PLI}(\alpha)$ is indeed a contiguous interval, then maximizing and minimizing $\phi(f_{\theta})$ across models $f_{\theta}$ in the empirical Rashomon set in Eq \ref{eq:like-a-rashomon-1} yields the same interval. 

\subsection{Unbiased Estimates of CMR\label{subsec:Unbiased-CMR}}

We claim in Section \ref{subsec:mcr-impute} that both

\[
\hat{e}_{\text{match}}(f)=\frac{1}{n(n-1)}\sum_{i=1}^{n}\sum_{j\neq i}\frac{1(\mathbf{X}_{2[j,\cdot]}=\mathbf{X}_{2[i,\cdot]})}{\mathbb{P}(X_{2}=\mathbf{X}_{2[i,\cdot]})}\times L\{f,(\mathbf{y}_{[j]},\mathbf{X}_{1[i,\cdot]},\mathbf{X}_{2[j,\cdot]})\}.
\]
and
\[
\hat{e}_{\text{weight}}(f)=\frac{1}{n(n-1)}\sum_{i=1}^{n}\sum_{j\neq i}\frac{\mathbb{P}(X_{1}=\mathbf{X}_{1[i,\cdot]}|X_{2}=\mathbf{X}_{2[j,\cdot]})}{\mathbb{P}(X_{1}=\mathbf{X}_{1[i,\cdot]})}\times L\{f,(\mathbf{y}_{[j]},\mathbf{X}_{1[i,\cdot]},\mathbf{X}_{2[j,\cdot]})\},
\]
are unbiased for
\begin{align*}
e_{\text{cond}}(f) & =\mathbb{E}_{X_{2}}\mathbb{E}\left[L\{f,(Y^{(b)},X_{1}^{(a)},X_{2}^{(b)})\}|X_{2}^{(a)}=X_{2}^{(b)},X_{2}\right].
\end{align*}

To show that $\hat{e}_{\text{match}}(f)$ is unbiased, we first note that each summation term in $\hat{e}_{\text{match}}(f)$ has the same expectation. Following the notation in Section \ref{subsec:model-reliance}, let $Z^{(a)}=(Y^{(a)},X_{1}^{(a)},X_{2}^{(a)})$ and $Z^{(b)}=(Y^{(b)},X_{1}^{(b)},X_{2}^{(b)})$ be independent random variables following the same distribution as $Z=(Y,X_{1},X_{2})$. The expectation of $\hat{e}_{\text{match}}(f)$ is
\begin{align*}
\mathbb{E}\hat{e}_{\text{match}}(f)= & \mathbb{E}\left[\frac{1(X_{2}^{(a)}=X_{2}^{(b)})}{p_{x_{2}}(X_{2}^{(a)})}\times L\{f,(Y^{(b)},X_{1}^{(a)},X_{2}^{(b)})\}\right]\\
= & \mathbb{E}_{X_{2}^{(a)}}\mathbb{E}\left[\frac{1(X_{2}^{(a)}=X_{2}^{(b)})}{p_{x_{2}}(X_{2}^{(a)})}\times L\{f,(Y^{(b)},X_{1}^{(a)},X_{2}^{(b)})\}|X_{2}^{(a)}\right]\\
= & \mathbb{E}_{X_{2}^{(a)}}\left\{ p_{x_{2}}(X_{2}^{(a)})\mathbb{E}\left[\frac{1}{p_{x_{2}}(X_{2}^{(a)})}\times L\{f,(Y^{(b)},X_{1}^{(a)},X_{2}^{(b)})\}|X_{2}^{(a)}=X_{2}^{(b)},X_{2}^{(a)}\right]+0\right\} \\
= & \mathbb{E}_{X_{2}^{(a)}}\mathbb{E}\left[L\{f,(Y^{(b)},X_{1}^{(a)},X_{2}^{(b)})\}|X_{2}^{(a)}=X_{2}^{(b)},X_{2}^{(a)}\right]\\
= & e_{\text{cond}}(f).
\end{align*}
To show that $\hat{e}_{\text{weight}}(f)$ is unbiased, we similarly note that each summation term in $\hat{e}_{\text{weight}}(f)$ has the same expectation. Without loss of generality, we show the result for discrete variables $(Y,X_{1},X_{2})$. Let $\mathcal{Y}_{x_{2}}$ be the domain of $Y$ conditional on the event that $X_{2}=x_{2}$. The expectation of $\hat{e}_{\text{weight}}(f)$ is\textcolor{red}{}
\begin{align*}
\mathbb{E}\hat{e}_{\text{weight}}(f) & =\sum_{x_{2}^{(b)}\in\mathcal{X}_{2}}\,\sum_{y^{(b)}\in\mathcal{Y}_{x_{2}^{(b)}}}\,\sum_{x_{1}^{(a)}\in\mathcal{X}_{1}}\left[L\{f,(y^{(b)},x_{1}^{(a)},x_{2}^{(b)})\}\left\{ \frac{\mathbb{P}(X_{1}=x_{1}^{(a)}|X_{2}=x_{2}^{(b)})}{\mathbb{P}(X_{1}=x_{1}^{(a)})}\right\} \right.\\
 & \hspace{2cm}\left.\times\vphantom{\sum_{y\in Y}}\mathbb{P}(X_{1}=x_{1}^{(a)})\mathbb{P}(Y=y^{(b)},X_{2}=x_{2}^{(b)})\right]\\
 & =\sum_{x_{2}^{(b)}\in\mathcal{X}_{2}}\mathbb{P}(X_{2}=x_{2}^{(b)})\sum_{y^{(b)}\in\mathcal{Y}_{x_{2}^{(b)}}}\,\sum_{x_{1}^{(a)}\in\mathcal{X}_{1}}\left[L\{f,(y^{(b)},x_{1}^{(a)},x_{2}^{(b)})\}\right.\\
 & \hspace{2cm}\left.\times\mathbb{P}(X_{1}=x_{1}^{(a)}|X_{2}=x_{2}^{(b)})\mathbb{P}(Y=y^{(b)}|X_{2}=x_{2}^{(b)})\right]\\
 & =\mathbb{E}_{X_{2}^{(b)}}\mathbb{E}\left[\int L\{f,(Y^{(b)},X_{1}^{(a)},X_{2}^{(b)})\}|X_{2}^{(a)}=X_{2}^{(b)},X_{2}^{(b)})\right]\\
 & =e_{\text{cond}}(f).
\end{align*}

\section{Proofs for Statistical Results\label{sec:Proofs}}

We present proofs for our statistical results in this section, and conclude by presenting proofs for our computational results in Appendix \ref*{subsec:Proofs-for-binary}.

\subsection{Lemma Relating Empirical and Population Rashomon Sets\label{subsec:Lemma-relating-empirical}}

\textcolor{black}{Throughout the remaining proofs, it will be useful to express the definition of population $\epsilon$-Rashomon sets in terms of the expectation of a single loss function, rather than a comparison of two loss functions. To do this, we simply introduce the ``standardized'' loss function $\tilde{L}$, defined as 
\begin{equation}
\tilde{L}(f,z):=L(f,z)-L(f_{\text{ref}},z).\label{eq:L-tilde-reference}
\end{equation}
Above, recall from Section \ref{sec:notation} that $L(f,z)$ denotes $L(f,(y,x_{1},x_{2}))$ for $z=(y,x_{1},x_{2})$. Because we assume $f_{\text{ref}}$ is prespecified and fixed, we omit notation for $f_{\text{ref}}$ in the definition of $\tilde{L}$. We can now write}

\textcolor{black}{
\begin{align*}
\mathcal{R}(\epsilon) & =\{f_{\text{ref}}\}\cup\left\{ f\in\mathcal{F}\text{ : }\mathbb{E}L(f,Z)\leq\mathbb{E}L(f_{\text{ref}},Z)+\epsilon\right\} \\
 & =\{f_{\text{ref}}\}\cup\left\{ f\in\mathcal{F}\text{ : }\mathbb{E}\tilde{L}(f,Z)\leq\epsilon\right\} ,
\end{align*}
and, similarly,}

\textcolor{black}{
\begin{align*}
\hat{\mathcal{R}}(\epsilon) & =\{f_{\text{ref}}\}\cup\left\{ f\in\mathcal{F}\text{ : }\hat{\mathbb{E}}\tilde{L}(f,Z)\leq\epsilon\right\} .
\end{align*}
}

\textcolor{black}{With this definition, the following lemma allows us to limit the probability that a given model $f_{1}\in\mathcal{R}(\epsilon)$ is excluded from an empirical Rashomon set.}
\begin{lemma}
\textcolor{black}{\label{lem:hoeffding}For $\epsilon\in\mathbb{R}$ and $\delta\in(0,1)$, let $\epsilon_{1}':=\epsilon+2B_{\text{ref}}\sqrt{\frac{\log(\delta^{-1})}{2n}}$, and let $f_{1}\in\mathcal{R}(\epsilon)$ denote a specific, possibly unknown prediction model. If $f_{1}$ satisfies Assumption \ref{assu:bnd-ref}, then }

\textcolor{black}{
\[
\mathbb{P}\{f_{1}\in\hat{\mathcal{R}}(\epsilon_{1}')\}\geq1-\delta.
\]
}
\end{lemma}

\begin{proof}
\textcolor{black}{If $f_{\text{ref}}$ and $f_{1}$ are the same function, then the result holds trivially. Otherwise, the proof follows from Hoeffding's inequality }(Theorem 2 of \citealp{hoeffding1963_bounded_sums_inequalities})\textcolor{black}{. First, note that if $f_{1}$ satisfies Assumption \ref{assu:bnd-ref}, then $\tilde{L}(f_{1})$ is bounded within an interval of length $2B_{\text{ref}}$. Applying this in line \ref{eq:apply-abs-value-bound-hoeff}, below, we see that}
\begin{align}
\mathbb{P}\{f_{1}\notin\hat{\mathcal{R}}(\epsilon_{1}')\} & =\mathbb{P}\left[\hat{\mathbb{E}}\tilde{L}(f_{1},Z)>\epsilon_{1}'\right] &  & \text{from }f_{1}\notin\{f_{\text{ref}}\}\nonumber \\
 & =\mathbb{P}\left[\hat{\mathbb{E}}\tilde{L}(f_{1},Z)-\epsilon>2B_{\text{ref}}\sqrt{\frac{\log\left(\delta^{-1}\right)}{2n}}\right] &  & \text{from definition of }\epsilon_{1}'\\
 & \leq\mathbb{P}\left[\hat{\mathbb{E}}\tilde{L}(f_{1},Z)-\mathbb{E}\tilde{L}(f_{1},Z)>2B_{\text{ref}}\sqrt{\frac{\log\left(\delta^{-1}\right)}{2n}}\right] &  & \text{from }\mathbb{E}\tilde{L}(f_{1},Z)\leq\epsilon\nonumber \\
 & \leq\exp\left\{ -\frac{2n}{(2B_{\text{ref}})^{2}}\left[2B_{\text{ref}}\sqrt{\frac{\log\left(\delta^{-1}\right)}{2n}}\right]^{2}\right\}  &  & \text{from Hoeffding's inequality}\label{eq:apply-abs-value-bound-hoeff}\\
 & =\delta.\label{eq:delta-e0-1}
\end{align}

For the inequality used in Line \ref{eq:apply-abs-value-bound-hoeff}, see Theorem 2 of \citealp{hoeffding1963_bounded_sums_inequalities}.
\end{proof}

\subsection{Lemma to Transform Between Bounds}

The following lemma will help us translate from bounds for variables to bounds for differences and ratios of those variables. We will apply this lemma to transform from bounds on empirical losses to bounds on empirical model reliance, defined either in terms of a ratio or in terms of a difference.\textcolor{blue}{{} }
\begin{lemma}
\label{lem:ratio-diff-bound}Let $X,Z,\mu_{X},\mu_{Z},k_{X},k_{Z}\in\mathbb{R}$ be cons\textcolor{black}{tants satisfying $\left|Z-\mu_{Z}\right|\leq k_{Z}$ and $\left|X-\mu_{X}\right|\leq k_{X}$, then}

\textcolor{black}{
\begin{equation}
\left|(Z-X)-(\mu_{Z}-\mu_{X})\right|\leq q_{\text{difference}}(k_{Z},k_{X}),\label{eq:diff-bound}
\end{equation}
}where $q_{\text{difference}}$ is the function
\begin{equation}
q_{\text{difference}}(k_{Z},k_{X}):=k_{Z}+k_{X}.\label{eq:q-diff-def}
\end{equation}

\textcolor{black}{Further, if there exists constants $b_{\text{orig}}$ and $B_{\text{switch}}$ such that $0<b_{\text{orig}}\leq X,\mu_{X}$ and $Z,\mu_{Z}\leq B_{\text{switch}}<\infty$, then
\begin{align}
\left|\frac{Z}{X}-\frac{\mu_{Z}}{\mu_{X}}\right|\leq q_{\text{ratio}}(k_{Z},k_{X}),\label{eq:ratio-bound}
\end{align}
where $q_{\text{ratio}}$ is the function
\begin{equation}
q_{\text{ratio}}(k_{Z},k_{X}):=\frac{B_{\text{switch}}}{b_{\text{orig}}}-\frac{B_{\text{switch}}-k_{Z}}{b_{\text{orig}}+k_{X}}.\label{eq:q-ratio-def}
\end{equation}
}
\end{lemma}

\begin{proof}
\textcolor{black}{Showing Eq \ref{eq:diff-bound},
\begin{align*}
\left|(Z-X)-(\mu_{Z}-\mu_{X})\right| & \leq\left|Z-\mu_{Z}\right|+\left|\mu_{X}-X\right|\\
 & \leq k_{Z}+k_{X}.
\end{align*}
}

\textcolor{black}{Showing Eq \ref{eq:ratio-bound}, let $A_{Z}=\max(Z,\mu_{Z})$, $a_{X}=\min(X,\mu_{X})$, $d_{Z}=|Z-\mu_{Z}|$, and $d_{X}=|X-\mu_{X}|$. This implies that $\max(X,\mu_{X})=a_{X}+d_{X}$ and $\min(Z,\mu_{Z})=A_{Z}-d_{Z}$. Thus, $\frac{Z}{X}$ and $\frac{\mu_{Z}}{\mu_{X}}$ are both bounded within the interval
\[
\left[\frac{\min(Z,\mu_{Z})}{\max(X,\mu_{X})},\frac{\max(Z,\mu_{Z})}{\min(X,\mu_{X})}\right]=\left[\frac{A_{Z}-d_{Z}}{a_{X}+d_{X}},\frac{A_{Z}}{a_{X}}\right],
\]
which implies}

\textcolor{black}{
\begin{align}
\left|\frac{Z}{X}-\frac{\mu_{Z}}{\mu_{X}}\right| & \leq\frac{A_{Z}}{a_{X}}-\frac{A_{Z}-d_{Z}}{a_{X}+d_{X}}.\label{eq:Az-dz}
\end{align}
}

\textcolor{black}{Taking partial derivatives of the right-hand side, we get}

\textcolor{black}{
\begin{align*}
\frac{\partial}{\partial a_{X}}\left(\frac{A_{Z}}{a_{X}}-\frac{A_{Z}-d_{Z}}{a_{X}+d_{X}}\right) & =\frac{-A_{Z}}{a_{X}^{2}}+\frac{A_{Z}-d_{Z}}{(a_{X}+d_{X})^{2}}\leq0,\\
\frac{\partial}{\partial A_{Z}}\left(\frac{A_{Z}}{a_{X}}-\frac{A_{Z}-d_{Z}}{a_{X}+d_{X}}\right) & =\frac{1}{a_{X}}-\frac{1}{a_{X}+d_{X}}\geq0,\\
\frac{\partial}{\partial d_{X}}\left(\frac{A_{Z}}{a_{X}}-\frac{A_{Z}-d_{Z}}{a_{X}+d_{X}}\right) & =\frac{A_{Z}-d_{Z}}{(a_{X}+d_{X})^{2}}>0,\\
\text{and }\frac{\partial}{\partial d_{Z}}\left(\frac{A_{Z}}{a_{X}}-\frac{A_{Z}-d_{Z}}{a_{X}+d_{X}}\right) & =\frac{1}{a_{X}+d_{X}}>0.
\end{align*}
}

\textcolor{black}{So the right-hand side of \ref{eq:Az-dz} is maximized when $d_{Z},d_{X}$, and $A_{Z}$ are maximized, and when $a_{X}$ is minimized. Thus, in the case where $|Z-\mu_{Z}|\leq k_{Z}$; $|X-\mu_{X}|\leq k_{X}$; $0<b_{\text{orig}}\leq X,\mu_{X}$; and $Z,\mu_{Z}\leq B_{\text{switch}}<\infty$, we have}

\textcolor{black}{
\begin{align*}
\left|\frac{Z}{X}-\frac{\mu_{Z}}{\mu_{X}}\right| & \leq\frac{A_{Z}}{a_{X}}-\frac{A_{Z}-d_{Z}}{a_{X}+d_{X}}\\
 & \leq\frac{B_{\text{switch}}}{b_{\text{orig}}}-\frac{B_{\text{switch}}-k_{Z}}{b_{\text{orig}}+k_{X}}.
\end{align*}
}

\end{proof}

\subsection{Proof of Theorem \ref{thm:mcr-conserve-bounds}\label{subsec:Proof-mcr-conserve-bounds}}
\begin{proof}
We proceed in 4 steps.
$\hspace{1em}$

\subsubsection[Step 1]{Step 1: Show that $\mathbb{P}\left[\widehat{MR}(f_{+,\epsilon})\leq\widehat{MCR}_{+}(\epsilon_{\text{out}})\right]\geq1-\frac{\delta}{3}$. }

Consider the event that 
\begin{equation}
\widehat{MR}(f_{+,\epsilon})\leq\widehat{MCR}_{+}(\epsilon_{\text{out}}).\label{eq:first-max}
\end{equation}
Eq \ref{eq:first-max} will always hold if $f_{+,\epsilon}\in\hat{\mathcal{R}}(\epsilon_{\text{out}})$, since $\widehat{MCR}_{+}(\epsilon_{\text{out}})$ upper bounds the empirical model reliance for models in $\hat{\mathcal{R}}(\epsilon_{\text{out}})$ by definition. Applying the above reasoning in Line \ref{eq:apply-greater-sets}, below, we get
\begin{align}
\mathbb{P}\left[\widehat{MR}(f_{+,\epsilon})>\widehat{MCR}_{+}(\epsilon_{\text{out}})\right] & \leq\mathbb{P}\left[f_{+,\epsilon}\notin\hat{\mathcal{R}}(\epsilon_{\text{out}})\right]\label{eq:apply-greater-sets}\\
 & \leq\frac{\delta}{3} &  & \text{from }\epsilon_{\text{out}}\text{ definition and Lemma \ref{lem:hoeffding}}.\label{eq:delta-e0}
\end{align}

\subsubsection[Step 2]{Step 2: Conditional on $\widehat{MR}(f_{+,\epsilon})\leq\widehat{MCR}_{+}(\epsilon_{\text{out}})$, Upper Bound $\mred{MR}(f_{+,\epsilon})$ by $\widehat{MCR}_{+}(\epsilon_{\text{out}})$ Added to an Error Term.\label{subsec:Step-2:-Conditional}}

When Eq \ref{eq:first-max} holds we have,
\begin{eqnarray}
\widehat{MR}(f_{+,\epsilon}) & \leq & \ensuremath{\widehat{MCR}_{+}(\epsilon_{\text{out}})}\nonumber \\
\mred{\widehat{MR}(f_{+,\epsilon})} & \leq & \ensuremath{\widehat{MCR}_{+}(\epsilon_{\text{out}})}+\{MR(f_{+,\epsilon})-MR(f_{+,\epsilon})\}\nonumber \\
MR(f_{+,\epsilon}) & \leq & \widehat{MCR}_{+}(\epsilon_{\text{out}})+[MR(f_{+,\epsilon})-\mred{\widehat{MR}(f_{+,\epsilon})}].\label{eq:mr-diff-inequality}
\end{eqnarray}

\subsubsection[Step 3]{Step 3: Probabilistically Bound the Error Term from Step 2.}

Next we show that the bracketed term in Line \ref{eq:mr-diff-inequality} is less than or equal to $\mathcal{Q}_{\text{out}}$ with high proba\textcolor{black}{bility. For $k\in\mathbb{R}$, let $q_{\text{difference}}$ and $q_{\text{ratio}}$ be defined as in Eqs \ref{eq:q-diff-def} and \ref{eq:q-ratio-def}. Let $q:\mathbb{R\rightarrow\mathbb{R}}$ be the function such that $q(k)=q_{\text{ratio}}\left(k,\frac{k}{\sqrt{2}}\right)$. Then}
\begin{align*}
\mathcal{Q}_{\text{out}} & =\frac{B_{\text{switch}}}{b_{\text{orig}}}-\frac{B_{\text{switch}}-B_{\text{ind}}\sqrt{\frac{\log(6\delta^{-1})}{n}}}{b_{\text{orig}}+B_{\text{ind}}\sqrt{\frac{\log(6\delta^{-1})}{2n}}}\\
 & =q_{\text{ratio}}\left(B_{\text{ind}}\sqrt{\frac{\log(6\delta^{-1})}{n}},B_{\text{ind}}\sqrt{\frac{\log(6\delta^{-1})}{2n}}\right)\\
 & =q\left(B_{\text{ind}}\sqrt{\frac{\log(6\delta^{-1})}{n}}\right).
\end{align*}

Applying this relation below, we have
\begin{align}
 & \mathbb{P}\left[MR(f_{+,\epsilon})-\widehat{MR}(f_{+,\epsilon})>\mathcal{Q}_{\text{out}}\right]\label{eq:start2deltas}\\
 & \hspace{1em}\leq\mathbb{P}\left[\left|MR(f_{+,\epsilon})-\widehat{MR}(f_{+,\epsilon})\right|>q\left(B_{\text{ind}}\sqrt{\frac{\log(6\delta^{-1})}{n}}\right)\right]\nonumber \\
 & \hspace{1em}\leq\mathbb{P}\left[\left\{ \left|\hat{e}_{\text{orig}}(f_{+,\epsilon})-e_{\text{orig}}(f_{+,\epsilon})\right|>B_{\text{ind}}\sqrt{\frac{\log(6\delta^{-1})}{2n}}\right\} \right.\nonumber \\
 & \hspace{1em}\hspace{1.6cm}\left.\bigcup\left\{ \left|\hat{e}_{\text{switch}}(f_{+,\epsilon})-e_{\text{switch}}(f_{+,\epsilon})\right|>B_{\text{ind}}\sqrt{\frac{\log(6\delta^{-1})}{n}}\right\} \right] &  & \text{from Lemma }\ref{lem:ratio-diff-bound}\nonumber \\
 & \hspace{1em}\leq\mathbb{P}\left[\left|\hat{e}_{\text{orig}}(f_{+,\epsilon})-e_{\text{orig}}(f_{+,\epsilon})\right|>B_{\text{ind}}\sqrt{\frac{\log\left(6\delta^{-1}\right)}{2n}}\right]\nonumber \\
 & \hspace{1em}\hspace{1.6cm}+\mathbb{P}\left[\left|\hat{e}_{\text{switch}}(f_{+,\epsilon})-e_{\text{switch}}(f_{+,\epsilon})\right|>B_{\text{ind}}\sqrt{\frac{\log\left(6\delta^{-1}\right)}{n}}\right] &  & \text{from the Union bound}\nonumber \\
 & \hspace{1em}\leq2\exp\left\{ -\frac{2n}{(B_{\text{ind}}-0)^{2}}\left[B_{\text{ind}}\sqrt{\frac{\log\left(6\delta^{-1}\right)}{2n}}\right]^{2}\right\} \nonumber \\
 & \hspace{1em}\hspace{1.6cm}+2\exp\left\{ -\frac{n}{(B_{\text{ind}}-0)^{2}}\left[B_{\text{ind}}\sqrt{\frac{\log\left(6\delta^{-1}\right)}{n}}\right]^{2}\right\}  &  & {\text{from Hoeffding's bound}\atop \text{for U-statistics}\text{\ensuremath{\hphantom{\text{from ibo}}}}}\label{eq:recall-apply-hoeff}\\
 & \hspace{1em}=\frac{2\delta}{6}+\frac{2\delta}{6}=\frac{2\delta}{3}.\label{eq:final2deltas}
\end{align}
\textcolor{red}{}

\textcolor{black}{In Line \ref{eq:recall-apply-hoeff}, above, recall that $\hat{e}_{\text{orig}}(f_{+,\epsilon})$ and $\hat{e}_{\text{switch}}(f_{+,\epsilon})$ are both U-statistics. Note that $\mathbb{E}\left[\hat{e}_{\text{switch}}(f_{+,\epsilon})\right]=e_{\text{switch}}(f_{+,\epsilon})$ because $\hat{e}_{\text{switch}}(f_{+,\epsilon})$ is an average of terms, and each term has expectation equal to $e_{\text{switch}}(f_{+,\epsilon})$. For the same reason, $\mathbb{E}\left[\hat{e}_{\text{orig}}(f_{+,\epsilon})\right]=e_{\text{orig}}(f_{+,\epsilon})$. This allows us to apply }Eq 5.7 of\textcolor{black}{{} }\citealp{hoeffding1963_bounded_sums_inequalities} (see also Eq 1 on page 201 of \citealp{serfling1980approximation}\textcolor{black}{, in Theorem A) to obtain Line \ref{eq:recall-apply-hoeff}.}

\textcolor{black}{Alternatively, if we instead define model reliance as $MR_{\text{difference}}(f)=e_{\text{switch}}(f)-e_{\text{orig}}(f)$ (see Appendix \ref{sec:Discussion---Ratios}), define empirical model reliance as $\widehat{MR}_{\text{difference}}(f):=\hat{e}_{\text{switch}}(f)-\hat{e}_{\text{orig}}(f)$, and define}

\textcolor{black}{
\[
\mathcal{Q}_{\text{out},\text{difference}}:=\left(1+\frac{1}{\sqrt{2}}\right)B_{\text{ind}}\sqrt{\frac{\log(6\delta^{-1})}{n}}=q_{\text{difference}}\left(B_{\text{ind}}\sqrt{\frac{\log(6\delta^{-1})}{n}},B_{\text{ind}}\sqrt{\frac{\log(6\delta^{-1})}{2n}}\right),
\]
then the same proof holds without Assumption \ref{assu:bnd-avg} if we replace $MR$, $\widehat{MR},$ $\mathcal{Q}_{\text{out}}$ respectively with $MR_{\text{difference}}$, $\widehat{MR}_{\text{difference}}$, $\mathcal{Q}_{\text{out},\text{difference}}$, and redefine $q:\mathbb{R}\rightarrow\mathbb{R}$ as the function $q(k)=q_{\text{difference}}\left(k,\frac{k}{\sqrt{2}}\right)$.}

\textcolor{black}{Eqs \ref{eq:start2deltas}-\ref{eq:final2deltas} also hold if we replace $\hat{e}_{\text{switch}}$ throughout with $\hat{e}_{\text{divide}}$, including in Assumption \ref{assu:bnd-avg}, since the same bound can be used for both $\hat{e}_{\text{switch}}$ and $\hat{e}_{\text{divide}}$ (}Eq 5.7 of \citealp{hoeffding1963_bounded_sums_inequalities}; see also Theorem A on page 201 of \citealp{serfling1980approximation}\textcolor{black}{).}\textcolor{teal}{}

\subsubsection[Step 4]{Step 4: Combine Results to Show Eq \ref{eq:result-for-f+}}

Finally, we connect the above results to show Eq \ref{eq:result-for-f+}. We know from Eq \ref{eq:delta-e0} that Eq \ref{eq:first-max} holds with high probability. Eq \ref{eq:first-max} implies Eq \ref{eq:mr-diff-inequality}, which bounds $MCR_{+}(\epsilon)=MR(f_{+,\epsilon})$ up to a bracketed residual term. We also know from Eq \ref{eq:final2deltas} that, with high probability, the residual term in Eq \ref{eq:mr-diff-inequality} is less than $\mathcal{Q}_{\text{out}}=q\left(B_{\text{ind}}\sqrt{\frac{\log(6\delta^{-1})}{n}}\right)$. Putting this together, we can show Eq \ref{eq:result-for-f+}:
\begin{align}
 & \mathbb{P}\left(MCR_{+}(\epsilon)>\ensuremath{\widehat{MCR}_{+}(\epsilon_{\text{out}})}+\mathcal{Q}_{\text{out}}\right)\nonumber \\
 & \hspace{1cm}=\mathbb{P}\left(MR(f_{+,\epsilon})>\ensuremath{\widehat{MCR}_{+}(\epsilon_{\text{out}})}+\mathcal{Q}_{\text{out}}\right)\nonumber \\
 & \hspace{1cm}\leq\mathbb{P}\left[\left(\widehat{MR}(f_{+,\epsilon})>\widehat{MCR}_{+}(\epsilon_{\text{out}})\right)\bigcup\left(MR(f_{+,\epsilon})-\widehat{MR}(f_{+,\epsilon})>\mathcal{Q}_{\text{out}}\right)\right] &  & \text{from Step 2}\nonumber \\
 & \hspace{1cm}\leq\mathbb{P}\left[\widehat{MR}(f_{+,\epsilon})>\widehat{MCR}_{+}(\epsilon_{\text{out}})\right]+\mathbb{P}\left[MR(f_{+,\epsilon})-\widehat{MR}(f_{+,\epsilon})>\mathcal{Q}_{\text{out}}\right]\nonumber \\
 & \hspace{1cm}\leq\frac{\delta}{3}+\frac{2\delta}{3}=\delta. &  & \text{from Steps 1 \& 3}\label{eq:outer-bnd-f+}
\end{align}

This completes the proof for Eq \ref{eq:result-for-f+}. For Eq \ref{eq:result_for_f-} we can use the same approach, shown below for completeness. Analogous to Eq \ref{eq:delta-e0}, we have
\begin{align*}
\mathbb{P}\left[\widehat{MR}(f_{-,\epsilon})<\widehat{MCR}_{-}(\epsilon_{\text{out}})\right] & \leq\frac{\delta}{3}.
\end{align*}

Analogous to Eq \ref{eq:mr-diff-inequality}, when $\widehat{MR}(f_{-,\epsilon})\geq\widehat{MR}(\hat{f}_{-,\epsilon_{\text{out}}})$ we have
\begin{eqnarray*}
\widehat{MR}(f_{-,\epsilon}) & \geq & \widehat{MCR}_{-}(\epsilon_{\text{out}})\\
\widehat{MR}(f_{-,\epsilon}) & \geq & \widehat{MCR}_{-}(\epsilon_{\text{out}})+\{MR(f_{-,\epsilon})-MR(f_{-,\epsilon})\}\\
MR(f_{-,\epsilon}) & \geq & \widehat{MCR}_{-}(\epsilon_{\text{out}})-\left[\widehat{MR}(f_{-,\epsilon})-MR(f_{-,\epsilon})\right].
\end{eqnarray*}

Analogous to Eq \ref{eq:final2deltas}, we have
\begin{equation}
\mathbb{P}\left[\widehat{MR}(f_{-,\epsilon})-MR(f_{-,\epsilon})>q\left(B_{\text{ind}}\sqrt{\frac{\log(6\delta^{-1})}{n}}\right)\right]\leq\frac{2\delta}{3}.\label{eq:q-reuse-1}
\end{equation}

Finally, analogous to Eq \ref{eq:outer-bnd-f+}, we have
\begin{align}
 & \mathbb{P}\left(MCR_{-}(\epsilon)<\ensuremath{\widehat{MCR}_{-}(\epsilon_{\text{out}})}-\mathcal{Q}_{\text{out}}\right)\nonumber \\
 & \hspace{1cm}\leq\mathbb{P}\left[\left(\widehat{MR}(f_{-,\epsilon})<\widehat{MCR}_{-}(\epsilon_{\text{out}})\right)\bigcup\left(\widehat{MR}(f_{-,\epsilon})-MR(f_{-,\epsilon})>q\left(B_{\text{ind}}\sqrt{\frac{\log(6\delta^{-1})}{n}}\right)\right)\right]\label{eq:q-reuse-2}\\
 & \hspace{1cm}\leq\mathbb{P}\left[\widehat{MR}(f_{-,\epsilon})<\widehat{MCR}_{-}(\epsilon_{\text{out}})\right]+\mathbb{P}\left[\widehat{MR}(f_{-,\epsilon})-MR(f_{-,\epsilon})>q\left(B_{\text{ind}}\sqrt{\frac{\log(6\delta^{-1})}{n}}\right)\right]\nonumber \\
 & \hspace{1cm}\leq\frac{\delta}{3}+\frac{2\delta}{3}=\delta.\nonumber 
\end{align}

\textcolor{black}{Again, the same proof holds without Assumption \ref{assu:bnd-avg} if we replace $MR$, $\widehat{MR},$ $\mathcal{Q}_{\text{out}}$ respectively with $MR_{\text{difference}}$, $\widehat{MR}_{\text{difference}}$, $\mathcal{Q}_{\text{out},\text{difference}}$, and redefine $q$ as the function satisfying $q(k)=q_{\text{difference}}\left(k,\frac{k}{\sqrt{2}}\right)$ in Eqs \ref{eq:q-reuse-1} \& \ref{eq:q-reuse-2}. }
\end{proof}

\subsection{\textcolor{black}{Proof of Corollary \ref{cor:best-in-class}\label{subsec:Proof-cor-best-in-class}}}
\begin{proof}
\textcolor{black}{By definition, $MR(f_{-,\epsilon_{\text{best}}})\leq MR(f^{\star})\leq MR(f_{+,\epsilon_{\text{best}}})$. Applying this relation in Line \ref{eq:mr-leq-mr}, below, we see
\begin{align}
 & \mathbb{P}\left(MR(f^{\star})\in\left[\widehat{MCR}_{-}\left(\epsilon_{\text{best}}\right)-\mathcal{Q}_{\text{best}},\hspace{1em}\widehat{MCR}_{+}\left(\epsilon_{\text{best}}\right)+\mathcal{Q}_{\text{best}}\right]\right)\nonumber \\
 & \hspace{1cm}=1-\mathbb{P}\left\{ MR(f^{\star})<\widehat{MCR}_{-}\left(\epsilon_{\text{best}}\right)-\mathcal{Q}_{\text{best}}\:\mred{\bigcup}\:MR(f^{\star})>\widehat{MCR}_{+}\left(\epsilon_{\text{best}}\right)+\mathcal{Q}_{\text{best}}\right\} \nonumber \\
 & \hspace{1cm}\geq1-\mathbb{P}\left\{ MR(f^{\star})<\widehat{MCR}_{-}\left(\epsilon_{\text{best}}\right)-\mathcal{Q}_{\text{best}}\right\} \mred{-}\mathbb{P}\left\{ MR(f^{\star})>\widehat{MCR}_{+}\left(\epsilon_{\text{best}}\right)+\mathcal{Q}_{\text{best}}\right\} \nonumber \\
 & \hspace{1cm}\geq1-\mathbb{P}\left\{ \mred{MR(f_{-,\epsilon})}<\widehat{MCR}_{-}\left(\epsilon_{\text{best}}\right)-\mathcal{Q}_{\text{best}}\right\} -\mathbb{P}\left\{ \mred{MR(f_{+,\epsilon})}>\widehat{MCR}_{+}\left(\epsilon_{\text{best}}\right)+\mathcal{Q}_{\text{best}}\right\} \label{eq:mr-leq-mr}\\
 & \hspace{1cm}\geq1-\frac{\delta}{2}-\frac{\delta}{2}\hspace{1.5cm}\text{\text{from Theorem \ref{thm:mcr-conserve-bounds}}}.\label{eq:applied-thm3}
\end{align}
}

\textcolor{black}{To apply Theorem \ref{thm:mcr-conserve-bounds} in Line \ref{eq:applied-thm3}, above, we note that $\mathcal{Q}_{\text{best}}$ and $\epsilon_{\text{best}}$ are equivalent to the definitions of $\mathcal{Q}_{\text{out}}$ and $\epsilon_{\text{out}}$ in Theorem \ref{thm:mcr-conserve-bounds}, but with $\delta$ replaced by $\frac{\delta}{2}$. }

\textcolor{black}{Alternatively, if we define model reliance as $MR_{\text{difference}}(f)=e_{\text{switch}}(f)-e_{\text{orig}}(f)$ (see Appendix \ref{sec:Discussion---Ratios}), and define empirical model reliance as $\widehat{MR}_{\text{difference}}(f)=\hat{e}_{\text{switch}}(f)-\hat{e}_{\text{orig}}(f)$, then let}

\textcolor{black}{
\[
\mathcal{Q}_{\text{best},\text{difference}}:=\left(1+\frac{1}{\sqrt{2}}\right)B_{\text{ind}}\sqrt{\frac{\log(12\delta^{-1})}{n}}.
\]
The term $\mathcal{Q}_{\text{best},\text{difference}}$ is equivalent to $\mathcal{Q}_{\text{out},\text{difference}}$ but with $\delta$ replaced with $\frac{\delta}{2}$. Under this difference-based definition of model reliance, Theorem \ref{thm:mcr-conserve-bounds} holds without Assumption \ref{assu:bnd-avg} if we replace $\mathcal{Q}_{\text{out}}$ with $\mathcal{Q}_{\text{out},\text{difference}}$ (see Section \ref{subsec:Proof-mcr-conserve-bounds}), and so we can apply this altered version of Theorem \ref{thm:mcr-conserve-bounds} in Line \ref{eq:applied-thm3}. Thus, Theorem \ref{cor:best-in-class} also holds without Assumption \ref{assu:bnd-avg} if we replace $MR$, $\widehat{MR}$, and $\mathcal{Q}_{\text{best}}$ respectively with $MR_{\text{difference}}$, $\widehat{MR}_{\text{difference}}$, and $\mathcal{Q}_{\text{best},\text{difference}}$.}
\end{proof}

\subsection{\textcolor{black}{Proof of Theo}rems \ref{thm:MR-unif} \& \ref{thm:mcr-consistent}\label{subsec:Proof-of-learning-bnd}}
We begin by proving Theorem \ref{thm:MR-unif}, along with related results. We then apply these results to show Theorem \ref{thm:mcr-consistent}.

\subsubsection{Proof of Theorem \textcolor{black}{\ref{thm:MR-unif}}, and Other Limits on Estimation Error, Based on Covering Number\label{subsec:Proof-of-other-cover-limits}}

The following theorem uses the covering number based on $r$-margin-expectation-covers to jointly bound empirical losses for any function $f\in\mathcal{F}$. Theorem \textcolor{black}{\ref{thm:MR-unif} in the main text follows directly from Eq \ref{eq:unif-ratio}, below.}
\begin{theorem}
\label{thm:unif-covering}If Assumptions \textcolor{black}{\ref{assu:bnd-ind},} \textcolor{black}{\ref{assu:bnd-ref}} and \textcolor{black}{\ref{assu:bnd-avg}} hold for all $f\in\mathcal{F}$, then for any $r>0$
\begin{align}
\mathbb{P}_{D}\left[\sup_{f\in\mathcal{F}}\left|\hat{e}_{\text{orig}}(f)-e_{\text{orig}}(f)\right|>B_{\text{ind}}\sqrt{\frac{\log(2\delta^{-1}\mathcal{N}\left(\mathcal{F},r\right))}{\mred{2}n}}+2r\right] & \leq\delta,\label{eq:unif-e0}\\
\mathbb{P}_{D}\left[\sup_{f\in\mathcal{F}}\left|\hat{\mathbb{E}}\tilde{L}(f,Z)-\mathbb{E}\tilde{L}(f,Z)\right|>2B_{\text{ref}}\sqrt{\frac{\log(2\delta^{-1}\mathcal{N}\left(\mathcal{F},r\right))}{\mred{2}n}}+2r\right] & \leq\delta,\label{eq:rashomon-ref-unif}\\
\mathbb{P}_{D}\left[\sup_{f\in\mathcal{F}}\left|\hat{e}_{\text{switch}}(f)-e_{\text{switch}}(f)\right|>B_{\text{ind}}\sqrt{\frac{\log(2\delta^{-1}\mathcal{N}\left(\mathcal{F},r\right))}{n}}+2r\right] & \leq\delta,\label{eq:unif-e1}\\
\mathbb{P}\left[\sup_{f\in\mathcal{F}}\left|\frac{\hat{e}_{\text{orig}}(f)}{\hat{e}_{\text{switch}}(f)}-\frac{e_{\text{orig}}(f)}{e_{\text{switch}}(f)}\right|>\text{\ensuremath{\mathcal{Q}}}_{4}\right] & \leq\delta,\label{eq:unif-ratio}\\
\mathbb{P}_{D}\left[\sup_{f\in\mathcal{F}}\left|\left\{ \hat{e}_{\text{switch}}(f)-\hat{e}_{\text{orig}}(f)\right\} -\left\{ e_{\text{switch}}(f)-e_{\text{orig}}(f)\right\} \right|>\mathcal{Q}_{4,\text{difference}}\right] & \leq\delta,\label{eq:unif-diff}
\end{align}

where
\begin{align}
\mathcal{Q}_{4}:=q_{\text{ratio}} & \left(B_{\text{ind}}\sqrt{\frac{\log(4\delta^{-1}\mathcal{N}\left(\mathcal{F},r\sqrt{2}\right))}{n}}+2r\sqrt{2},B_{\text{ind}}\sqrt{\frac{\log(4\delta^{-1}\mathcal{N}\left(\mathcal{F},r\right))}{2n}}+2r\right),\label{eq:Q4-ratio-def}\\
\mathcal{Q}_{4,\text{difference}}:=q_{\text{difference}} & \left(B_{\text{ind}}\sqrt{\frac{\log(4\delta^{-1}\mathcal{N}\left(\mathcal{F},r\sqrt{2}\right))}{n}}+2r\sqrt{2},B_{\text{ind}}\sqrt{\frac{\log(4\delta^{-1}\mathcal{N}\left(\mathcal{F},r\right))}{2n}}+2r\right),\label{eq:Q4-diff-def}
\end{align}
and $q_{\text{ratio}}$ and $q_{\text{difference}}$ are defined as in Lemma \ref{lem:ratio-diff-bound}. For Eq \ref{eq:unif-diff}, the result is unaffected if we remove Assumption \textcolor{black}{\ref{assu:bnd-avg}}.
\end{theorem}

\subsubsection{Proof of Eq \ref{eq:unif-e0}}
\begin{proof}
Let $\mathcal{G}_{r}$ be a $r$-margin-expectation-cover for $\mathcal{F}$ of size $\mathcal{N}\left(\mathcal{F},r\right)$. Let $D_{p}$ denote the population distribution, let $D_{s}$ be the sample distribution, and let $D^{\star}$ be the uniform mixture of $D_{p}$ and $D_{s}$, i.e., for any $z\in\mathcal{Z}$,
\begin{equation}
\mathbb{P}_{D^{\star}}(Z\leq z)=\frac{1}{2}\mathbb{P}_{D_{p}}(Z\leq z)+\frac{1}{2}\mathbb{P}_{D_{s}}(Z\leq z).\label{eq:unif-example}
\end{equation}

Unless otherwise stated, we take expectations and probabilities with respect to $D_{p}$. Since $\mathcal{G}_{r}$ is a $r$-margin-expectation-cover, we know that for any $f\in\mathcal{F}$ we can find a function $g\in\mathcal{G}_{r}$ such that $\mathbb{E}_{D^{\star}}\left|L(g,Z)-L(f,Z)\right|=\mathbb{E}_{D^{\star}}\left|\tilde{L}(g,Z)-\tilde{L}(f,Z)\right|\leq r$, and\textcolor{blue}{}
\begin{align}
\left|\hat{\mathbb{E}}L(f,Z)-\mathbb{E}L(f,Z)\right| & =\left|\hat{\mathbb{E}}L(f,Z)-\mathbb{E}L(f,Z)+\left\{ \hat{\mathbb{E}}L(g,Z)-\hat{\mathbb{E}}L(g,Z)\right\} +\left\{ \mathbb{E}L(g,Z)-\mathbb{E}L(g,Z)\right\} \right|\label{eq:unif-restart-1}\\
 & \leq\left|\hat{\mathbb{E}}L(g,Z)-\mathbb{E}L(g,Z)\right|+\left|\mathbb{E}L(g,Z)-\mathbb{E}L(f,Z)\right|+\left|\hat{\mathbb{E}}L(f,Z)-\hat{\mathbb{E}}L(g,Z)\right|\nonumber \\
 & \leq\left|\hat{\mathbb{E}}L(g,Z)-\mathbb{E}L(g,Z)\right|+\mathbb{E}_{D_{p}}\left|L(g,Z)-L(f,Z)\right|+\mathbb{E}_{D_{s}}\left|L(f,Z)-L(g,Z)\right|\nonumber \\
 & =\left|\hat{\mathbb{E}}L(g,Z)-\mathbb{E}L(g,Z)\right|+2\mathbb{E}_{D^{\star}}\left|L(g,Z)-L(f,Z)\right|\nonumber \\
 & \leq\left|\hat{\mathbb{E}}L(g,Z)-\mathbb{E}L(g,Z)\right|+2r.\nonumber 
\end{align}

Applying the above relation in Line \ref{eq:apply-result} below, we have
\begin{align}
 & \mathbb{P}\left(\sup_{f\in\mathcal{F}}\left|\hat{\mathbb{E}}L(f,Z)-\mathbb{E}L(f,Z)\right|>B_{\text{ind}}\sqrt{\frac{\log(2\delta^{-1}\mathcal{N}\left(\mathcal{F},r\right))}{2n}}+2r\right)\nonumber \\
 & \hspace{1cm}=\mathbb{P}\left(\mred{\exists f\in\mathcal{F}}\text{ : }\left|\hat{\mathbb{E}}L(f,Z)-\mathbb{E}L(f,Z)\right|>B_{\text{ind}}\sqrt{\frac{\log(2\delta^{-1}\mathcal{N}\left(\mathcal{F},r\right))}{2n}}+2r\right)\nonumber \\
 & \hspace{1cm}\leq\mathbb{P}\left(\exists g\in\mathcal{G}_{r}\text{ : }\left|\hat{\mathbb{E}}L(g,Z)-\mathbb{E}L(g,Z)\right|+2r>B_{\text{ind}}\sqrt{\frac{\log(2\delta^{-1}\mathcal{N}\left(\mathcal{F},r\right))}{2n}}+2r\right)\label{eq:apply-result}\\
 & \hspace{1cm}=\mathbb{P}\left(\mred{\bigcup_{g\in\mathcal{G}_{r}}}\left|\hat{\mathbb{E}}L(g,Z)-\mathbb{E}L(g,Z)\right|>B_{\text{ind}}\sqrt{\frac{\log(2\delta^{-1}\mathcal{N}\left(\mathcal{F},r\right))}{2n}}\right)\nonumber \\
 & \hspace{1cm}\leq\mred{\sum_{g\in\mathcal{G}_{r}}}\mathbb{P}\left(\left|\hat{\mathbb{E}}L(g,Z)-\mathbb{E}L(g,Z)\right|>B_{\text{ind}}\sqrt{\frac{\log(2\delta^{-1}\mathcal{N}\left(\mathcal{F},r\right))}{2n}}\right)\hspace{1.25cm}\text{from the Union bound}\nonumber \\
 & \hspace{1cm}\leq\mathcal{N}\left(\mathcal{F},r\right)2\exp\left[-\frac{2n}{(B_{\text{ind}})^{2}}\left\{ B_{\text{ind}}\sqrt{\frac{\log(2\delta^{-1}\mathcal{N}\left(\mathcal{F},r\right))}{2n}}\right\} ^{2}\right]\hspace{0.45cm}\text{\ensuremath{\text{from Hoeffding's inequality}}}\label{eq:unif-apply-hoef}\\
 & \hspace{1cm}=\delta.\label{eq:unif-restart-2}
\end{align}
To apply Hoeffding's inequality (Theorem 2 of \citealp{hoeffding1963_bounded_sums_inequalities}) in Line \ref{eq:unif-apply-hoef}, above, we use the fact that $L(g,Z)$ is bounded within an interval of length $B_{\text{ind}}$.
\end{proof}

\subsubsection{Proof of Eq \ref{eq:rashomon-ref-unif}}
\begin{proof}
The proof for Eq \ref{eq:rashomon-ref-unif} is nearly identical to the proof for Eq \ref{eq:unif-e0}. Simply replacing $L$ and $B_{\text{ind}}$ respectively with $\tilde{L}$ and $(2B_{\text{ref}})$ in Eqs \ref{eq:unif-restart-1}-\ref{eq:unif-restart-2} yields a valid proof for Eq \ref{eq:rashomon-ref-unif}.
\end{proof}

\subsubsection{Proof of Eq \ref{eq:unif-e1}}
\begin{proof}
Let $F_{D}$ denote the cumulative distribution function for a distribution $D$. Let $\tilde{D}_{p}$ be the distribution such that
\[
F_{\tilde{D}_{p}}(Y=y,X_{1}=x_{1},X_{2}=x_{2})=F_{D_{p}}(Y=y,X_{2}=x_{2})F_{D_{p}}(X_{1}=x_{1}).
\]

Let $\tilde{D}_{s}$ be the distribution satisfying
\[
\mathbb{P}_{\tilde{D}_{s}}(Y=y,X_{1}=x_{1},X_{2}=x_{2})=\frac{1}{n(n-1)}\sum_{i=1}^{n}\sum_{j\neq i}\mathbf{1}(\mathbf{y}_{[j]}=y,\mathbf{X}_{1[i,\cdot]}=x_{1},\mathbf{X}_{2[j,\cdot]}=x_{2}).
\]

Let $\tilde{D}^{\star}$ be the uniform mixture of $\tilde{D}_{p}$ and $\tilde{D}_{s}$, as in Eq \ref{eq:unif-example}. Replacing $e_{\text{orig}}$, $\hat{e}_{\text{orig}}$, $D_{p}$, $D_{s}$, and $D^{\star}$ respectively with $e_{\text{switch}}$, $\hat{e}_{\text{switch}}$, $\tilde{D}_{p}$, $\tilde{D}_{s}$, and $\tilde{D}^{\star}$, we can follow the same steps as in the proof for Eq \ref{eq:unif-e0}. For any $f\in\mathcal{F}$, we know that there exists a function $g\in\mathcal{G}_{r}$ satisfying $\mathbb{E}_{\tilde{D}^{\star}}\left|L(g,Z)-L(f,Z)\right|\leq r$, which implies
\begin{align*}
\left|\hat{e}_{\text{switch}}(f)-e_{\text{switch}}(f)\right| & =\left|\hat{e}_{\text{switch}}(f)-e_{\text{switch}}(f)+\left\{ \hat{e}_{\text{switch}}(g)-\hat{e}_{\text{switch}}(g)\right\} +\left\{ e_{\text{switch}}(g)-e_{\text{switch}}(g)\right\} \right|\\
 & \leq\left|\hat{e}_{\text{switch}}(g)-e_{\text{switch}}(g)\right|+\mathbb{E}_{\tilde{D}_{p}}\left|L(g,Z)-L(f,Z)\right|+\mathbb{E}_{\tilde{D}_{s}}\left|L(f,Z)-L(g,Z)\right|\\
 & =\left|\hat{e}_{\text{switch}}(g)-e_{\text{switch}}(g)\right|+2\mathbb{E}_{\tilde{D}^{\star}}\left|L(g,Z)-L(f,Z)\right|\\
 & \leq\left|\hat{e}_{\text{switch}}(g)-e_{\text{switch}}(g)\right|+2r.
\end{align*}

As a result,
\begin{align}
 & \mathbb{P}\left(\sup_{f\in\mathcal{F}}\left|\hat{e}_{\text{switch}}(f)-e_{\text{switch}}(f)\right|>B_{\text{ind}}\sqrt{\frac{\log(2\delta^{-1}\mathcal{N}\left(\mathcal{F},r\right))}{n}}+2r\right)\nonumber \\
 & \hspace{1cm}\leq\mathbb{P}\left(\mred{\exists g\in\mathcal{G}_{r}\text{ : }\left|\hat{e}_{\text{switch}}(g)-e_{\text{switch}}(g)\right|+2r}>B_{\text{ind}}\sqrt{\frac{\log(2\delta^{-1}\mathcal{N}\left(\mathcal{F},r\right))}{n}}+2r\right)\nonumber \\
 & \hspace{1cm}\leq\mred{\sum_{g\in\mathcal{G}_{r}}}\mathbb{P}\left(\left|\hat{e}_{\text{switch}}(g)-e_{\text{switch}}(g)\right|>B_{\text{ind}}\sqrt{\frac{\log(2\delta^{-1}\mathcal{N}\left(\mathcal{F},r\right))}{n}}\right)\nonumber \\
 & \hspace{1cm}\leq\mathcal{N}\left(\mathcal{F},r\right)2\exp\left[-\frac{n}{(B_{\text{ind}}-0)^{2}}\left\{ B_{\text{ind}}\sqrt{\frac{\log(2\delta^{-1}\mathcal{N}\left(\mathcal{F},r\right))}{n}}\right\} ^{2}\right]\label{eq:hoeffding-bound-ustat-uniform}\\
 & \hspace{1cm}=\delta.\nonumber 
\end{align}
In Line \ref{eq:hoeffding-bound-ustat-uniform}, above, we\textcolor{black}{{} apply }Eq 5.7 of\textcolor{black}{{} }\citealp{hoeffding1963_bounded_sums_inequalities} (see also Eq 1 on page 201 of \citealp{serfling1980approximation}\textcolor{black}{, in Theorem A), in the same way as in Eq \ref{eq:recall-apply-hoeff}.}
\end{proof}

\subsubsection{Proof for Eq \ref{eq:unif-ratio}}
\begin{proof}
We apply Lemma \ref{lem:ratio-diff-bound} and Eq \ref{eq:Q4-ratio-def} in Line \ref{eq:q4-lr-apply}, below, to obtain\textcolor{blue}{}
\begin{align}
\hspace{0.5cm} & \mathbb{P}\left[\sup_{f\in\mathcal{F}}\left|\frac{\hat{e}_{\text{orig}}(f)}{\hat{e}_{\text{switch}}(f)}-\frac{e_{\text{orig}}(f)}{e_{\text{switch}}(f)}\right|>\mathcal{Q}_{4}\right]\label{eq:run-start}\\
 & =\mathbb{P}\left(\exists f\in\mathcal{F}\text{ : }\left|\frac{\hat{e}_{\text{orig}}(f)}{\hat{e}_{\text{switch}}(f)}-\frac{e_{\text{orig}}(f)}{e_{\text{switch}}(f)}\right|>\mathcal{Q}_{4}\right)\nonumber \\
 & \leq\mathbb{P}\left(\left\{ \exists f\in\mathcal{F}\text{ : }\left|\hat{e}_{\text{orig}}(f)-e_{\text{orig}}(f)\right|>B_{\text{ind}}\sqrt{\frac{\log(4\delta^{-1}\mathcal{N}\left(\mathcal{F},r\right))}{2n}}+2r\right\} \right.\label{eq:q4-lr-apply}\\
 & \hspace{1cm}\hspace{1cm}\bigcup\left.\left\{ \exists f\in\mathcal{F}\text{ : }\left|\hat{e}_{\text{switch}}(f)-e_{\text{switch}}(f)\right|>B_{\text{ind}}\sqrt{\frac{\log(4\delta^{-1}\mathcal{N}\left(\mathcal{F},r\sqrt{2}\right))}{n}}+2r\sqrt{2}\right\} \right)\nonumber \\
 & =\mathbb{P}\left(\mred{\ensuremath{\sup_{f\in\mathcal{F}}}}\left|\hat{e}_{\text{orig}}(f)-e_{\text{orig}}(f)\right|>B_{\text{ind}}\sqrt{\frac{\log(4\delta^{-1}\mathcal{N}\left(\mathcal{F},r\right))}{2n}}+2r\right)\nonumber \\
 & \hspace{1cm}\hspace{1cm}+\mathbb{P}\left(\mred{\ensuremath{\sup_{f\in\mathcal{F}}}}\left|\hat{e}_{\text{switch}}(f)-e_{\text{switch}}(f)\right|>B_{\text{ind}}\sqrt{\frac{\log(4\delta^{-1}\mathcal{N}\left(\mathcal{F},r\sqrt{2}\right))}{n}}+2r\sqrt{2}\right)\nonumber \\
 & \leq\frac{\delta}{2}+\frac{\delta}{2}.\hspace{8cm}\hspace{2cm}\hspace{1cm}\text{from Eqs \ref{eq:unif-e0} and \ref{eq:unif-e1}}\label{eq:first-run}
\end{align}
\end{proof}

\subsubsection{Proof for Eq \ref{eq:unif-diff}}
\begin{proof}
Finally, to show \ref{eq:unif-diff}, we apply the same steps as in Eqs \ref{eq:run-start} through \ref{eq:first-run}. We apply Eq \ref{eq:Q4-diff-def} \& Lemma \ref{lem:ratio-diff-bound} to obtain
\begin{align*}
 & \mathbb{P}\left[\sup_{f\in\mathcal{F}}\left|\left\{ \hat{e}_{\text{switch}}(f)-\hat{e}_{\text{orig}}(f)\right\} -\left\{ e_{\text{switch}}(f)-e_{\text{orig}}(f)\right\} \right|>\mathcal{Q}_{4,\text{difference}}\right]\\
 & \leq\mathbb{P}\left(\left\{ \exists f\in\mathcal{F}\text{ : }\left|\hat{e}_{\text{orig}}(f)-e_{\text{orig}}(f)\right|>B_{\text{ind}}\sqrt{\frac{\log(4\delta^{-1}\mathcal{N}\left(\mathcal{F},r\right))}{2n}}+2r\right\} \right.\\
 & \hspace{1cm}\hspace{1cm}\bigcup\left.\left\{ \exists f\in\mathcal{F}\text{ : }\left|\hat{e}_{\text{switch}}(f)-e_{\text{switch}}(f)\right|>B_{\text{ind}}\sqrt{\frac{\log(4\delta^{-1}\mathcal{N}\left(\mathcal{F},r\sqrt{2}\right))}{n}}+2r\sqrt{2}\right\} \right)\\
 & \leq\frac{\delta}{2}+\frac{\delta}{2}.
\end{align*}
\end{proof}

\subsubsection{Implementing Theorem \ref{thm:unif-covering} to Show Theorem \ref{thm:mcr-consistent}\label{subsec:inner-MCR-estimates}}
\begin{proof}
Consider the event that
\begin{equation}
\exists\hat{f}_{+,\epsilon_{\text{in}}}\in\argmax_{f\in\hat{\mathcal{R}}(\epsilon_{\text{in}})}\widehat{MR}(f)\text{ such that }MCR_{+}(\epsilon)<MR\left(\hat{f}_{+,\epsilon_{\text{in}}}\right).\label{eq:learning-mcr-event}
\end{equation}

A brief outline of our proof for Eq \ref{eq:result-inner-for-f+} is as follows. We expect Eq \ref{eq:learning-mcr-event} to be unlikely due to the fact that $\epsilon_{\text{in}}<\epsilon$. If Eq \ref{eq:learning-mcr-event} does not hold, then the only way that $MCR_{+}(\epsilon)<\widehat{MCR}_{+}\left(\epsilon_{\text{in}}\right)-\mathcal{Q}_{\text{in}}$ holds is if there exists $\hat{f}_{+,\epsilon_{\text{in}}}\in\argmax_{f\in\hat{\mathcal{R}}(\epsilon_{\text{in}})}\widehat{MR}(f)$ which has an empirical MR that differs from its population-level MR by at least $\mathcal{Q}_{\text{in}}$.

To show that Eq \ref{eq:learning-mcr-event} is unlikely, we apply Theorem \ref{thm:unif-covering}:
\begin{align}
 & \mathbb{P}\left(\exists\hat{f}_{+,\epsilon_{\text{in}}}\in\argmax_{f\in\hat{\mathcal{R}}(\epsilon_{\text{in}})}\widehat{MR}(f)\text{ : }MCR_{+}(\epsilon)<MR\left(\hat{f}_{+,\epsilon_{\text{in}}}\right)\right)\nonumber \\
 & \hspace{0.22cm}\leq\mathbb{P}\left(\exists f\in\hat{\mathcal{R}}(\epsilon_{\text{in}})\text{ : }MCR_{+}(\epsilon)<MR\left(f\right)\right)\nonumber \\
 & \hspace{0.22cm}=\mathbb{P}\left(\exists f\in\hat{\mathcal{R}}(\epsilon_{\text{in}})\backslash f_{\text{ref}}\text{ : }MCR_{+}(\epsilon)<MR\left(f\right)\right) &  & \text{by }MCR_{+}(\epsilon)\geq MR(f_{\text{ref}})\nonumber \\
 & \hspace{0.22cm}\leq\mathbb{P}\left(\exists f\in\hat{\mathcal{R}}(\epsilon_{\text{in}})\backslash f_{\text{ref}}\text{ : }\mathbb{E}\tilde{L}(f,Z)>\epsilon\right) &  & \text{by \ensuremath{MCR_{+}(\epsilon)} Def}\nonumber \\
 & \hspace{0.22cm}=\mathbb{P}\left(\exists f\in\mathcal{F},\mathbb{E}\tilde{L}(f,Z)>\epsilon\text{ : }\hat{\mathbb{E}}\tilde{L}(f,Z)\leq\epsilon_{\text{in}}\right) &  & \text{by \ensuremath{\hat{\mathcal{R}}(\epsilon)} Def}\nonumber \\
 & \hspace{0.22cm}=\mathbb{P}\left(\exists f\in\mathcal{F},\mathbb{E}\tilde{L}(f,Z)>\epsilon\text{ : }\vphantom{\sqrt{\frac{\log(4\delta^{-1}\mathcal{N}\left(\mathcal{F},r\right))}{2n}}}\right.\nonumber \\
 & \hspace{2cm}\left.\hat{\mathbb{E}}\tilde{L}(f,Z)-\epsilon\leq-2B_{\text{ref}}\sqrt{\frac{\log(4\delta^{-1}\mathcal{N}\left(\mathcal{F},r\right))}{2n}}-2r\right) &  & \text{by \ensuremath{\epsilon_{\text{in}}} Def}\\
 & \hspace{0.22cm}\leq\mathbb{P}\left(\exists f\in\mathcal{F},\mathbb{E}\tilde{L}(f,Z)>\epsilon\text{ : }\vphantom{\sqrt{\frac{\log(4\delta^{-1}\mathcal{N}\left(\mathcal{F},r\right))}{2n}}}\right.\nonumber \\
 & \hspace{2cm}\left.\hat{\mathbb{E}}\tilde{L}(f,Z)-\mred{\mathbb{E}\tilde{L}(f,Z)}\leq-2B_{\text{ref}}\sqrt{\frac{\log(4\delta^{-1}\mathcal{N}\left(\mathcal{F},r\right))}{2n}}-2r\right)\nonumber \\
 & \hspace{0.22cm}\leq\mathbb{P}\left(\sup_{f\in\mathcal{F}}|\hat{\mathbb{E}}\tilde{L}(f,Z)-\mathbb{E}\tilde{L}(f,Z)|\geq2B_{\text{ref}}\sqrt{\frac{\log(4\delta^{-1}\mathcal{N}\left(\mathcal{F},r\right))}{2n}}+2r\right)\nonumber \\
 & \hspace{0.22cm}=\frac{\delta}{2} &  & \text{by Thm \ref{thm:unif-covering}.}\label{eq:apply-learning-bound}
\end{align}

If Eq \ref{eq:learning-mcr-event} does not hold, we have
\begin{align}
MCR_{+}(\epsilon) & \geq MR\left(\hat{f}_{+,\epsilon_{\text{in}}}\right) &  & \text{for all }\hat{f}_{+,\epsilon_{\text{in}}}\in\argmax_{f\in\hat{\mathcal{R}}(\epsilon_{\text{in}})}\widehat{MR}(f)\nonumber \\
 & =\widehat{MR}\left(\hat{f}_{+,\epsilon_{\text{in}}}\right)-\left\{ \widehat{MR}\left(\hat{f}_{+,\epsilon_{\text{in}}}\right)-MR\left(\hat{f}_{+,\epsilon_{\text{in}}}\right)\right\}  &  & \text{for all }\hat{f}_{+,\epsilon_{\text{in}}}\in\argmax_{f\in\hat{\mathcal{R}}(\epsilon_{\text{in}})}\widehat{MR}(f)\nonumber \\
 & =\widehat{MCR}_{+}\left(\epsilon_{\text{in}}\right)-\left\{ \widehat{MR}\left(\hat{f}_{+,\epsilon_{\text{in}}}\right)-MR\left(\hat{f}_{+,\epsilon_{\text{in}}}\right)\right\}  &  & \text{for all }\hat{f}_{+,\epsilon_{\text{in}}}\in\argmax_{f\in\hat{\mathcal{R}}(\epsilon_{\text{in}})}\widehat{MR}(f)\nonumber \\
 & \geq\widehat{MCR}_{+}\left(\epsilon_{\text{in}}\right)-\sup_{f\in\mathcal{F}}|\widehat{MR}(f)-MR(f)|.\label{eq:lower+}
\end{align}

\textcolor{black}{Let $q_{\text{ratio}}$ and $q_{\text{difference}}$ be defined as in Lemma \ref{lem:ratio-diff-bound}. Then }

\textcolor{black}{
\begin{align}
\mathcal{Q}_{\text{in}} & =\frac{B_{\text{switch}}}{b_{\text{orig}}}-\frac{B_{\text{switch}}-\left\{ B_{\text{ind}}\sqrt{\frac{\log(8\delta^{-1}\mathcal{N}\left(\mathcal{F},r\sqrt{2}\right))}{n}}+2r\sqrt{2}\right\} }{b_{\text{orig}}+\left\{ B_{\text{ind}}\sqrt{\frac{\log(8\delta^{-1}\mathcal{N}\left(\mathcal{F},r\right))}{2n}}+2r\right\} }\nonumber \\
 & =q_{\text{ratio}}\left(B_{\text{ind}}\sqrt{\frac{\log(8\delta^{-1}\mathcal{N}\left(\mathcal{F},r\sqrt{2}\right))}{n}}+2r\sqrt{2},B_{\text{ind}}\sqrt{\frac{\log(8\delta^{-1}\mathcal{N}\left(\mathcal{F},r\right))}{2n}}+2r\right)\label{eq:Q3-plug-in}
\end{align}
Theorem \ref{thm:unif-covering} implies that the sup term in Eq \ref{eq:lower+} is less than $\mathcal{Q}_{\text{in}}$ with probability at least $1-\frac{\delta}{2}$. Now, examining the left-hand side of Eq }\ref{eq:result-inner-for-f+}\textcolor{black}{, we see}
\begin{align}
 & \mathbb{P}\left(MCR_{+}(\epsilon)<\widehat{MCR}_{+}\left(\epsilon_{\text{in}}\right)-\mathcal{Q}_{\text{in}}\right)\nonumber \\
 & \hspace{1cm}\leq\mathbb{P}\left[\left\{ \exists\hat{f}_{+,\epsilon_{\text{in}}}\in\argmax_{f\in\hat{\mathcal{R}}(\epsilon_{\text{in}})}\widehat{MR}(f)\text{ : }MCR_{+}(\epsilon)<MR\left(\hat{f}_{+,\epsilon_{\text{in}}}\right)\right\} \right.\nonumber \\
 & \hspace{1cm}\hspace{1cm}\hspace{1cm}\bigcup\left.\left\{ \sup_{f\in\mathcal{F}}|\widehat{MR}(f)-MR(f)|>\mathcal{Q}_{\text{in}}\right\} \right]\hspace{1.66cm}\text{from Eq \ref{eq:lower+}}\nonumber \\
 & \hspace{1cm}\leq\mathbb{P}\left[\exists\hat{f}_{+,\epsilon_{\text{in}}}\in\argmax_{f\in\hat{\mathcal{R}}(\epsilon_{\text{in}})}\widehat{MR}(f)\text{ : }MCR_{+}(\epsilon)<MR\left(\hat{f}_{+,\epsilon_{\text{in}}}\right)\right]\nonumber \\
 & \hspace{1cm}\hspace{1cm}\hspace{1cm}\mred{+}\mathbb{P}\left[\sup_{f\in\mathcal{F}}|\widehat{MR}(f)-MR(f)|>\mathcal{Q}_{\text{in}}\right]\text{\hspace{2cm}from the Union bound}\nonumber \\
 & \hspace{1cm}=\frac{\delta}{2}+\frac{\delta}{2}\hspace{2.55cm}\text{from Eq \ref{eq:apply-learning-bound}, Eq \ref{eq:Q3-plug-in}, \& Theorem \ref{thm:unif-covering}}.\label{eq:prob-combine-precise}
\end{align}

This completes the proof for Eq \ref{eq:result-inner-for-f+}.

\textcolor{black}{Alternatively, if we have defined model reliance as $MR(f)=e_{\text{switch}}(f)-e_{\text{orig}}(f)$ (see Appendix \ref{sec:Discussion---Ratios}), with $\widehat{MR}(f)=\hat{e}_{\text{switch}}(f)-\hat{e}_{\text{orig}}(f)$, and}

\textcolor{black}{
\begin{align*}
\mathcal{Q}_{\text{in},\text{difference}} & =B_{\text{ind}}\left\{ \sqrt{\frac{\log(8\delta^{-1}\mathcal{N}\left(\mathcal{F},r\sqrt{2}\right))}{n}}+\sqrt{\frac{\log(8\delta^{-1}\mathcal{N}\left(\mathcal{F},r\right))}{2n}}\right\} +2r(\sqrt{2}+1)\\
 & =q_{\text{difference}}\left(B_{\text{ind}}\sqrt{\frac{\log(8\delta^{-1}\mathcal{N}\left(\mathcal{F},r\sqrt{2}\right))}{n}}+2r\sqrt{2},B_{\text{ind}}\sqrt{\frac{\log(8\delta^{-1}\mathcal{N}\left(\mathcal{F},r\right))}{2n}}+2r\right),
\end{align*}
then same proof of Eq }\ref{eq:result-inner-for-f+}\textcolor{black}{{} holds without Assumption \ref{assu:bnd-avg} if we replace $\mathcal{Q}_{\text{in}}$ with $\mathcal{Q}_{\text{in},\text{difference}}$, and apply Eq \ref{eq:unif-diff} in Eq \ref{eq:prob-combine-precise}.}

\textcolor{black}{For Eq \ref{eq:result_inner_for_f-} we can use} the same approach. Consider the event that
\begin{equation}
\exists\hat{f}_{-,\epsilon_{\text{in}}}\in\argmin_{f\in\hat{\mathcal{R}}(\epsilon_{\text{in}})}\widehat{MR}(f)\text{ : }MCR_{-}(\epsilon)>MR\left(\hat{f}_{-,\epsilon_{\text{in}}}\right).\label{eq:upper--event}
\end{equation}

Applying steps analogous to those used to derive Eq \ref{eq:apply-learning-bound}, we have
\begin{align*}
 & \mathbb{P}\left(\exists\hat{f}_{-,\epsilon_{\text{in}}}\in\argmin_{f\in\hat{\mathcal{R}}(\epsilon_{\text{in}})}\widehat{MR}(f)\text{ : }MCR_{-}(\epsilon)>MR\left(\hat{f}_{-,\epsilon_{\text{in}}}\right)\right)\\
 & \hspace{3cm}\leq\hspace{0.25cm}\mathbb{P}\left(\exists f\in\mathcal{F},\mathbb{E}\tilde{L}(f,Z)>\epsilon\text{ : }\hat{\mathbb{E}}\tilde{L}(f,Z)\leq\epsilon_{\text{in}}\right)\hspace{0.25cm}\leq\hspace{0.25cm}\frac{\delta}{2}.
\end{align*}

Analogous to \ref{eq:lower+}, when Eq \ref{eq:upper--event} does not hold, we have have
\begin{align*}
MCR_{-}(\epsilon) & \leq MR\left(\hat{f}_{-,\epsilon_{\text{in}}}\right) &  & \text{for all }\hat{f}_{-,\epsilon_{\text{in}}}\in\argmin_{f\in\hat{\mathcal{R}}(\epsilon_{\text{in}})}\widehat{MR}(f)\\
 & =\widehat{MR}\left(\hat{f}_{-,\epsilon_{\text{in}}}\right)+\left\{ MR\left(\hat{f}_{-,\epsilon_{\text{in}}}\right)-\widehat{MR}\left(\hat{f}_{-,\epsilon_{\text{in}}}\right)\right\}  &  & \text{for all }\hat{f}_{-,\epsilon_{\text{in}}}\in\argmin_{f\in\hat{\mathcal{R}}(\epsilon_{\text{in}})}\widehat{MR}(f)\\
 & =\widehat{MCR}_{-}\left(\epsilon_{\text{in}}\right)+\left\{ MR\left(\hat{f}_{-,\epsilon_{\text{in}}}\right)-\widehat{MR}\left(\hat{f}_{-,\epsilon_{\text{in}}}\right)\right\}  &  & \text{for all }\hat{f}_{-,\epsilon_{\text{in}}}\in\argmin_{f\in\hat{\mathcal{R}}(\epsilon_{\text{in}})}\widehat{MR}(f)\\
 & \leq\widehat{MCR}_{-}\left(\epsilon_{\text{in}}\right)+\sup_{f\in\mathcal{F}}|MR(f)-\widehat{MR}(f)|
\end{align*}

Finally, analogous to Eq \ref{eq:prob-combine-precise}, 
\begin{align}
 & \mathbb{P}\left(MCR_{-}(\epsilon)>\widehat{MCR}\left(\epsilon_{\text{in}}\right)+\mathcal{Q}_{\text{in}}\right)\nonumber \\
 & \hspace{1cm}\leq\mathbb{P}\left[\left\{ \exists\hat{f}_{-,\epsilon_{\text{in}}}\in\argmin_{f\in\hat{\mathcal{R}}(\epsilon_{\text{in}})}\widehat{MR}(f)\text{ : }MCR_{-}(\epsilon)>MR\left(\hat{f}_{-,\epsilon_{\text{in}}}\right)\right\} \right.\nonumber \\
 & \hspace{1cm}\hspace{1cm}\hspace{1cm}\bigcup\left.\left\{ \sup_{f\in\mathcal{F}}|\widehat{MR}(f)-MR(f)|>\mathcal{Q}_{\text{in}}\right\} \right]\nonumber \\
 & \hspace{1cm}\leq\mathbb{P}\left[\exists\hat{f}_{-,\epsilon_{\text{in}}}\in\argmin_{f\in\hat{\mathcal{R}}(\epsilon_{\text{in}})}\widehat{MR}(f)\text{ : }MCR_{-}(\epsilon)>MR\left(\hat{f}_{-,\epsilon_{\text{in}}}\right)\right]\nonumber \\
 & \hspace{1cm}\hspace{1cm}\hspace{1cm}\mred{+}\mathbb{P}\left[\sup_{f\in\mathcal{F}}|\widehat{MR}(f)-MR(f)|>\mathcal{Q}_{\text{in}}\right]\nonumber \\
 & \hspace{1cm}=\frac{\delta}{2}+\frac{\delta}{2}.\label{eq:upper--final}
\end{align}

\textcolor{black}{Under the difference-based definition of model reliance (see Appendix \ref{sec:Discussion---Ratios}), the same proof for Eq \ref{eq:result_inner_for_f-} holds without Assumption \ref{assu:bnd-avg} if we replace $MR$, $\widehat{MR}$, \& $\mathcal{Q}_{\text{in}}$ respectively with $MR_{\text{difference}}$, $\widehat{MR}_{\text{difference}}$, \& $\mathcal{Q}_{\text{in},\text{difference}}$, and apply Eq \ref{eq:unif-diff} in Eq \ref{eq:upper--final}.}
\end{proof}

\subsection{Proof of Proposition \ref{rem:simplified-thm3-general}, and Corollary for a Unique Best-in-class Model.\label{subsec:Proof-of-Propositions-general}}

We first introduce a lemma to describe the performance of any individual model in the population $\epsilon$-Rashomon set.
\begin{lemma}
\textcolor{black}{\label{lem:f1-range}Let $\epsilon_{1}':=2B_{\text{ref}}\sqrt{\frac{\log(\delta^{-1})}{2n}}$, and let the functions $\hat{\phi}_{-}$ and $\hat{\phi}_{+}$ be defined as in Proposition \ref{rem:simplified-thm3-general}. Given a function $f_{1}\in\mathcal{R}(\epsilon)$, if Assumption \ref{assu:bnd-ref} holds for $f_{1}$, then 
\[
\mathbb{P}\left\{ \phi(f_{1})\in\left[\hat{\phi}_{-}(\epsilon_{1}'),\hat{\phi}_{+}(\epsilon_{1}')\right]\right\} \geq1-\delta.
\]
}
\end{lemma}

\begin{proof}
\textcolor{black}{Consider the event that 
\begin{equation}
\phi(f_{1})\in\left[\hat{\phi}_{-}(\epsilon_{1}'),\hat{\phi}_{+}(\epsilon_{1}')\right].\label{eq:first-max-1}
\end{equation}
Eq \ref{eq:first-max-1} will always hold if $f_{1}\in\hat{\mathcal{R}}(\epsilon_{1}')$, since} the interval $\left[\hat{\phi}_{-}(\epsilon_{1}'),\hat{\phi}_{+}(\epsilon_{1}')\right]$ contains $\phi(f)$ for any $f\in\hat{\mathcal{R}}(\epsilon_{1}')$\textcolor{black}{{} by} definition. Thus,
\begin{align*}
\mathbb{P}\left\{ \phi(f_{1})\notin[\left[\hat{\phi}_{-}(\epsilon_{1}'),\hat{\phi}_{+}(\epsilon_{1}')\right]\right\}  & \leq\mathbb{P}\left\{ f_{1}\notin\hat{\mathcal{R}}(\epsilon_{1}')\right\} \\
 & \leq\delta &  & \text{from Lemma \ref{lem:hoeffding}}.
\end{align*}
\end{proof}

\subsubsection{Proof of Proposition \ref{rem:simplified-thm3-general}}
\begin{proof}
Let $f_{-,\epsilon,\phi}\in\argmin_{f\in\mathcal{R}\left(\epsilon\right)}\phi(f)$ and $f_{+,\epsilon,\phi}\in\argmax_{f\in\mathcal{R}\left(\epsilon\right)}\phi(f)$ respectively denote functions that attain the lowest and highest values of $\phi(f)$ among models $f\in\mathcal{R}\left(\epsilon\right)$. Applying the definitions of $f_{-,\epsilon,\phi}$ and $f_{+,\epsilon,\phi}$ in Line \ref{eq:a-interval}, below, we have
\begin{align}
 & \mathbb{P}\left(\,\left\{ \phi(f):f\in\mathcal{R}\left(\epsilon\right)\right\} \not\subset[\hat{\phi}_{-}(\epsilon'),\hat{\phi}_{+}(\epsilon')]\,\right)\nonumber \\
 & \hspace{0.7cm}=\mathbb{P}\left(\left[\phi(f_{-,\epsilon,\phi}),\phi(f_{+,\epsilon,\phi})\right]\not\subset[\hat{\phi}_{-}(\epsilon'),\hat{\phi}_{+}(\epsilon')]\right)\label{eq:a-interval}\\
 & \hspace{0.7cm}=\mathbb{P}\left(\phi(f_{-,\epsilon,\phi})\notin[\hat{\phi}_{-}(\epsilon'),\hat{\phi}_{+}(\epsilon')]\;\;\;\bigcup\;\;\phi(f_{+,\epsilon,\phi})\notin[\hat{\phi}_{-}(\epsilon'),\hat{\phi}_{+}(\epsilon')]\right)\nonumber \\
 & \hspace{0.7cm}\leq\mathbb{P}\left(\phi(f_{-,\epsilon,\phi})\notin[\hat{\phi}_{-}(\epsilon'),\hat{\phi}_{+}(\epsilon')]\right)+\mathbb{P}\left(\phi(f_{+,\epsilon,\phi})\notin[\hat{\phi}_{-}(\epsilon'),\hat{\phi}_{+}(\epsilon')]\right)\nonumber \\
 & \hspace{0.7cm}\leq\frac{\delta}{2}+\frac{\delta}{2}=\delta\hspace{1.5cm}\text{from Lemma \ref{lem:f1-range}, and the definition of }\epsilon'=\epsilon+2B_{\text{ref}}\sqrt{\frac{\log(2\delta^{-1})}{2n}}.\nonumber 
\end{align}
\end{proof}

\subsubsection{Corollary for a Unique Best-in-Class Model}

When the best-in-class model is unique, it can be described by the corollary below.
\begin{corollary}
\label{rem:simplified-thm3-best}Let $\hat{\phi}_{-}(\epsilon_{0}'):=\min_{f\in\hat{\mathcal{R}}\left(\epsilon_{1}'\right)}\phi(f)$ and $\hat{\phi}_{+}(\epsilon_{1}'):=\max_{f\in\hat{\mathcal{R}}\left(\epsilon_{1}'\right)}\phi(f)$, where $\epsilon_{0}':=2B_{\text{ref}}\sqrt{\frac{\log(\delta^{-1})}{2n}}$. Let $f^{\star}\in\argmin_{f\in\mathcal{F}}e_{\text{orig}}(f)$ be the prediction model that uniquely attains the lowest possible expected loss. If $f^{\star}$ satisfies Assumption \ref{assu:bnd-ref}, then 
\[
\mathbb{P}\{\phi(f^{\star})\in[\hat{\phi}_{-}(\epsilon_{1}'),\hat{\phi}_{+}(\epsilon_{1}')]\}\geq1-\delta.
\]
\end{corollary}

\begin{proof}
Since $f^{\star}\in\mathcal{R}(0)$, Corollary \ref{rem:simplified-thm3-best} follows immediately from Lemma \ref{lem:f1-range}.

Notice that by assuming $f^{\star}$ is unique, we can use the threshold $\epsilon_{0}':=2B_{\text{ref}}\sqrt{\frac{\log(\delta^{-1})}{2n}}$, which is lower than the threshold of $\epsilon'=\epsilon+2B_{\text{ref}}\sqrt{\frac{\log(2\delta^{-1})}{2n}}$ with $\epsilon=0$, as in Proposition \ref{rem:simplified-thm3-general}. In this way, assuming uniqueness allows a stronger statement than the one in Proposition \ref{rem:simplified-thm3-general}.
\end{proof}

\subsection{\textcolor{black}{Absolute Losses versus Relative Losses in the Definition of the Rashomon Set\label{subsec:absolute-rashomon}}}

\textcolor{black}{In this paper we primarily define Rashomon sets as the models that perform well }\textcolor{black}{\emph{relative }}\textcolor{black}{to a reference model $f_{\text{ref}}$. We can also study an alternate formulation of Rashomon sets by replacing the relative loss $\tilde{L}$ with the non-standardized loss $L$ throughout. This results in a new interpretation of the Rashomon set $\mathcal{R}(\epsilon_{\text{abs}},f_{\text{ref}},\mathcal{F})=\{f_{\text{ref}}\}\cup\{f\in\mathcal{F}\text{ : }\mathbb{E}L(f,Z)\leq\epsilon_{\text{abs}}\}$ as the union of $f_{\text{ref}}$ and the subset of models with }\textcolor{black}{\emph{absolute}}\textcolor{black}{{} loss $L$ no higher than $\epsilon_{\text{abs}}$, for $\epsilon_{\text{abs}}>0$. }The process of computing empirical MCR is largely unaffected by whether $L$ or \textcolor{black}{$\tilde{L}$} is used, as it is simple to transform from one optimization problem to the other.

\textcolor{black}{We still require the explicit inclusion of $f_{\text{ref}}$ in empirical and population Rashomon sets to ensure that they are nonempty. However, in many cases, this inclusion becomes redundant when interpreting a Rashomon set (e.g., when $\epsilon\geq0$, and $\mathbb{E}L(f_{\text{ref}},Z)\leq\epsilon_{\text{abs}}$).}

\textcolor{black}{Under the replacement of $\tilde{L}$ with $L$, we also replace Assumption \ref{assu:bnd-ref} with Assumption \ref{assu:bnd-ind} (whenever this is not redundant), and replace $2B_{\text{ref}}$ with $B_{\text{ind}}$ in the definitions of $\epsilon_{\text{out}}$, $\epsilon_{\text{best}}$, $\epsilon_{\text{in}}$, $\epsilon'$ and $\epsilon_{1}'$ in Theorem \ref{thm:mcr-conserve-bounds}, Corollary \ref{cor:best-in-class}, Theorem \ref{thm:mcr-consistent}, Proposition \ref{rem:simplified-thm3-general}, and Corollary \ref{rem:simplified-thm3-best}. This is because the motivation for the $2B_{\text{ref}}$ term is that $\tilde{L}(f_{1})$ is bounded within an interval of length $2B_{\text{ref}}$ when $f_{1}$ satisfies Assumption \ref{assu:bnd-ref}. However, under Assumption \ref{assu:bnd-ind}, $L(f_{1})$ is bounded within an interval of length $B_{\text{ind}}$.}

\subsection{Proof of Proposition \ref{thm:Linear-models}\label{subsec:Proof-lin-models-thm}}
\begin{proof}
To show Eq \ref{eq:linear-models-pop} we start with $e_{\text{orig}}(f_{\beta})$,
\begin{align*}
e_{\text{orig}}(f_{\beta}) & =\mathbb{E}[\{Y-X_{1}'\beta_{1}-X_{2}'\beta_{2}\}^{2}]\\
 & =\mathbb{E}[\{(Y-X_{2}'\beta_{2})-X_{1}'\beta_{1}\}^{2}]\\
 & =\mathbb{E}[(Y-X_{2}'\beta_{2})^{2}]-2\mathbb{E}[(Y-X_{2}'\beta_{2})X_{1}']\beta_{1}+\beta_{1}'\mathbb{E}[X_{1}X_{1}']\beta_{1}.
\end{align*}

For $e_{\text{switch}}(f_{\beta})$, we can follow the same steps as above:
\begin{align*}
e_{\text{switch}}(f_{\beta}) & =\mathbb{E}_{Y^{(b)},X_{1}^{(a)},X_{2}^{(b)}}[\{Y^{(b)}-X_{1}^{(a)'}\beta_{1}-X_{2}^{(b)'}\beta_{2}\}^{2}]\\
 & =\mathbb{E}[(Y^{(b)}-X_{2}^{(b)'}\beta_{2})^{2}]-2\mathbb{E}[Y^{(b)}-X_{2}^{(b)'}\beta_{2}]\mathbb{E}[X_{1}^{(a)'}]\beta_{1}+\beta_{1}'\mathbb{E}[X_{1}^{(a)}X_{1}^{(a)'}]\beta_{1}.
\end{align*}

Since $(Y^{(b)},X_{1}^{(b)},X_{2}^{(b)})$ and $(Y^{(a)},X_{1}^{(a)},X_{2}^{(a)})$ each have the same distribution as $(Y,X_{1},X_{2})$, we can omit the superscript notation to show Eq \ref{eq:linear-models-pop}:
\begin{align*}
e_{\text{switch}}(f_{\beta}) & =\mathbb{E}[(Y-X_{2}'\beta_{2})^{2}]-2\mathbb{E}[Y-X_{2}'\beta_{2}]\mathbb{E}[X_{1}']\beta_{1}+\beta_{1}'\mathbb{E}[X_{1}X_{1}']\beta_{1}\\
e_{\text{switch}}(f_{\beta}) & =e_{\text{orig}}(f_{\beta})-2\mathbb{E}[Y-X_{2}'\beta_{2}]\mathbb{E}[X_{1}']\beta_{1}+2\mathbb{E}[(Y-X_{2}'\beta_{2})X_{1}']\beta_{1}\\
e_{\text{switch}}(f_{\beta}) & =e_{\text{orig}}(f_{\beta})+2\text{Cov}(Y-X_{2}'\beta_{2},X_{1})\beta_{1}\\
e_{\text{switch}}(f_{\beta}) & =e_{\text{orig}}(f_{\beta})+2\text{Cov}(Y,X_{1})\beta_{1}-2\beta_{2}\text{Cov}(X_{2},X_{1})\beta_{1}.\\
\end{align*}

Dividing both sides by  $e_{\text{orig}}(f_{\beta})$ gives the desired result.

Next, we can use a similar approach to show Eq \ref{eq:linear-models-finite}:
\begin{align}
\hat{e}_{\text{switch}}(f_{\beta}) & =\frac{1}{n(n-1)}\sum_{i=1}^{n}\sum_{j\neq i}(\mathbf{y}_{[j]}-\mathbf{X}_{2[j,\cdot]}\beta_{2}-\mathbf{X}_{1[i,\cdot]}\beta_{1})^{2}\nonumber \\
n(n-1)\hat{e}_{\text{switch}}(f_{\beta}) & =\sum_{i=1}^{n}\sum_{j\neq i}\left\{ (\mathbf{y}_{[j]}-\mathbf{X}_{2[j,\cdot]}\beta_{2})^{2}-2(\mathbf{y}_{[j]}-\mathbf{X}_{2[j,\cdot]}\beta_{2})(\mathbf{X}_{1[i,\cdot]}\beta_{1})+(\mathbf{X}_{1[i,\cdot]}\beta_{1})^{2}\right\} \nonumber \\
 & =(n-1)\sum_{i=1}^{n}(\mathbf{y}_{[i]}-\mathbf{X}_{2[i,\cdot]}\beta_{2})^{2}\nonumber \\
 & \hspace{1cm}-2\left\{ \sum_{i=1}^{n}\sum_{j\neq i}(\mathbf{X}_{1[i,\cdot]}\beta_{1})(\mathbf{y}_{[j]}-\mathbf{X}_{2[j,\cdot]}\beta_{2})\right\} +(n-1)\sum_{i=1}^{n}(\mathbf{X}_{1[i,\cdot]}\beta_{1})^{2}.\label{eq:break-apart-linear-sum}
\end{align}

Focusing on the term in braces,
\begin{align}
 & \sum_{i=1}^{n}\sum_{\mred{j\neq i}}(\mathbf{X}_{1[i,\cdot]}\beta_{1})(\mathbf{y}_{[j]}-\mathbf{X}_{2[j,\cdot]}\beta_{2})\nonumber \\
 & \hspace{1.5cm}=\sum_{i=1}^{n}\sum_{\mred{j=1}}^{n}(\mathbf{X}_{1[i,\cdot]}\beta_{1})(\mathbf{y}_{[j]}-\mathbf{X}_{2[j,\cdot]}\beta_{2})-\sum_{\mred{i=1}}^{n}(\mathbf{X}_{1[i,\cdot]}\beta_{1})(\mathbf{y}_{[i]}-\mathbf{X}_{2[i,\cdot]}\beta_{2})\nonumber \\
 & \hspace{1.5cm}=\sum_{i=1}^{n}(\mathbf{X}_{1[i,\cdot]}\beta_{1})\sum_{j=1}^{n}(\mathbf{y}_{[j]}-\mathbf{X}_{2[j,\cdot]}\beta_{2})-\sum_{i=1}^{n}(\mathbf{X}_{1[i,\cdot]}\beta_{1})(\mathbf{y}_{[i]}-\mathbf{X}_{2[i,\cdot]}\beta_{2})\nonumber \\
 & \hspace{1.5cm}=\left\{ (\mathbf{X}_{1}\beta_{1})'\mathbf{1}_{n}\right\} \left\{ \mathbf{1}_{n}'(\mathbf{y}-\mathbf{X}_{2}\beta_{2})\right\} -(\mathbf{X}_{1}\beta_{1})'(\mathbf{y}-\mathbf{X}_{2}\beta_{2})\label{eq:linear-alg-rep}\\
 & \hspace{1.5cm}=(\mathbf{X}_{1}\beta_{1})'(\mathbf{1}_{n}\mathbf{1}_{n}'-\mathbf{I}_{n})(\mathbf{y}-\mathbf{X}_{2}\beta_{2}).\nonumber 
\end{align}

Plugging this into Eq \ref{eq:break-apart-linear-sum}, and applying the sample linear algebra representations as in Eq \ref{eq:linear-alg-rep}, we get
\begin{align*}
n(n-1)\hat{e}_{\text{switch}}(f_{\beta}) & =(n-1)\|\mathbf{y}-\mathbf{X}_{2}\beta_{2}\|_{2}^{2}\\
 & \hspace{1cm}-2(\mathbf{X}_{1}\beta_{1})'(\mathbf{1}_{n}\mathbf{1}_{n}'-\mathbf{I}_{n})(\mathbf{y}-\mathbf{X}_{2}\beta_{2})\\
 & \hspace{1cm}+(n-1)\|\mathbf{X}_{1}\beta_{1}\|_{2}^{2}\\
n\hat{e}_{\text{switch}}(f_{\beta}) & =\|\mathbf{y}-\mathbf{X}_{2}\beta_{2}\|_{2}^{2}\\
 & \hspace{1cm}-2(\mathbf{X}_{1}\beta_{1})'\mathbf{W}(\mathbf{y}-\mathbf{X}_{2}\beta_{2})\\
 & \hspace{1cm}+\|\mathbf{X}_{1}\beta_{1}\|_{2}^{2}\\
 & =\mathbf{y}'\mathbf{y}-2\mathbf{y}'\mathbf{X}_{2}\beta_{2}+\beta_{2}'\mathbf{X}_{2}'\mathbf{X}_{2}\beta_{2}\\
 & \hspace{1cm}-2\beta_{1}'\mathbf{X}_{1}'\mathbf{W}\mathbf{y}+2\beta_{1}'\mathbf{X}_{1}'\mathbf{W}\mathbf{X}_{2}\beta_{2}\\
 & \hspace{1cm}+\beta_{1}'\mathbf{X}_{1}'\mathbf{X}_{1}\beta_{1}\\
 & =\mathbf{y}'\mathbf{y}-2\left[\begin{array}{c}
\mathbf{X}_{1}'\mathbf{W}\mathbf{y}\\
\mathbf{X}_{2}'\mathbf{y}
\end{array}\right]^{'}\beta+\beta'\left[\begin{array}{cc}
\mathbf{X}_{1}'\mathbf{X}_{1} & \mathbf{X}_{1}'\mathbf{W}\mathbf{X}_{2}\\
\mathbf{X}_{2}'\mathbf{W}\mathbf{X}_{1} & \mathbf{X}_{2}'\mathbf{X}_{2}
\end{array}\right]\beta.
\end{align*}
\end{proof}

\subsection{Proof of Proposition \ref{thm:VI_TRT}\label{sec:Proof-VI-TRT}}
\begin{proof}
First we consider $e_{\text{orig}}(f_{0})$. We briefly recall that the notation $f_{0}(t,c)$ refers to the \emph{true} conditional expectation function for \emph{both} potential outcomes $Y_{1},Y_{0}$, rather than the expectation for $Y_{0}$ alone. 

Under the assumption that $(Y_{1},Y_{0})\perp T|C$, we have $f_{0}(t,c)=\mathbb{E}(Y|C=c,T=t)=\mathbb{E}(Y_{t}|C=c)$. Applying this, we see that

\begin{eqnarray}
e_{\text{orig}}(f_{0}) & = & \mathbb{E}L(f_{0},(Y,T,C))\nonumber \\
 & = & \mathbb{E}L(f_{0},(Y_{T},T,C))\nonumber \\
 & = & \mathbb{E}_{T}\mathbb{E}_{C|T}\mathbb{E}_{Y_{T}|C}[\{Y_{T}-\mathbb{E}(Y_{T}|C)\}^{2}]\nonumber \\
 & = & \mathbb{E}_{T}\mathbb{E}_{C|T}\text{Var}(Y_{T}|C)\nonumber \\
 & = & q\mathbb{E}_{C|T=0}\text{Var}(Y_{0}|C)+p\mathbb{E}_{C|T=1}\text{Var}(Y_{1}|C),\label{eq:def-e0}
\end{eqnarray}
where $p:=\mathbb{P}(T=1)$ and $q:=\mathbb{P}(T=0)$.

Now we consider $e_{\text{switch}}(f_{0})$. Let $(Y_{0}^{(a)},Y_{1}^{(a)},T^{(a)},C^{(a)})$ and $(Y_{0}^{(b)},Y_{1}^{(b)},T^{(b)},C^{(b)})$ be a pair of independent random variable vectors, each with the same distribution as $(Y_{0},Y_{1},T,C)$. Then

\begin{eqnarray*}
e_{\text{switch}}(f_{0}) & = & \mathbb{E}_{T^{(b)},T^{(a)},C^{(b)},Y_{T^{(b)}}^{(b)}}[\{Y_{T^{(b)}}^{(b)}-f_{0}(T^{(a)},C^{(b)})\}^{2}]\\
 & = & \mathbb{E}_{T^{(b)},T^{(a)},C^{(b)},Y_{T^{(b)}}^{(b)}}[\{Y_{T^{(b)}}^{(b)}-\mathbb{E}(Y_{T^{(a)}}|C=C^{(b)})\}^{2}]\\
 & = & \mathbb{E}_{T^{(b)},T^{(a)}}\mathbb{E}_{C^{(b)}|T^{(b)}}\mathbb{E}_{Y_{T^{(b)}}^{(b)}|C^{(b)}}[\{Y_{T^{(b)}}^{(b)}-\mathbb{E}(Y_{T^{(a)}}|C=C^{(b)})\}^{2}].
\end{eqnarray*}

First we expand the outermost expectation, over $T^{(b)},T^{(a)}$:
\begin{align}
 & e_{\text{switch}}(f_{0})\nonumber \\
 & \hspace{1em}=\sum_{i,j\in\{0,1\}}\mathbb{P}(T^{(b)}=i,T^{(a)}=j)\mathbb{E}_{C^{(b)}|T^{(b)}=i}\mathbb{E}_{Y_{i}^{(b)}|C^{(b)}}[\{Y_{i}^{(b)}-\mathbb{E}(Y_{j}|C=C^{(b)})\}^{2}].\label{eq:sum-1st}
\end{align}

Since $T^{(b)}\perp T^{(a)}$, we can write

\begin{eqnarray*}
\mathbb{P}(T^{(b)}=i,T^{(a)}=j) & = & \mathbb{P}(T^{(b)}=i)\mathbb{P}(T^{(a)}=j)\\
 & = & p^{i+j}q^{2-i-j}.
\end{eqnarray*}

Plugging this into Eq \ref{eq:sum-1st} we get

\[
e_{\text{switch}}(f_{0})=\sum_{i,j\in\{0,1\}}p^{i+j}q^{2-i-j}\mathbb{E}_{C^{(b)}|T^{(b)}=i}\mathbb{E}_{Y_{i}^{(b)}|C^{(b)}}[\{Y_{i}^{(b)}-\mathbb{E}(Y_{j}|C=C^{(b)})\}^{2}].
\]

Since ($Y_{0}^{(b)},Y_{1}^{(b)},C^{(b)},T^{(b)})$ are the only random variables remaining, we can omit the superscript notation (e.g., replace $C^{(b)}$ with $C$) to get

\begin{eqnarray*}
e_{\text{switch}}(f_{0}) & = & \sum_{i,j\in\{0,1\}}p^{i+j}q^{2-i-j}\mathbb{E}_{C|T=i}\mathbb{E}_{Y_{i}|C}[\{Y_{i}-\mathbb{E}(Y_{j}|C)\}^{2}]\\
 & =: & \sum_{i,j\in\{0,1\}}A_{ij},
\end{eqnarray*}
where $A_{ij}=p^{i+j}q^{2-i-j}\mathbb{E}_{C|T=i}\mathbb{E}_{Y_{i}|C}[\{Y_{i}-\mathbb{E}(Y_{j}|C)\}^{2}]$. First, we consider $A_{00}$ and $A_{11}$:

\begin{eqnarray*}
A_{00} & = & q^{2}\mathbb{E}_{C|T=0}\mathbb{E}_{Y_{0}|C}[\{Y_{0}-\mathbb{E}(Y_{0}|C)\}^{2}]\\
 & = & q^{2}\mathbb{E}_{C|T=0}\text{Var}(Y_{0}|C),
\end{eqnarray*}
and, similarly, $A_{11}=p^{2}\mathbb{E}_{C|T=1}\text{Var}(Y_{1}|C)$.

Next we consider $A_{01}$ and $A_{10}$:
\begin{eqnarray*}
A_{01}: & = & pq\mathbb{E}_{C|T=0}\mathbb{E}_{Y_{0}|C}[\{Y_{0}-\mathbb{E}(Y_{1}|C)\}^{2}]\\
 & = & pq\mathbb{E}_{C|T=0}\left(\mathbb{E}(Y_{0}^{2}|C)-2\mathbb{E}(Y_{0}|C)\mathbb{E}(Y_{1}|C)+\mathbb{E}(Y_{1}|C)^{2}\right)\\
 & = & pq\mathbb{E}_{C|T=0}\left(\text{Var}(Y_{0}|C)+\mathbb{E}(Y_{0}|C)^{2}-2\mathbb{E}(Y_{0}|C)\mathbb{E}(Y_{1}|C)+\mathbb{E}(Y_{1}|C)^{2}\right)\\
 & = & pq\mathbb{E}_{C|T=0}\left(\text{Var}(Y_{0}|C)+\left[\mathbb{E}(Y_{1}|C)-\mathbb{E}(Y_{0}|C)\right]^{2}\right)\\
 & = & pq\mathbb{E}_{C|T=0}\left(\text{Var}(Y_{0}|C)+\text{CATE}(C)^{2}\right),
\end{eqnarray*}
and, following the same steps,

\[
A_{10}=pq\mathbb{E}_{C|T=1}\left(\text{Var}(Y_{1}|C)+\text{CATE}(C)^{2}\right).
\]

Plugging in $A_{00},A_{01},A_{10}$, and $A_{11}$ we get
\begin{eqnarray}
e_{\text{switch}}(f_{0}) & = & \left\{ A_{00}+A_{11}\right\} \nonumber \\
 &  & \hspace{1em}+\left[A_{01}+A_{10}\right]\nonumber \\
 & = & \left\{ q^{2}\mathbb{E}_{C|T=0}\text{Var}(Y_{0}|C)+p^{2}\mathbb{E}_{C|T=1}\text{Var}(Y_{1}|C)\right\} \nonumber \\
 &  & \hspace{1em}+\left[pq\mathbb{E}_{C|T=0}\left(\text{Var}(Y_{0}|C)+\text{CATE}(C)^{2}\right)+pq\mathbb{E}_{C|T=1}\left(\text{Var}(Y_{1}|C)+\text{CATE}(C)^{2}\right)\right]\nonumber \\
 & = & \left\{ q(q+p)\mathbb{E}_{C|T=0}\text{Var}(Y_{0}|C)+p(p+q)\mathbb{E}_{C|T=1}\text{Var}(Y_{1}|C)\right\} \label{eq:consolidate-p1}\\
 &  & \hspace{1em}+pq\left[\mathbb{E}_{C|T=0}\left(\text{CATE}(C)^{2}\right)+\mathbb{E}_{C|T=1}\left(\text{CATE}(C)^{2}\right)\right]\label{eq:consolidate-p2}\\
 & = & \left\{ e_{\text{orig}}(f_{0})\right\} \label{eq:pq1}\\
 &  & \hspace{1em}+\text{Var}(T)\left[\mathbb{E}_{C|T=0}\left(\text{CATE}(C)^{2}\right)+\mathbb{E}_{C|T=1}\left(\text{CATE}(C)^{2}\right)\right].\label{eq:var-pq-def-e0}
\end{eqnarray}

In Lines \ref{eq:consolidate-p1} and \ref{eq:consolidate-p2}, we consolidate terms involving $\mathbb{E}_{C|T=0}\text{Var}(Y_{0}|C)$ and $\mathbb{E}_{C|T=1}\text{Var}(Y_{1}|C)$. In Line \ref{eq:pq1}, we use $p+q=1$ to reduce Line \ref{eq:consolidate-p1} to the right-hand side of Eq \ref{eq:def-e0}. Finally, in Line \ref{eq:var-pq-def-e0}, we use $qp=\text{Var}(T)$. Dividing both sides by $e_{\text{orig}}(f_{0})=\mathbb{E}_{T,C}Var(Y|T,C)$ gives the desired result.
\end{proof}

\section{\textcolor{black}{\label{subsec:Proofs-for-binary}Proofs for Computational Results}}

\textcolor{black}{Almost all of the proofs in this section are unchanged if we replace $\hat{e}_{\text{switch}}(f)$ with $\hat{e}_{\text{divide}}(f)$ in our definitions of $\hat{h}_{-,\gamma}$, $\hat{h}_{+,\gamma}$, $\hat{g}_{-,\gamma}$, $\hat{g}_{+,\gamma}$, and $\widehat{MR}$. The only exception is in Appendix \ref{subsec:Proof-of-Remark-convexity}.}\textcolor{blue}{} 

Throughout the following proofs, we will make use of the fact that, for constants $a,b,c,d\in\mathbb{R}$ satisfying $a\geq c$, the relation $a+b\leq c+d$ implies
\begin{align}
a+b & \leq c+d\nonumber \\
a-c & \leq d-b\nonumber \\
0 & \leq d-b & \text{since }0\leq a-c\nonumber \\
b & \leq d.\label{eq:abcd}
\end{align}

We also make use of the fact that for any $\gamma_{1},\gamma_{2}\in\mathbb{R}$, the definitions of $\hat{g}_{+,\gamma_{1}}$ and $\hat{g}_{-,\gamma_{1}}$ imply 
\begin{equation}
\hat{h}_{+,\gamma_{1}}(\hat{g}_{+,\gamma_{1}})\leq\hat{h}_{+,\gamma_{1}}(\hat{g}_{+,\gamma_{2}}),\hspace{1em}\text{and}\hspace{1em}\hat{h}_{-,\gamma_{1}}(g_{-,\gamma_{1}})\leq\hat{h}_{-,\gamma_{1}}(g_{-,\gamma_{2}}).\label{eq:hg-minimized-at-gg}
\end{equation}

Finally, for any two values $\gamma_{1},\gamma_{2}\in\mathbb{R}$, we make use of the fact that
\begin{align}
\hat{h}_{+,\gamma_{1}}(f) & =\hat{e}_{\text{orig}}(f)+\gamma_{1}\hat{e}_{\text{switch}}(f)\nonumber \\
 & =\hat{e}_{\text{orig}}(f)+\gamma_{2}\hat{e}_{\text{switch}}(f)+\{\gamma_{1}\hat{e}_{\text{switch}}(f)-\gamma_{2}\hat{e}_{\text{switch}}(f)\}\nonumber \\
 & =\hat{h}_{+,\gamma_{2}}(f)+(\gamma_{1}-\gamma_{2})\hat{e}_{\text{switch}}(f),\label{eq:expand-h+1}
\end{align}

and, by the same steps,
\begin{equation}
\hat{h}_{-,\gamma_{1}}(f)=\hat{h}_{-,\gamma_{2}}(f)+(\gamma_{1}-\gamma_{2})\hat{e}_{\text{orig}}(f).\label{eq:expand-h-1}
\end{equation}

\subsection{Proof of Lemma \ref{lem:mcr--bound-compute} (Lower Bound for MR)\label{subsec:Proof-of-lwr-bnd-mcr-}}
\begin{proof}
We prove Lemma \ref{lem:mcr--bound-compute} in 2 parts.

\subsubsection[Part 1]{Part 1: Showing Eq \ref{eq:lower-bnd-mcr--thm} Holds for All $f\in\mathcal{F}$ Satisfying $\hat{e}_{\text{orig}}(f)\le\epsilon_{\text{abs}}$.}

If $\hat{h}_{-,\gamma}(\hat{g}_{-,\gamma})\geq0$, then for any function $f\in\mathcal{F}$ satisfying $\hat{e}_{\text{orig}}(f)\leq\epsilon_{\text{abs}}$ we know that
\begin{align}
\frac{1}{\epsilon_{\text{abs}}} & \leq\frac{1}{\hat{e}_{\text{orig}}(f)}\nonumber \\
\frac{\hat{h}_{-,\gamma}(\hat{g}_{-,\gamma})}{\epsilon_{\text{abs}}} & \leq\frac{\hat{h}_{-,\gamma}(\hat{g}_{-,\gamma})}{\hat{e}_{\text{orig}}(f)}.\label{eq:h-leq-0-1}
\end{align}

Now, for any $f\in\mathcal{F}$ satisfying $\hat{e}_{\text{orig}}(f)\leq\epsilon_{\text{abs}}$, the definition of $\hat{g}_{-,\gamma}$ implies that
\begin{align*}
\hat{h}_{-,\gamma}(f) & \geq\hat{h}_{-,\gamma}\left(\hat{g}_{-,\gamma}\right)\\
\gamma\hat{e}_{\text{orig}}(f)+\hat{e}_{\text{switch}}(f) & \geq\hat{h}_{-,\gamma}\left(\hat{g}_{-,\gamma}\right)\\
\gamma+\frac{\hat{e}_{\text{switch}}(f)}{\hat{e}_{\text{orig}}(f)} & \geq\frac{\hat{h}_{-,\gamma}\left(\hat{g}_{-,\gamma}\right)}{\hat{e}_{\text{orig}}(f)}\\
\gamma+\frac{\hat{e}_{\text{switch}}(f)}{\hat{e}_{\text{orig}}(f)} & \geq\frac{\hat{h}_{-,\gamma}\left(\hat{g}_{-,\gamma}\right)}{\epsilon_{\text{abs}}} & \text{from Eq \ref{eq:h-leq-0-1}}\\
\widehat{MR}(f) & \geq\frac{\hat{h}_{-,\gamma}\left(\hat{g}_{-,\gamma}\right)}{\epsilon_{\text{abs}}}-\gamma.
\end{align*}

\subsubsection[Part 2]{Part 2: Showing that, if $f=\hat{g}_{-,\gamma}$, and at Least One of the Inequalities in Condition \ref{cond:mcr--bound} Holds with Equality, then Eq \ref{eq:lower-bnd-mcr--thm} Holds with Equality. }

We consider each of the two inequalities in Condition \ref{cond:mcr--bound} separately. If $\hat{h}_{-,\gamma}(\hat{g}_{-,\gamma})=0$, then
\begin{align*}
0 & =\gamma\hat{e}_{\text{orig}}(\hat{g}_{-,\gamma})+\hat{e}_{\text{switch}}(\hat{g}_{-,\gamma})\\
\frac{-\hat{e}_{\text{switch}}(\hat{g}_{-,\gamma})}{\hat{e}_{\text{orig}}(\hat{g}_{-,\gamma})} & =\gamma.
\end{align*}

As a result 

\[
\frac{\hat{h}_{-,\gamma}(\hat{g}_{-,\gamma})}{\epsilon_{\text{abs}}}-\gamma=\frac{0}{\epsilon_{\text{abs}}}-\left\{ \frac{-\hat{e}_{\text{switch}}(\hat{g}_{-,\gamma})}{\hat{e}_{\text{orig}}(\hat{g}_{-,\gamma})}\right\} =\widehat{MR}(\hat{g}_{-,\gamma}).
\]

Alternatively, if $\hat{e}_{\text{orig}}(\hat{g}_{-,\gamma})=\epsilon_{\text{abs}}$, then
\[
\frac{\hat{h}_{-,\gamma}(\hat{g}_{-,\gamma})}{\epsilon_{\text{abs}}}-\gamma=\frac{\gamma\hat{e}_{\text{orig}}(\hat{g}_{-,\gamma})+\hat{e}_{\text{switch}}(\hat{g}_{-,\gamma})}{\hat{e}_{\text{orig}}(\hat{g}_{-,\gamma})}-\gamma=\gamma+\frac{\hat{e}_{\text{switch}}(\hat{g}_{-,\gamma})}{\hat{e}_{\text{orig}}(\hat{g}_{-,\gamma})}-\gamma=\widehat{MR}(\hat{g}_{-,\gamma}).
\]
\end{proof}

\subsection{Proof of Lemma \ref{lem:mcr--compute-mono} (Monotonicity for MR Lower Bound Binary Search)\label{subsec:Proof-mono-mcr--}}
\begin{proof}
We prove Lemma \ref{lem:mcr--compute-mono} in 3 parts.

\subsubsection[Part \ref{enu:thm-h-mono}]{Part \ref{enu:thm-h-mono}: $\hat{h}_{-,\gamma}(\hat{g}_{-,\gamma})$ is Monotonically Increasing in $\gamma$.}

Let $\gamma_{1},\gamma_{2}\in\mathbb{R}$ satisfy $\gamma_{1}<\gamma_{2}$\textcolor{black}{. We have assumed that $0<\hat{e}_{\text{orig}}(f)$ for any $f\in\mathcal{F}$}. Thus, for any $f\in\mathcal{F}$ we have
\begin{align}
\gamma_{1}\hat{e}_{\text{orig}}(f)+\hat{e}_{\text{switch}}(f) & <\gamma_{2}\hat{e}_{\text{orig}}(f)+\hat{e}_{\text{switch}}(f)\nonumber \\
\hat{h}_{-,\gamma_{1}}(f) & <\hat{h}_{-,\gamma_{2}}(f).\label{eq:h-mono-1}
\end{align}

Applying this, we have
\begin{align*}
\hat{h}_{-,\gamma_{1}}(\hat{g}_{-,\gamma_{1}}) & \leq\hat{h}_{-,\gamma_{1}}(\hat{g}_{-,\gamma_{2}}) &  & \text{ from Eq \ref{eq:hg-minimized-at-gg}}\\
 & \leq\hat{h}_{-,\gamma_{2}}(\hat{g}_{-,\gamma_{2}}) &  & \text{ from Eq \ref{eq:h-mono-1}}.
\end{align*}

This result is analogous to Lemma 3 from \citet{dinkelbach1967nonlinear_FP}.

\subsubsection[Part \ref{enu:thm-e0--1}]{Part \ref{enu:thm-e0--1}: $\hat{e}_{\text{orig}}(\hat{g}_{-,\gamma})$ is Monotonically Decreasing in $\gamma$.}

Let $\gamma_{1},\gamma_{2}\in\mathbb{R}$ satisfy $\gamma_{1}<\gamma_{2}$. Then
\begin{align*}
\hat{h}_{-,\gamma_{1}}(\hat{g}_{-,\gamma_{1}}) & \le\hat{h}_{-,\gamma_{1}}(\hat{g}_{-,\gamma_{2}}) &  & \text{from Eq \ref{eq:hg-minimized-at-gg}}\\
\hat{h}_{-,\gamma_{2}}(\hat{g}_{-,\gamma_{1}})+(\gamma_{1}-\gamma_{2})\hat{e}_{\text{orig}}(\hat{g}_{-,\gamma_{1}}) & \le\hat{h}_{-,\gamma_{2}}(\hat{g}_{-,\gamma_{2}})+(\gamma_{1}-\gamma_{2})\hat{e}_{\text{orig}}(\hat{g}_{-,\gamma_{2}}) &  & \text{from Eq \ref{eq:expand-h-1}}\\
(\gamma_{1}-\gamma_{2})\hat{e}_{\text{orig}}(\hat{g}_{-,\gamma_{1}}) & \le(\gamma_{1}-\gamma_{2})\hat{e}_{\text{orig}}(\hat{g}_{-,\gamma_{2}}) &  & \text{from Eqs \ref{eq:abcd} \& \ref{eq:hg-minimized-at-gg}}\\
\hat{e}_{\text{orig}}(\hat{g}_{-,\gamma_{1}}) & \geq\hat{e}_{\text{orig}}(\hat{g}_{-,\gamma_{2}}).
\end{align*}

\subsubsection[Part \ref{enu:thm-lowerbnd-mono}]{Part \ref{enu:thm-lowerbnd-mono}: $\left\{ \frac{\hat{h}_{-,\gamma}(\hat{g}_{-,\gamma})}{\epsilon_{\text{abs}}}-\gamma\right\} $ is Monotonically Decreasing in $\gamma$ in the Range Where $\hat{e}_{\text{orig}}(\hat{g}_{-,\gamma})\leq\epsilon_{\text{abs}}$ , and Increasing Otherwise.}

Suppose $\gamma_{1}<\gamma_{2}$ and $\hat{e}_{\text{orig}}(\hat{g}_{-,\gamma_{1}}),\hat{e}_{\text{orig}}(\hat{g}_{-,\gamma_{2}})\leq\epsilon_{\text{abs}}$. Then, from Eq \ref{eq:hg-minimized-at-gg},
\begin{align*}
\hat{h}_{-,\gamma_{2}}(\hat{g}_{-,\gamma_{2}}) & \leq\hat{h}_{-,\gamma_{2}}(\hat{g}_{-,\gamma_{1}})\\
\hat{h}_{-,\gamma_{2}}(\hat{g}_{-,\gamma_{2}}) & \leq\hat{h}_{-,\gamma_{1}}(\hat{g}_{-,\gamma_{1}})+(\gamma_{2}-\gamma_{1})\hat{e}_{\text{orig}}(\hat{g}_{-,\gamma_{1}}) &  & \text{from Eq \ref{eq:expand-h-1}}\\
\hat{h}_{-,\gamma_{2}}(\hat{g}_{-,\gamma_{2}}) & \leq\hat{h}_{-,\gamma_{1}}(\hat{g}_{-,\gamma_{1}})+(\gamma_{2}-\gamma_{1})\epsilon_{\text{abs}} &  & \text{from }\hat{e}_{\text{orig}}(\hat{g}_{-,\gamma_{1}})\leq\epsilon_{\text{abs}}\\
\frac{\hat{h}_{-,\gamma_{2}}(\hat{g}_{-,\gamma_{2}})}{\epsilon_{\text{abs}}}-\gamma_{2} & \leq\frac{\hat{h}_{-,\gamma_{1}}(\hat{g}_{-,\gamma_{1}})}{\epsilon_{\text{abs}}}-\gamma_{1}.
\end{align*}

Similarly, if $\gamma_{1}<\gamma_{2}$ and $\hat{e}_{\text{orig}}(\hat{g}_{-,\gamma_{1}}),\hat{e}_{\text{orig}}(\hat{g}_{-,\gamma_{2}})\geq\epsilon_{\text{abs}}$. Then, from Eq \ref{eq:hg-minimized-at-gg}
\begin{align*}
\hat{h}_{-,\gamma_{1}}(\hat{g}_{-,\gamma_{1}}) & \leq\hat{h}_{-,\gamma_{1}}(\hat{g}_{-,\gamma_{2}})\\
\hat{h}_{-,\gamma_{1}}(\hat{g}_{-,\gamma_{1}}) & \leq\hat{h}_{-,\gamma_{2}}(\hat{g}_{-,\gamma_{2}})+(\gamma_{1}-\gamma_{2})\hat{e}_{\text{orig}}(\hat{g}_{-,\gamma_{2}}) &  & \text{from Eq \ref{eq:expand-h-1}}\\
\hat{h}_{-,\gamma_{1}}(\hat{g}_{-,\gamma_{1}}) & \leq\hat{h}_{-,\gamma_{2}}(\hat{g}_{-,\gamma_{2}})+(\gamma_{1}-\gamma_{2})\epsilon_{\text{abs}} &  & \text{from }\hat{e}_{\text{orig}}(\hat{g}_{-,\gamma_{1}})\geq\epsilon_{\text{abs}}\\
\frac{\hat{h}_{-,\gamma_{1}}(\hat{g}_{-,\gamma_{1}})}{\epsilon_{\text{abs}}}-\gamma_{1} & \leq\frac{\hat{h}_{-,\gamma_{2}}(\hat{g}_{-,\gamma_{2}})}{\epsilon_{\text{abs}}}-\gamma_{2}.
\end{align*}
\end{proof}

\subsection{Proof of Proposition \ref{rem:(Convexity-for-} (Nonnegative Weights for MR Lower Bound Binary Search)\label{subsec:Proof-of-Remark-convexity}}

\begin{proof}
Let $\gamma_{1}:=\frac{1}{n-1}$. First we show that there exists a function $\hat{g}_{-,\gamma_{1}}$ minimizing $\hat{h}_{-,\gamma_{1}}$ such that $\widehat{MR}(\hat{g}_{-,\gamma_{1}})=1$. Let $D_{s}$ denote the sample distribution of the data, and let $D_{m}$ be the distribution satisfying
\begin{align*}
\mathbb{P}_{D_{m}}\{(Y,X_{1},X_{2})=(y,x_{1},x_{2})\} & =\mathbb{P}_{D_{s}}\{(Y,X_{2})=(y,x_{2})\}\times\mathbb{P}_{D_{s}}(X_{1}=x_{1})\\
 & =\frac{1}{n^{2}}\sum_{i=1}^{n}1(\mathbf{y}_{[i]}=y\text{ and }\mathbf{X}_{2[i]}=x_{2})\sum_{j=1}^{n}1(\mathbf{X}_{1[j]}=x_{1}).
\end{align*}

From $\gamma_{1}=\frac{1}{n-1}$ and Eq \ref{eq:h-expand-weighted}, we see that 
\begin{align*}
\hat{h}_{-,\gamma_{1}}(f) & =\sum_{i=1}^{n}\sum_{j=1}^{n}L\{f,(\mathbf{y}_{[i]},\mathbf{X}_{1[j]},\mathbf{X}_{2[i]})\}\times\left\{ \frac{\gamma_{1}1(i=j)}{n}+\frac{1(i\neq j)}{n(n-1)}\right\} \\
 & =\sum_{i=1}^{n}\sum_{j=1}^{n}L\{f,(\mathbf{y}_{[i]},\mathbf{X}_{1[j]},\mathbf{X}_{2[i]})\}\times\left\{ \frac{1}{n(n-1)}\right\} .\\
 & \propto\mathbb{E}_{D_{m}}L\{f,(Y,X_{1},X_{2})\}.
\end{align*}

Thus, from Condition \ref{cond:independence} of Proposition \ref{rem:(Convexity-for-}, we know there exists a function $\hat{g}_{-,\gamma_{1}}$ that minimizes $\hat{h}_{-,\gamma_{1}}$ with $\hat{g}_{-,\gamma_{1}}(x_{1}^{(a)},x_{2})=\hat{g}_{-,\gamma_{1}}(x_{1}^{(b)},x_{2})$ for any $x_{1}^{(a)},x_{1}^{(b)}\in\mathcal{X}_{1}$ and $x_{2}\in\mathcal{X}_{2}$. Condition \ref{cond:(Predictions-are-sufficient} of Proposition \ref{rem:(Convexity-for-} then implies that $L\{\hat{g}_{-,\gamma_{1}},(y,x_{1}^{(a)},x_{2})\}=L\{\hat{g}_{-,\gamma_{1}},(y,x_{1}^{(b)},x_{2})\}$ for any $x_{1}^{(a)},x_{1}^{(b)}\in\mathcal{X}_{1}$, $x_{2}\in\mathcal{X}_{2}$, and $y\in\mathcal{Y}$. We apply this result in Line \ref{eq:apply-same-loss}, below, to show that loss of model $\hat{g}_{-,\gamma_{1}}$ is unaffected by permuting $X_{1}$ within our sample:
\begin{align}
\hat{e}_{\text{switch}}(\hat{g}_{-,\gamma_{1}}) & =\frac{1}{n}\sum_{i=1}^{n}\frac{1}{n-1}\sum_{j\neq i}L\{\hat{g}_{-,\gamma_{1}},(\mathbf{y}_{[i]},\mathbf{X}_{1[j]},\mathbf{X}_{2[i]})\}\nonumber \\
 & =\frac{1}{n}\sum_{i=1}^{n}\frac{1}{n-1}\sum_{j\neq i}L\{\hat{g}_{-,\gamma_{1}},(\mathbf{y}_{[i]},\mathbf{X}_{1\mred{[i]}},\mathbf{X}_{2[i]})\}\label{eq:apply-same-loss}\\
 & =\frac{1}{n}\sum_{i=1}^{n}L\{\hat{g}_{-,\gamma_{1}},(\mathbf{y}_{[i]},\mathbf{X}_{1[i]},\mathbf{X}_{2[i]})\}\nonumber \\
 & =\hat{e}_{\text{orig}}(\hat{g}_{-,\gamma_{1}}).\nonumber 
\end{align}

It follows that $\widehat{MR}(\hat{g}_{-,\gamma_{1}})=1$. To show the result of Proposition \ref{rem:(Convexity-for-}, let $\gamma_{2}=0$. For any function $\hat{g}_{-,\gamma_{2}}$ minimizing $\hat{h}_{-,\gamma_{2}}$, we know that
\begin{align}
\hat{h}_{-,\gamma_{2}}(\hat{g}_{-,\gamma_{2}}) & \leq\hat{h}_{-,\gamma_{2}}(\hat{g}_{-,\gamma_{1}}) &  & \text{from the definition of }\hat{g}_{-,\gamma_{2}}\nonumber \\
0+\hat{e}_{\text{switch}}(\hat{g}_{-,\gamma_{2}}) & \leq0+\hat{e}_{\text{switch}}(\hat{g}_{-,\gamma_{1}}) &  & \text{from }\gamma_{2}=0\text{ and the definition of }\hat{h}_{-,\gamma_{2}}.\label{eq:gg2-leq-perm}
\end{align}

From $\gamma_{2}\leq\gamma_{1}$, and Part \ref{enu:thm-e0--1} of Lemma \ref{lem:mcr--compute-mono}, we know that
\begin{equation}
\hat{e}_{\text{orig}}(\hat{g}_{-,\gamma_{2}})\geq\hat{e}_{\text{orig}}(\hat{g}_{-,\gamma_{1}}).\label{eq:gg2-leq-stnd}
\end{equation}

Combining Eqs \ref{eq:gg2-leq-perm} and \ref{eq:gg2-leq-stnd}, we have 
\begin{equation}
\widehat{MR}(\hat{g}_{-,\gamma_{2}})=\frac{\hat{e}_{\text{switch}}(\hat{g}_{-,\gamma_{2}})}{\hat{e}_{\text{orig}}(\hat{g}_{-,\gamma_{2}})}\leq\frac{\hat{e}_{\text{switch}}(\hat{g}_{-,\gamma_{1}})}{\hat{e}_{\text{orig}}(\hat{g}_{-,\gamma_{1}})}=\widehat{MR}(\hat{g}_{-,\gamma_{1}})=1.\label{eq:mrg2-leq-1}
\end{equation}

Since $\hat{h}_{-,\gamma_{2}}(\hat{g}_{-,\gamma_{2}})=\hat{e}_{\text{switch}}(\hat{g}_{-,\gamma_{2}})\geq0$ by definition, Condition \ref{cond:mcr--bound} holds for $\gamma_{2}$, $\epsilon_{\text{abs}}$ and $\hat{g}_{-,\gamma_{2}}$ if and only if $\hat{e}_{\text{orig}}(\hat{g}_{-,\gamma_{2}})\leq\epsilon_{\text{abs}}$. This, combined with Eq \ref{eq:mrg2-leq-1}, completes the proof.

The same result does not necessarily hold if we replace $\hat{e}_{\text{switch}}$ with $\hat{e}_{\text{divide}}$ in our definitions of $\hat{h}_{-,\gamma}$, $\widehat{MR}$, and $\widehat{MCR}_{-}$. This is because $\hat{e}_{\text{divide}}$ does not correspond to the expectation over a distribution in which $X_{1}$ is independent of $X_{2}$ and $Y$, due to the fixed pairing structure used in $\hat{e}_{\text{divide}}$. Thus, Condition \ref{cond:independence} of Proposition \ref{rem:(Convexity-for-} will not apply.
\end{proof}

\subsection{Proof of Lemma \ref{lem:mcr+-compute-bound} (Upper Bound for MR)}
\begin{proof}
We prove Lemma \ref{lem:mcr+-compute-bound} in 2 parts.

\subsubsection[Part 1]{Part 1: Showing Eq \ref{eq:thm-bnd-mcr+} Holds for All $f\in\mathcal{F}$ Satisfying $\hat{e}_{\text{orig}}(f)\le\epsilon_{\text{abs}}$.}

If $\hat{h}_{+,\gamma}(\hat{g}_{+,\gamma})\geq0$, then for any function $f\in\mathcal{F}$ satisfying $\hat{e}_{\text{orig}}(f)\leq\epsilon_{\text{abs}}$ we know that 
\begin{align}
\frac{1}{\epsilon_{\text{abs}}} & \leq\frac{1}{\hat{e}_{\text{orig}}(f)}\nonumber \\
\frac{\hat{h}_{+,\gamma}(\hat{g}_{+,\gamma})}{\epsilon_{\text{abs}}} & \leq\frac{\hat{h}_{+,\gamma}(\hat{g}_{+,\gamma})}{\hat{e}_{\text{orig}}(f)}.\label{eq:h-leq-0}
\end{align}

Now, if $\gamma\le0$, then for any $f\in\mathcal{F}$ satisfying $\hat{e}_{\text{orig}}(f)\leq\epsilon_{\text{abs}}$, the definition of $\hat{g}_{+,\gamma}$ implies
\begin{align*}
\hat{h}_{+,\gamma}(f) & \geq\hat{h}_{+,\gamma}\left(\hat{g}_{+,\gamma}\right)\\
\hat{e}_{\text{orig}}(f)+\gamma\hat{e}_{\text{switch}}(f) & \geq\hat{h}_{+,\gamma}\left(\hat{g}_{+,\gamma}\right)\\
1+\gamma\frac{\hat{e}_{\text{switch}}(f)}{\hat{e}_{\text{orig}}(f)} & \geq\frac{\hat{h}_{+,\gamma}\left(\hat{g}_{+,\gamma}\right)}{\hat{e}_{\text{orig}}(f)}\\
1+\gamma\frac{\hat{e}_{\text{switch}}(f)}{\hat{e}_{\text{orig}}(f)} & \geq\frac{\hat{h}_{+,\gamma}\left(\hat{g}_{+,\gamma}\right)}{\epsilon_{\text{abs}}} & \text{from Eq \ref{eq:h-leq-0}}\\
1+\gamma\widehat{MR}(f) & \geq\frac{\hat{h}_{+,\gamma}\left(\hat{g}_{+,\gamma}\right)}{\epsilon_{\text{abs}}}\\
\widehat{MR}(f) & \leq\left\{ \frac{\hat{h}_{+,\gamma}(\hat{g}_{+,\gamma})}{\epsilon_{\text{abs}}}-1\right\} \gamma^{-1}.
\end{align*}

\subsubsection[Part 2]{Part 2: Showing that if $f=\hat{g}_{+,\gamma}$, and at Least One of the Enequalities in Condition \ref{cond:mcr+-bound} Holds with Equality, then Eq \ref{eq:thm-bnd-mcr+} Holds with Equality.}

We consider each of the two inequalities in Condition \ref{cond:mcr+-bound} separately. If $\hat{h}_{+,\gamma}(\hat{g}_{+,\gamma})=0$, then
\begin{align*}
0 & =\hat{e}_{\text{orig}}(\hat{g}_{+,\gamma})+\gamma\hat{e}_{\text{switch}}(\hat{g}_{+,\gamma})\\
-\gamma\hat{e}_{\text{switch}}(\hat{g}_{+,\gamma}) & =\hat{e}_{\text{orig}}(\hat{g}_{+,\gamma})\\
-\frac{\hat{e}_{\text{switch}}(\hat{g}_{+,\gamma})}{\hat{e}_{\text{orig}}(\hat{g}_{+,\gamma})} & =\frac{1}{\gamma}.
\end{align*}

As a result,
\begin{align*}
\left\{ \frac{\hat{h}_{+,\gamma}(\hat{g}_{+,\gamma})}{\epsilon_{\text{abs}}}-1\right\} \gamma^{-1} & =\left\{ \frac{0}{\epsilon_{\text{abs}}}-1\right\} \left\{ -\frac{\hat{e}_{\text{switch}}(\hat{g}_{+,\gamma})}{\hat{e}_{\text{orig}}(\hat{g}_{+,\gamma})}\right\} =\widehat{MR}(\hat{g}_{+,\gamma}).
\end{align*}

Alternatively, if $\hat{e}_{\text{orig}}(\hat{g}_{+,\gamma})=\epsilon_{\text{abs}}$, then
\begin{align*}
\left\{ \frac{\hat{h}_{+,\gamma}(\hat{g}_{+,\gamma})}{\epsilon_{\text{abs}}}-1\right\} \gamma^{-1}=\left\{ \frac{\hat{e}_{\text{orig}}(\hat{g}_{+,\gamma})+\gamma\hat{e}_{\text{switch}}(\hat{g}_{+,\gamma})}{\hat{e}_{\text{orig}}(\hat{g}_{+,\gamma})}-1\right\} \gamma^{-1} & =\left\{ 1+\gamma\frac{\hat{e}_{\text{switch}}(\hat{g}_{+,\gamma})}{\hat{e}_{\text{orig}}(\hat{g}_{+,\gamma})}-1\right\} \gamma^{-1}\\
 & =\widehat{MR}(\hat{g}_{+,\gamma}).
\end{align*}
\end{proof}

\subsection{Proof of Lemma \ref{lem:mcr+-compute-mono} (Monotonicity for MR Upper Bound Binary Search)}
\begin{proof}
We prove Lemma \ref{lem:mcr+-compute-mono} in 3 parts.

\subsubsection[Part \ref{enu:thm-h+-mono}]{Part \ref{enu:thm-h+-mono}: $\hat{h}_{+,\gamma}(\hat{g}_{+,\gamma})$ is Monotonically Increasing in $\gamma$.}

Let $\gamma_{1},\gamma_{2}\in\mathbb{R}$ satisfy $\gamma_{1}<\gamma_{2}$. We have assumed that $0\leq\hat{e}_{\text{switch}}(f)$ for any $f\in\mathcal{F}$. Thus, for any $f\in\mathcal{F}$ we have
\begin{align}
\hat{e}_{\text{orig}}(f)+\gamma_{1}\hat{e}_{\text{switch}}(f) & <\hat{e}_{\text{orig}}(f)+\gamma_{2}\hat{e}_{\text{switch}}(f)\nonumber \\
\hat{h}_{+,\gamma_{1}}(f) & <\hat{h}_{+,\gamma_{2}}(f).\label{eq:h+mono}
\end{align}

Applying this, we have
\begin{align*}
\hat{h}_{+,\gamma_{1}}(\hat{g}_{+,\gamma_{1}}) & \leq\hat{h}_{+,\gamma_{1}}(\hat{g}_{+,\gamma_{2}}) &  & \text{ from Eq \ref{eq:hg-minimized-at-gg}}\\
 & <\hat{h}_{+,\gamma_{2}}(\hat{g}_{+,\gamma_{2}}) &  & \text{ from Eq \ref{eq:h+mono}}.
\end{align*}

\subsubsection[Part \ref{enu:thm-e0-+1}]{Part \ref{enu:thm-e0-+1}: $\hat{e}_{\text{orig}}(\hat{g}_{+,\gamma})$ is Monotonically Decreasing in $\gamma$ for $\gamma\leq0$, and Condition \ref{cond:mcr+-bound} Holds for $\gamma=0$ and $\epsilon_{\text{abs}}\geq\min_{f\in\mathcal{F}}\hat{e}_{\text{orig}}(f)$.}

Let $\gamma_{1},\gamma_{2}\in\mathbb{R}$ satisfy $\gamma_{1}<\gamma_{2}\leq0$. Then
\begin{align}
\hat{h}_{+,\gamma_{1}}(\hat{g}_{+,\gamma_{1}}) & \le\hat{h}_{+,\gamma_{1}}(\hat{g}_{+,\gamma_{2}}) & \text{from Eq \ref{eq:hg-minimized-at-gg}}\nonumber \\
\hat{h}_{+,\gamma_{2}}(\hat{g}_{+,\gamma_{1}})+(\gamma_{1}-\gamma_{2})\hat{e}_{\text{switch}}(\hat{g}_{+,\gamma_{1}}) & \le\hat{h}_{+,\gamma_{2}}(\hat{g}_{+,\gamma_{2}})+(\gamma_{1}-\gamma_{2})\hat{e}_{\text{switch}}(g_{\gamma_{2}}) & \text{from Eq \ref{eq:expand-h+1}}\nonumber \\
(\gamma_{1}-\gamma_{2})\hat{e}_{\text{switch}}(\hat{g}_{+,\gamma_{1}}) & \le(\gamma_{1}-\gamma_{2})\hat{e}_{\text{switch}}(\hat{g}_{+,\gamma_{2}}) & \text{from Eqs \ref{eq:abcd} \& \ref{eq:hg-minimized-at-gg}}\nonumber \\
\hat{e}_{\text{switch}}(\hat{g}_{+,\gamma_{1}}) & \geq\hat{e}_{\text{switch}}(\hat{g}_{+,\gamma_{2}})\nonumber \\
\gamma_{2}\hat{e}_{\text{switch}}(\hat{g}_{+,\gamma_{1}}) & \leq\gamma_{2}\hat{e}_{\text{switch}}(\hat{g}_{+,\gamma_{2}}) & \text{from }\gamma_{2}\leq0.\label{eq:e1-g1-monotonic-gamma}
\end{align}

Now we are equipped to show the result that \textcolor{black}{$\hat{e}_{\text{orig}}(\hat{g}_{+,\gamma})$ is monotonically decreasing in $\gamma$ for $\gamma\leq0$}:
\begin{align}
\hat{h}_{+,\gamma_{2}}(\hat{g}_{+,\gamma_{2}}) & \le\hat{h}_{+,\gamma_{2}}(\hat{g}_{+,\gamma_{1}}) & \text{from Eq \ref{eq:hg-minimized-at-gg}}\nonumber \\
\hat{e}_{\text{orig}}(\hat{g}_{+,\gamma_{2}})+\gamma_{2}\hat{e}_{\text{switch}}(\hat{g}_{+,\gamma_{2}}) & \leq\hat{e}_{\text{orig}}(\hat{g}_{+,\gamma_{1}})+\gamma_{2}\hat{e}_{\text{switch}}(\hat{g}_{+,\gamma_{1}})\nonumber \\
\hat{e}_{\text{orig}}(\hat{g}_{+,\gamma_{2}}) & \leq\hat{e}_{\text{orig}}(\hat{g}_{+,\gamma_{1}}) & \text{from Eqs \ref{eq:abcd} \& \ref{eq:e1-g1-monotonic-gamma}}.\label{eq:finish-mono-e0p1}
\end{align}

To show that \textcolor{black}{Condition \ref{cond:mcr+-bound} holds for $\gamma=0$ and $\min_{f\in\mathcal{F}}\hat{e}_{\text{orig}}(f)\leq\epsilon_{\text{abs}}$, we first note that $h_{0,+}(g_{0,+})=\hat{e}_{\text{orig}}(g_{0,+})$, which is positive by assumption. Second, we note that}
\begin{align*}
\hat{e}_{\text{orig}}(g_{0,+})=h_{0,+}(g_{0,+})=\min_{f\in\mathcal{F}}h_{0,+}(f) & =\min_{f\in\mathcal{F}}\hat{e}_{\text{orig}}(f)\leq\epsilon_{\text{abs}}.
\end{align*}

\subsubsection[Part \ref{enu:thm-upperbnd-mono}]{Part \ref{enu:thm-upperbnd-mono}: $\left\{ \frac{\hat{h}_{+,\gamma}(\hat{g}_{+,\gamma})}{\epsilon_{\text{abs}}}-1\right\} \gamma^{-1}$ is Monotonically Increasing in $\gamma$ in the Range Where $\hat{e}_{\text{orig}}(\hat{g}_{+,\gamma})\leq\epsilon_{\text{abs}}$ and $\gamma<0$, and Decreasing in the Range Where $\hat{e}_{\text{orig}}(\hat{g}_{+,\gamma})>\epsilon_{\text{abs}}$ and $\gamma<0$.}

\textcolor{black}{To prove the first result, suppose that $\gamma_{1}<\gamma_{2}<0$ and $\hat{e}_{\text{orig}}(\hat{g}_{+,\gamma_{1}}),\hat{e}_{\text{orig}}(\hat{g}_{+,\gamma_{2}})\leq\epsilon_{\text{abs}}$. This implies}
\begin{align}
\frac{1}{\gamma_{2}} & <\frac{1}{\gamma_{1}}\nonumber \\
\frac{\hat{e}_{\text{orig}}(\hat{g}_{+,\gamma_{1}})-\epsilon_{\text{abs}}}{\gamma_{2}} & >\frac{\hat{e}_{\text{orig}}(\hat{g}_{+,\gamma_{1}})-\epsilon_{\text{abs}}}{\gamma_{1}}.\label{eq:inv-gamma-e-neg}
\end{align}

\textcolor{black}{Then, starting with Eq \ref{eq:hg-minimized-at-gg},
\begin{align*}
\hat{h}_{+,\gamma_{2}}(\hat{g}_{+,\gamma_{2}}) & \leq\hat{h}_{+,\gamma_{2}}(\hat{g}_{+,\gamma_{1}})\\
\hat{h}_{+,\gamma_{2}}(\hat{g}_{+,\gamma_{2}}) & \leq\gamma_{2}\hat{e}_{\text{switch}}(\hat{g}_{+,\gamma_{1}})+\hat{e}_{\text{orig}}(\hat{g}_{+,\gamma_{1}})\\
\frac{\hat{h}_{+,\gamma_{2}}(\hat{g}_{+,\gamma_{2}})-\epsilon_{\text{abs}}}{\gamma_{2}} & \geq\hat{e}_{\text{switch}}(\hat{g}_{+,\gamma_{1}})+\frac{\hat{e}_{\text{orig}}(\hat{g}_{+,\gamma_{1}})-\epsilon_{\text{abs}}}{\gamma_{2}}\\
 & \geq\hat{e}_{\text{switch}}(\hat{g}_{+,\gamma_{1}})+\frac{\hat{e}_{\text{orig}}(\hat{g}_{+,\gamma_{1}})-\epsilon_{\text{abs}}}{\mred{\gamma_{1}}} & \text{from Eq \ref{eq:inv-gamma-e-neg}}\\
 & =\frac{\hat{h}_{+,\gamma_{1}}(\hat{g}_{+,\gamma_{1}})-\epsilon_{\text{abs}}}{\gamma_{1}}.
\end{align*}
}

\textcolor{black}{Dividing both sides of the above equation by $\epsilon_{\text{abs}}$ proves that }\textbf{ $\left\{ \frac{\hat{h}_{+,\gamma}(\hat{g}_{+,\gamma})}{\epsilon_{\text{abs}}}-1\right\} \gamma^{-1}$} is monotonically increasing in $\gamma$ in the range where $\hat{e}_{\text{orig}}(\hat{g}_{+,\gamma})\leq\epsilon_{\text{abs}}$ and $\gamma<0$.

\textcolor{black}{To prove the second result we proceed in the same way. Suppose that $\gamma_{1}<\gamma_{2}<0$ and $\hat{e}_{\text{orig}}(\hat{g}_{+,\gamma_{1}}),\hat{e}_{\text{orig}}(\hat{g}_{+,\gamma_{2}})\geq\epsilon_{\text{abs}}$. This implies}
\begin{align}
\frac{1}{\gamma_{2}} & <\frac{1}{\gamma_{1}}\nonumber \\
\frac{\hat{e}_{\text{orig}}(\hat{g}_{+,\gamma_{2}})-\epsilon_{\text{abs}}}{\gamma_{2}} & <\frac{\hat{e}_{\text{orig}}(\hat{g}_{+,\gamma_{2}})-\epsilon_{\text{abs}}}{\gamma_{1}}.\label{eq:gamma-inv-e-pos}
\end{align}

\textcolor{black}{Then, starting with Eq \ref{eq:hg-minimized-at-gg},
\begin{align*}
\hat{h}_{+,\gamma_{1}}(\hat{g}_{+,\gamma_{1}}) & \leq\hat{h}_{+,\gamma_{1}}(\hat{g}_{+,\gamma_{2}})\\
\hat{h}_{+,\gamma_{1}}(\hat{g}_{+,\gamma_{1}}) & \leq\gamma_{1}\hat{e}_{\text{switch}}(\hat{g}_{+,\gamma_{2}})+\hat{e}_{\text{orig}}(\hat{g}_{+,\gamma_{2}})\\
\frac{\hat{h}_{+,\gamma_{1}}(\hat{g}_{+,\gamma_{1}})-\epsilon_{\text{abs}}}{\gamma_{1}} & \geq\hat{e}_{\text{switch}}(\hat{g}_{+,\gamma_{2}})+\frac{\hat{e}_{\text{orig}}(\hat{g}_{+,\gamma_{2}})-\epsilon_{\text{abs}}}{\gamma_{1}}\\
 & \geq\hat{e}_{\text{switch}}(\hat{g}_{+,\gamma_{2}})+\frac{\hat{e}_{\text{orig}}(\hat{g}_{+,\gamma_{2}})-\epsilon_{\text{abs}}}{\mred{\gamma_{2}}} & \text{from Eq \ref{eq:gamma-inv-e-pos}}\\
 & =\frac{\hat{h}_{+,\gamma_{1}}(\hat{g}_{+,\gamma_{2}})-\epsilon_{\text{abs}}}{\gamma_{2}}.
\end{align*}
}

\textcolor{black}{Diving both sides of the above equation by $\epsilon_{\text{abs}}$ proves that }$\left[\left\{ \frac{\hat{h}_{+,\gamma}(\hat{g}_{+,\gamma})}{\epsilon_{\text{abs}}}-1\right\} \gamma^{-1}\right]$ is monotonically decreasing in $\gamma$ in the range where $\hat{e}_{\text{orig}}(\hat{g}_{+,\gamma})>\epsilon_{\text{abs}}$ and $\gamma<0$.
\end{proof}

\subsection{Proof of Remark \ref{rem:For-any-LM-comp} (Tractability of Empirical MCR for Linear Model Classes)}
\begin{proof}
To show Remark \ref{rem:For-any-LM-comp}, we apply Proposition \ref{thm:Linear-models} to see that
\begin{align*}
 & \xi_{\text{orig}}\hat{e}_{\text{orig}}(f_{\beta})+\xi_{\text{switch}}\hat{e}_{\text{switch}}(f_{\beta})\\
 & \hspace{1em}=\frac{\xi_{\text{orig}}}{n}\left|\left|\mathbf{y}-\mathbf{X}\beta\right|\right|_{2}^{2}+\xi_{\text{switch}}\hat{e}_{\text{switch}}(f_{\beta})\\
 & \hspace{1em}=\frac{\xi_{\text{orig}}}{n}\left(\mathbf{y}'\mathbf{y}-2\mathbf{y}'\mathbf{X}\beta+\beta'\mathbf{X}'\mathbf{X}\beta\right)\\
 & \hspace{2cm}+\frac{\xi_{\text{switch}}}{n}\left\{ \mathbf{y}'\mathbf{y}-2\left[\begin{array}{c}
\mathbf{X}_{1}'\mathbf{W}\mathbf{y}\\
\mathbf{X}_{2}'\mathbf{y}
\end{array}\right]^{'}\beta+\beta'\left[\begin{array}{cc}
\mathbf{X}_{1}'\mathbf{X}_{1} & \mathbf{X}_{1}'\mathbf{W}\mathbf{X}_{2}\\
\mathbf{X}_{2}'\mathbf{W}\mathbf{X}_{1} & \mathbf{X}_{2}'\mathbf{X}_{2}
\end{array}\right]\beta\right\} \\
 & \hspace{1em}\propto_{\beta}-2\mathbf{q}'\beta+\beta'\mathbf{Q}\beta.
\end{align*}
\end{proof}

\subsection{\label{subsec:Proof-upper-loss-linear}Proof of Lemma \ref{lem:-lm-get-BB} (Loss Upper Bound for Linear Models)}
\begin{proof}
Under the conditions in Lemma \ref{lem:-lm-get-BB} and Eq \ref{eq:lm-constr}, we can construct an upper bound on $L(f_{\beta},(y,x))=(y-x'\beta)^{2}$ by either maximizing or minimizing $x'\beta$. First, we consider the maximization problem
\begin{equation}
\max_{\beta,x\in\mathbb{R}^{p}}x'\beta\text{ subject to }x'\mathbf{M}_{\text{lm}}^{-1}x\leq r_{\mathcal{X}}\text{ and }\beta'\mathbf{M}_{\text{lm}}\beta\le r_{\text{lm}}.\label{eq:max-prob-lm}
\end{equation}

We can see that both constraints hold with equality at the solution to this problem. Next, we apply the change of variables $\tilde{x}=\frac{1}{\sqrt{r_{\mathcal{X}}}}\mathbf{D}^{\frac{-1}{2}}\mathbf{U}'x$ and $\tilde{\beta}=\frac{1}{\sqrt{r_{\text{lm}}}}\mathbf{D}^{\frac{1}{2}}\mathbf{U}'\beta$, where $\mathbf{U}\mathbf{D}\mathbf{U}'=\mathbf{M}_{\text{lm}}$ is the eigendecomposition of $\mathbf{M}_{\text{lm}}$. We obtain
\begin{align*}
 & \max_{\tilde{\beta},\tilde{x}\in\mathbb{R}^{p}}\tilde{x}'\tilde{\beta}\sqrt{r_{\mathcal{X}}r_{\text{lm}}}\text{ subject to }\tilde{x}'\tilde{x}=1\text{ and }\tilde{\beta}'\tilde{\beta}=1,
\end{align*}
\textcolor{black}{which has an optimal objective value equal to $\sqrt{r_{\mathcal{X}}r_{\text{lm}}}$. By negating the objective in Eq \ref{eq:max-prob-lm}, we see that the minimum possible value of $x'\beta$, subject to the constraints in }Eq \ref{eq:lm-constr} and Lemma \ref{lem:-lm-get-BB}\textcolor{black}{, is found at $-\sqrt{r_{\mathcal{X}}r_{\text{lm}}}$. Thus, we know that
\[
L(f,(y,x_{1},x_{2}))\leq\max\left[\left\{ \left(\min_{y\in\mathcal{Y}}y\right)-\sqrt{r_{\mathcal{X}}r_{\text{lm}}}\right\} ^{2},\left\{ \left(\max_{y\in\mathcal{Y}}y\right)+\sqrt{r_{\mathcal{X}}r_{\text{lm}}}\right\} ^{2}\right],
\]
}for any $(y,x_{1},x_{2})\in(\mathcal{Y}\times\mathcal{X}_{1}\times\mathcal{X}_{2})$.
\end{proof}

\subsection{Proof of Lemma \ref{lem:BB-kernel} (Loss Upper Bound for Regression in a RKHS)}

This proofs follows a similar structure as the proof in Section \ref{subsec:Proof-upper-loss-linear}. From the assumptions of Lemma \ref{lem:BB-kernel}, we know from Eq \ref{eq:f_k_norm_def} that the largest possible output from a model $f_{\alpha}\in\mathcal{F}_{\mathbf{D},r_{k}}$ is
\begin{align*}
 & \mu+\max_{x\in\mathbb{R}^{p},\alpha\in\mathbb{R}^{R}}\sum_{i=1}^{R}k(x,\mathbf{D}_{[i,\cdot]})\alpha_{[i]} &  & \text{ subject to }v(x)'\mathbf{K}_{\mathbf{D}}^{-1}v(x)\leq r_{\mathbf{D}}\text{ and }\alpha'\mathbf{K}_{\mathbf{D}}\alpha\leq r_{k}\\
= & \mu+\max_{x\in\mathbb{R}^{p},\alpha\in\mathbb{R}^{R}}v(x)'\alpha &  & \text{ subject to }v(x)'\mathbf{K}_{\mathbf{D}}^{-1}v(x)\leq r_{\mathbf{D}}\text{ and }\alpha'\mathbf{K}_{\mathbf{D}}\alpha\leq r_{k}\\
\leq & \mu+\max_{\mathbf{z},\alpha\in\mathbb{R}^{R}}\mathbf{z}'\alpha &  & \text{ subject to }\mathbf{z}'\mathbf{K}_{\mathbf{D}}^{-1}\mathbf{z}\leq r_{\mathbf{D}}\text{ and }\alpha'\mathbf{K}_{\mathbf{D}}\alpha\leq r_{k}.
\end{align*}

The above problem can be solved in the same way as Eq \ref{eq:max-prob-lm}, and has a solution at $\left(\mu+\sqrt{r_{\mathcal{\mathbf{D}}}r_{k}}\right)$. The smallest possible model output will similarly be lower bounded by $-\left(\mu+\sqrt{r_{\mathcal{\mathbf{D}}}r_{k}}\right)$. Thus, $B_{\text{ind}}$ is less than or equal to 

\textcolor{black}{
\[
\max\left[\left\{ \min_{y\in\mathcal{Y}}\left(y\right)-\left(\mu+\sqrt{r_{\mathcal{\mathbf{D}}}r_{k}}\right)\right\} ^{2},\left\{ \max_{y\in\mathcal{Y}}\left(y\right)+\left(\mu+\sqrt{r_{\mathcal{\mathbf{D}}}r_{k}}\right)\right\} ^{2}\right].
\]
}

\bibliography{overfit}

\end{document}